\documentclass[12pt, amssymb,amsmath,aps,showpacs,floatfix,nofootinbib,showpacs,balancelastpage]{revtex4}
\usepackage{hangcaption,fancybox,feynmp}
\usepackage{indentfirst}
\usepackage{graphicx}
\usepackage{epsfig,subfigure,psfrag}
\usepackage{mathrsfs}
\usepackage{color}

\begin{document}

\include{newcmd}
%%%%%%%%%%%%%%%%%%%%%%%%%%%%%%%%%%%

%\title{Lepton and bottom quark forward-backward asymmetry at the LHC}
\title{One-side forward-backward asymmetry at the LHC}
\author{You-kai Wang$^{1}$ \footnote{E-mail:wangyk@pku.edu.cn}, Bo Xiao$^{1}$
\footnote{E-mail:homenature@pku.edu.cn},
 and Shou-hua Zhu$^{1,2}$ \footnote{E-mail:shzhu@pku.edu.cn} }

\affiliation{ $ ^1$ Institute of Theoretical Physics $\&$ State Key
Laboratory of Nuclear Physics and Technology, Peking University,
Beijing 100871, China \\
$ ^2$ Center for High Energy Physics, Peking University, Beijing
100871, China }

\date{\today}

\begin{abstract}

Forward-backward asymmetry $A_{\rm FB}$ is an essential observable
to study the nature of coupling in the standard model and physics
beyond the standard model, as shown at LEP and Tevatron. As a
proton-proton collider, the LHC does not have the preferred
direction contrary to her counterpart, namely, LEP and Tevatron.
Therefore $A_{\rm FB}$ is not applicable at the LHC. However for the
proton the momentum of valence quark is usually larger than that of
the sea quark. Utilizing this feature we have defined a so-called
one-side forward-backward asymmetry $A_{\rm OFB}$ for the top quark
pair production at LHC in the previous work. In this paper we extend
our studies to
 the charged leptons and bottom quarks as the final states. Our numerical results show that at the LHC
 $A_{\rm OFB}$ can be utilized to study the nature of the couplings once enough events are collected.

\end{abstract}


\maketitle

\section{Introduction}

At high energy colliders  discovering a new particle is not enough,
one of the most important subsequent tasks is how to pin down its
properties, for example, spin, nature of the coupling, and so on.
Based on this information the internal quantum structure can be
scrutinized and the possible subtle deviation may be found.   In
practice once enough data sample is collected, the forward-backward
asymmetry ($A_{\rm FB}$) for a specific final state can be measured
and compared with theoretical prediction.

$A_{\rm FB}$, or sometimes called the charge asymmetry if $CP$
conservation is assumed, is an interesting experimental observable.
The primary definition of the $A_{\rm FB}$ is
\begin{equation}
A_{\rm FB}\equiv\frac{N(\cos\theta>0)-N(\cos\theta<0)}
{N(\cos\theta>0)+N(\cos\theta<0)}, \label{fba1}
\end{equation}
where $\theta$ is the polar angle between the final state particle
and the beam line. The polar angle in Eq.(\ref{fba1}) can be defined
in different frames, such as Collins-Soper frame for the lepton pair
production in the Drell-Yan processes, lab frame, and $t\bar{t}$
rest frame for top pair production process at Tevatron. In the
$t\bar{t}$ rest frame, Eq.(\ref{fba1}) can be transformed as
\begin{equation}
A_{\rm FB}=\frac{\sigma(\Delta Y>0)-\sigma(\Delta
Y<0)}{\sigma(\Delta Y>0)+\sigma(\Delta Y<0)}, \label{fba2}
\end{equation}
where $\Delta Y\equiv Y_{t}-Y_{\bar{t}}$ is the difference of
rapidity of the top and antitop quark, which is invariant under
$t\bar{t}$ or $p\bar{p}$ rest frame. Here the use of anti-top quark
information implies that $CP$ conservation of the top and antitop
quark is assumed.

In some sense $A_{\rm FB}$ is a measure to study the angular
distributions of the specific final particle. The distribution is
determined by the nature of the couplings among the initial and
final particles with the intermediate particle in a certain theory.
Currently the successful theory which can describe the data is the
standard model (SM). There are good reasons to expect physics beyond
the SM (BSM), which usually predict new particles and/or new
couplings. Such new particles and/or couplings can be firstly
detected via $A_{\rm FB}$ measurements, namely, the deviation from
the SM prediction. Therefore $A_{\rm FB}$ is a useful tool to test
SM and even to discover BSM.

Up to now, $A_{\rm FB}$ for many final particles, e.g., charged
leptons, bottom quark, and top quark have been measured at different
colliders, say SLD, LEP and Tevatron. Generally speaking, the
measurements are in excellent agreement with SM predictions. However
there are some anomaly  for the bottom quark at LEP and for the top
quark at Tevatron. The measurements and theoretical predictions are
listed in Table \ref{Afbtable}.

\begin{table}[htb]
\caption{\label{Afbtable}Measurements and theoretical predictions
(in bracket) of $A_{\rm FB}$ for bottom and top quark. Here $p\bar
p$ and $t\bar t $ represent measurements in the lab and the
center-of-mass frame of the top quark pair respectively. } \center
\small{
\begin{tabular}{ccc}\hline\hline
& Bottom & Top\\
LEP&$0.0992\pm0.0016 \ (0.10324\pm0.00088)$ \cite{2005ema}& $\cdots$
\\\hline Tevatron & $\cdots$ &
\begin{tabular}{ccc}
CDF $t\bar{t}$&$0.158\pm0.072\pm0.017$ &$(0.058\pm0.009)$ \cite{CDFAfb53}\\
CDF $p\bar{p}$&$0.150\pm0.050\pm0.024$ &$(0.038\pm0.006)$ \cite{CDFAfb53}\\
D0 $t\bar{t}$&$0.08\pm0.04\pm0.01 $&$(1^{+2}_{-1}\%)$ \cite{D0AFB43}\\
\end{tabular} \\
\hline\hline
\end{tabular}
}
\end{table}

Both $A_{\rm FB}$ measurements have a deviation about 2 $\sigma$
from SM predictions. It is obvious that only less than $3\sigma$
deviation is inadequate to conclude the failure of the SM. However
it is interesting to explore the implications of the deviations both
in the SM and the BSM
\cite{Almeida:2008ug,Ahrens:2010zv,Frampton:2009rk,Shu:2009xf,Chivukula:2010fk,
Jung:2009jz, Cheung:2009ch,
Cao:2010zb,Djouadi:2009nb,Jung:2009pi,Cao:2009uz,Barger:2010mw,
Arhrib:2009hu, Xiao:2010hm,Bauer:2010iq, Xiao:2010ph,
Dorsner:2009mq, Chen:2010hm}. The present experimental results still
have too large uncertainties to make a clear judgement. So the
cross-check of these measurements in the more powerful collider are
extremely necessary.

The large hadron collider (LHC) is the most hopeful machine to make
this cross-check and even discovery, because it has the larger
production rate and most importantly, the more powerful
reconstruction capacity of both the bottom and top quark.
Unfortunately, unlike the $e^+e^-$ collider, LEP, or $p\bar{p}$
collider, Tevatron, the $pp$ collider, LHC, does not have preferred
direction in the laboratory frame. The definition of  $A_{\rm FB}$
in Eqs.(\ref{fba1}) and (\ref{fba2}) are not applicable here.
Forward-backward asymmetry at $pp$ collider has already been
discussed in the literature\cite{Langacker:1984dc, Dittmar:1996my,
Petriello:2008zr, Li:2009xh, Diener:2009ee, Diener:2010sy}. $A_{\rm
FB}$ in these papers are mostly used for exploiting a possible
massive $Z'$ boson. For example, $A_{\rm FB}$ can be defined
as\cite{Diener:2009ee, Diener:2010sy}
\begin{equation}
A_{\rm FB}=\frac{\int[F(y)-B(y)]dy}{\int[F(y)+B(y)]dy}
\end{equation}
where $F(y)$ is the number of forward events with pseudorapidity
$|\eta_f|>|\eta_{\bar{f}}|$ and $B(y)$ is the number of backward
events with pseudorapidity $|\eta_f|<|\eta_{\bar{f}}|$ for a given
$Z'$ rapidity, $y$. In a previous paper, we proposed a new
definition of forward-backward asymmetry, namely, the one-side
forward-backward asymmetry $A_{\rm OFB}$, to study the
forward-backward asymmetry at the LHC\cite{Wang:2010du}. The basic
idea is that valence quark momentum is averagely larger than that of
sea quark in the proton. Once the $z$ direction momentum of the
final states is required to be larger than a specific value, the
partonic forward-backward asymmetry will be kept.

$A_{\rm OFB}$ is defined as
\begin{equation}
A_{\rm OFB}=\frac{F_- + B_-}{F_+ +B_+}\equiv\frac{\sigma^A}{\sigma}
\label{AOFB}
\end{equation}
with
\begin{equation} F_\pm= \left. \left(\sigma( \Delta Y>0)\pm \sigma(\Delta
Y<0)\right)\right|_{P_{f^+f^-}^z>P^z_{\rm cut}, M_{f^+f^-}>M_{\rm
cut}}
\end{equation}
\begin{equation} B_\pm= \left. \left(\sigma(\Delta
Y<0)\pm \sigma(\Delta Y>0)\right)\right|_{P_{f^+f^-}^z<-P^z_{\rm
cut}, M_{f^+f^-}>M_{\rm cut}}
\end{equation}
where $P_{f^+f^-}^z$ is the final particle pair's $z$ direction
momentum and $M_{f^+f^-}$ is the invariant mass of the final
particle pair.

By adopting some kinematic cuts, especially cuts on $P_{f^+f^-}^z$,
the forward-backward asymmetry generated at the partonic level can
be kept even after the convolution with parton distribution
functions. The $A_{\rm OFB}$ can be an efficient tool in
investigating the forward-backward asymmetry at the LHC. In
principle, all the forward backward asymmetry measured in the left
right asymmetric beam collides, eg., $e^+ e^-$ or $p \bar{p}$, can
now be cross-checked at the left right symmetric $pp$ beam collider,
LHC. In this paper, we will extend our previous study to various
final state cases\cite{Wang:2010du}.

As shown in Eq.(\ref{AOFB}), the precise momentum measurement at $z$
direction is essential for $A_{\rm OFB}$. At the LHC, the momentum
of charged leptons are the most precisely measured quantities.  Thus
it is quite natural to study first the $A_{\rm OFB}$ for charged
leptons at the LHC. In the SM, the charged lepton pair can be
generated via $s$ channel $Z$ and/or $\gamma^*$ electroweak (EW)
diagrams, and these tree-level diagrams can contribute  to $A_{\rm
OFB}$ because the couplings of left- and right-handed fermions with
gauge boson Z  are different. At the LHC, besides the $s$ channel
$Z$ and/or $\gamma^*$ induced EW diagrams, bottom quarks are also
produced via the strong interaction. The contributions to $A_{\rm
OFB}^b$ in QCD starts from the next-to-leading order, namely, at
$\mathscr{O}(\alpha_S^3)$ \cite{Kuhn:1998jr, Kuhn:1998kw}. The
situation is similar to that of top quark pair production
\cite{Wang:2010du}. Away from Z-pole, the EW contributions to
$A_{\rm OFB}^b$ is much less than that of QCD ones. In order to
study $A_{\rm OFB}^b$  arising from the EW source, we have to select
events around the Z-pole.

The paper is organized as following. In Sec. \ref{two}, the charged
lepton one-side forward backward asymmetry $A_{\rm OFB}^\ell$ at the
LHC is calculated. As the charged lepton momentum can be precisely
measured,  $A_{\rm OFB}^\ell$ can be a test ground of the newly
proposed one-side forward-backward asymmetry. In Sec. \ref{three},
$A_{\rm OFB}^b$ is calculated at the NLO in QCD. In section
\ref{four},  $A_{\rm OFB}^b$ is calculated in the vicinity of the
$Z$ pole in order to study the EW origin of forward-backward
asymmetry. Section \ref{five} contains our conclusions and
discussions.

\section{Charged Lepton one-side forward-backward asymmetry $A_{\rm OFB}^\ell$ \label{two}}

At the LHC, the main production mechanisms of the charged leptons
(electron/muon/tau) at partonic level are $q \bar q \rightarrow
Z/\gamma^* \rightarrow l^+ l^-$, similar to those at LEP and
Tevatron. Because the couplings among Z boson and left- or
right-handed fermions are different, even at leading order
$\mathscr{O}(\alpha^2)$, forward-backward asymmetry is non-zero
\footnote{For $e^+e^- \rightarrow e^+e^-$ the $A_{\rm FB}\neq 0$
arises also from the interference between s- and t-channel QED
diagrams. }. The measurements of $A_{\rm FB}^\ell$ (cf. Eq.
\ref{fba1}) at LEP and Tevatron  are in good agreement with the SM
predictions. At the LHC there is no preferred direction in lab
frame. The one-side forward-backward asymmetry for $l^+l^-$
production process is defined as Eq.(\ref{AOFB}), where
\begin{equation} F_\pm= \left. \left(\sigma( \Delta Y>0)\pm \sigma(\Delta
Y<0)\right)\right|_{P_{l^+l^-}^z>P^z_{\rm cut}}
\end{equation}
\begin{equation} B_\pm= \left. \left(\sigma(\Delta
Y<0)\pm \sigma(\Delta Y>0)\right)\right|_{P_{l^+l^-}^z<-P^z_{\rm
cut}}.
\end{equation}
Here $\Delta Y=Y_{l^+}-Y_{l^-}$ is the difference of rapidity of the
charged leptons, which is invariant along the boost in beam
directions. $P_{l^+l^-}^z$ is the $z$ direction momentum of the
lepton pair in the laboratory frame.

At the $pp$ collider LHC, for the subprocess $q\bar{q}\to l^+l^-$,
the momentum of the valence quark q is usually larger than that of
the sea quark $\bar{q}$. If taking the momentum of q as the positive
z direction, we will get $P_{l^+l^-}^z>0$. However there is the
possibility that momentum of valence quark is less than that of sea
quark. In this case, $P_{l^+l^-}^z < 0$. This will induce the
opposite contribution to asymmetric cross section. Moreover, the
valence quark can symmetrically come from the other proton. The
usual $A_{\rm FB}$ is strictly equal to zero. The asymmetric cross
section of the partonic processes can survive only if we just
observe one-side $l^+l^-$ events, for example $P_{l^+l^-}^z>0$ or
$P_{l^+l^-}^z> P_{\rm cut}^z$. The usual forward cross section
$\sigma(\Delta Y>0)$ and backward cross section $\sigma(\Delta Y<0)$
can be calculated after imposing the $P_{l^+l^-}^z$ cut. So the
forward-backward asymmetry in this side is $F_-/F_+$. If we evaluate
the opposite side events, namely, $P_{l^+l^-}^z < -P_{cut}^z$, the
forward-backward asymmetry in this side is $B_-/B_+$. At the LHC the
consistence between these two forward-backward asymmetries can be
checked. Moreover if we define $A_{\rm OFB}$ in Eq.(\ref{AOFB}), the
statistics will be doubled. Besides keeping the forward-backward
asymmetry at partonic level, $P^z_{cut}$ has other advantages for
example increasing the significance to observe the forward-backward
asymmetry.

In our calculations here, $F_{\pm}$ and $B_{\pm}$ are calculated at
the leading order $\mathscr{O}(\alpha^2)$. Because of the small mass
compared with the collider beam energy, three charged leptons will
have similar signatures although they will be measured
(reconstructed) by different methods. Limited by the coverage of the
real detector, the charged lepton is required to satisfy
$|\eta|<2.4$ \cite{Ball:2007zza}. In massless limit, $\eta=Y$.

\begin{figure}[htbp]
\centerline{\hbox{
\includegraphics[height=5cm]
{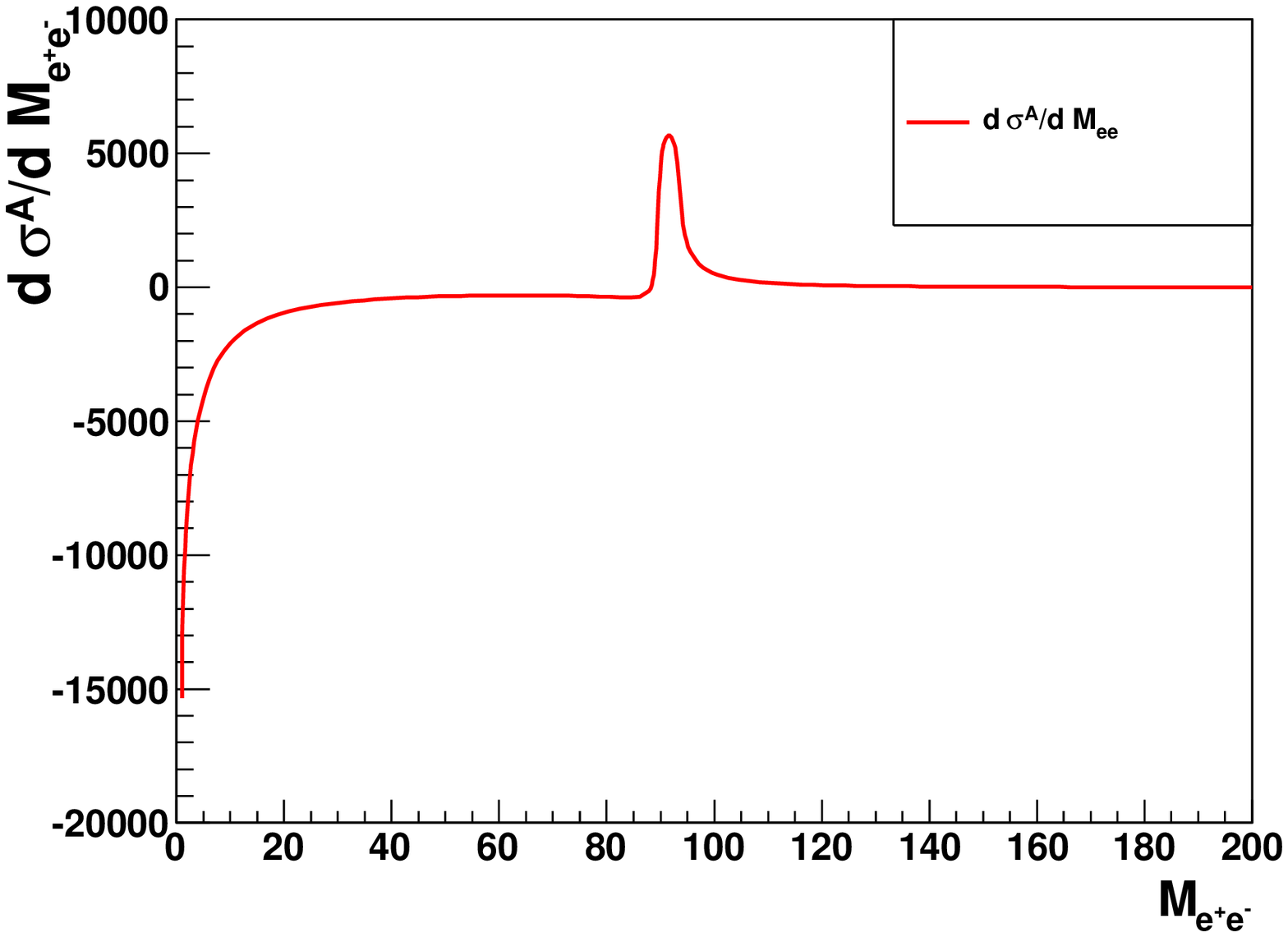}
\includegraphics[height=5cm]
{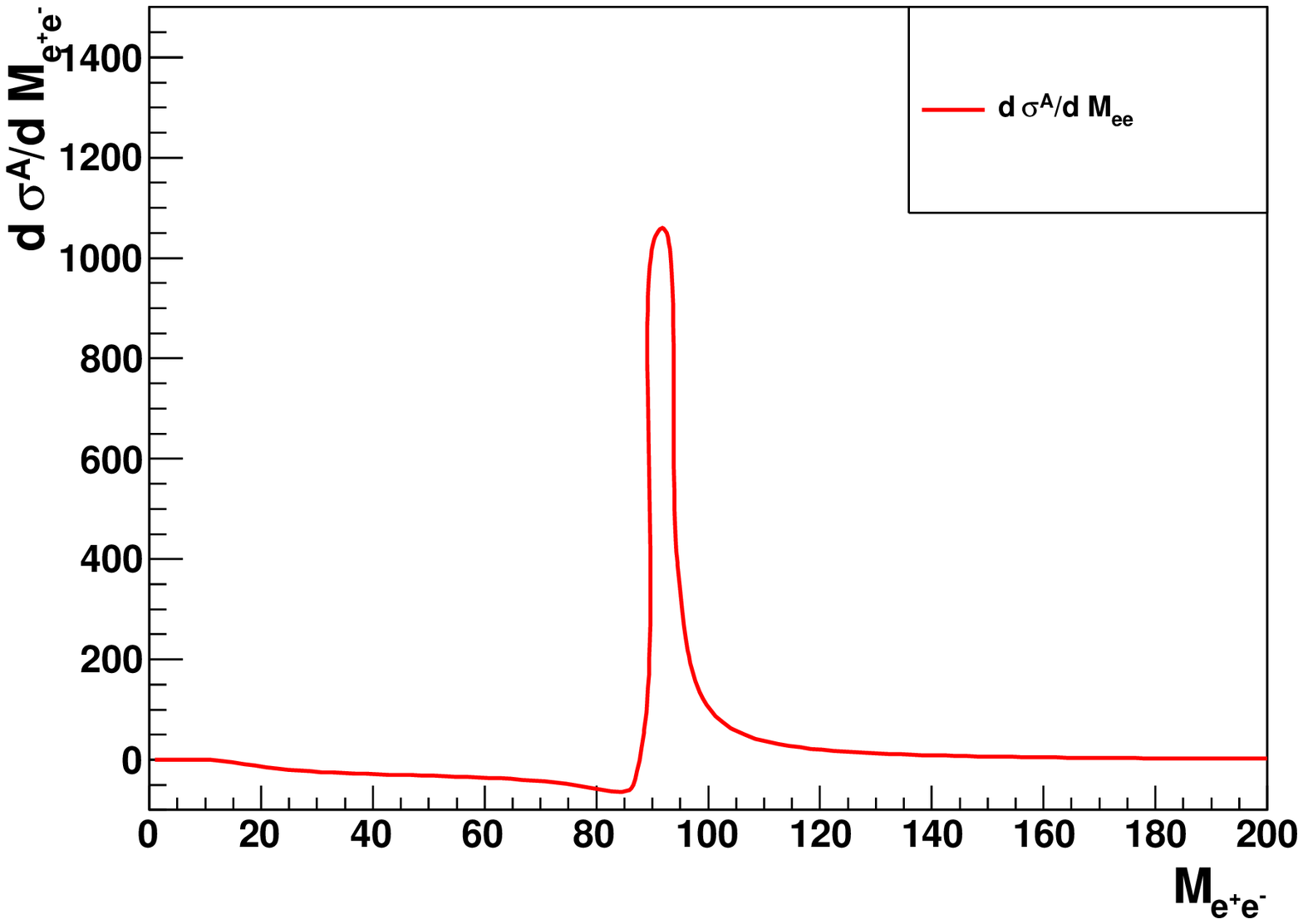}}} \centerline{\hbox{
\includegraphics[height=5cm]
{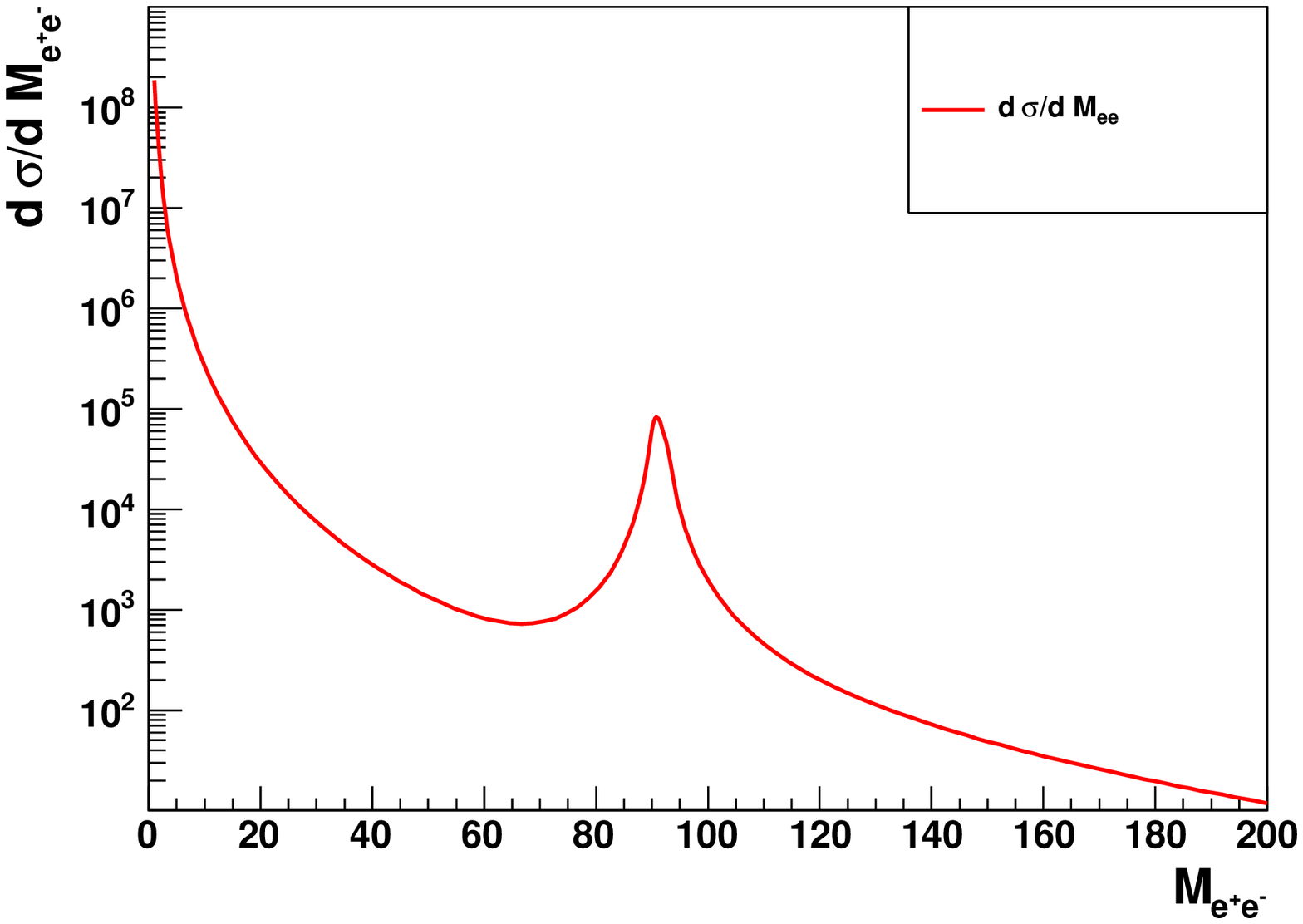}
\includegraphics[height=5cm]
{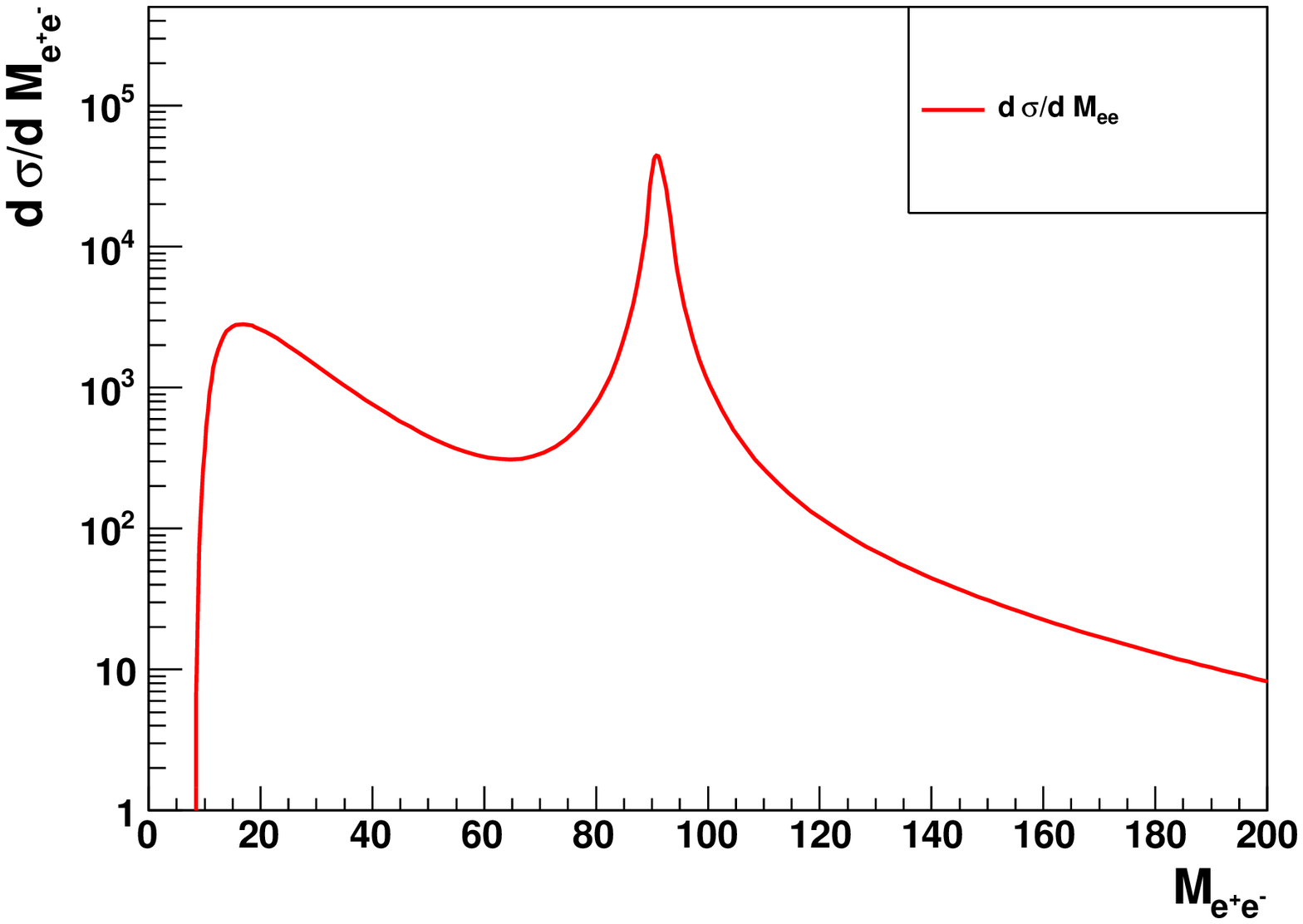}}} \centerline{\hbox{
\includegraphics[height=5cm]
{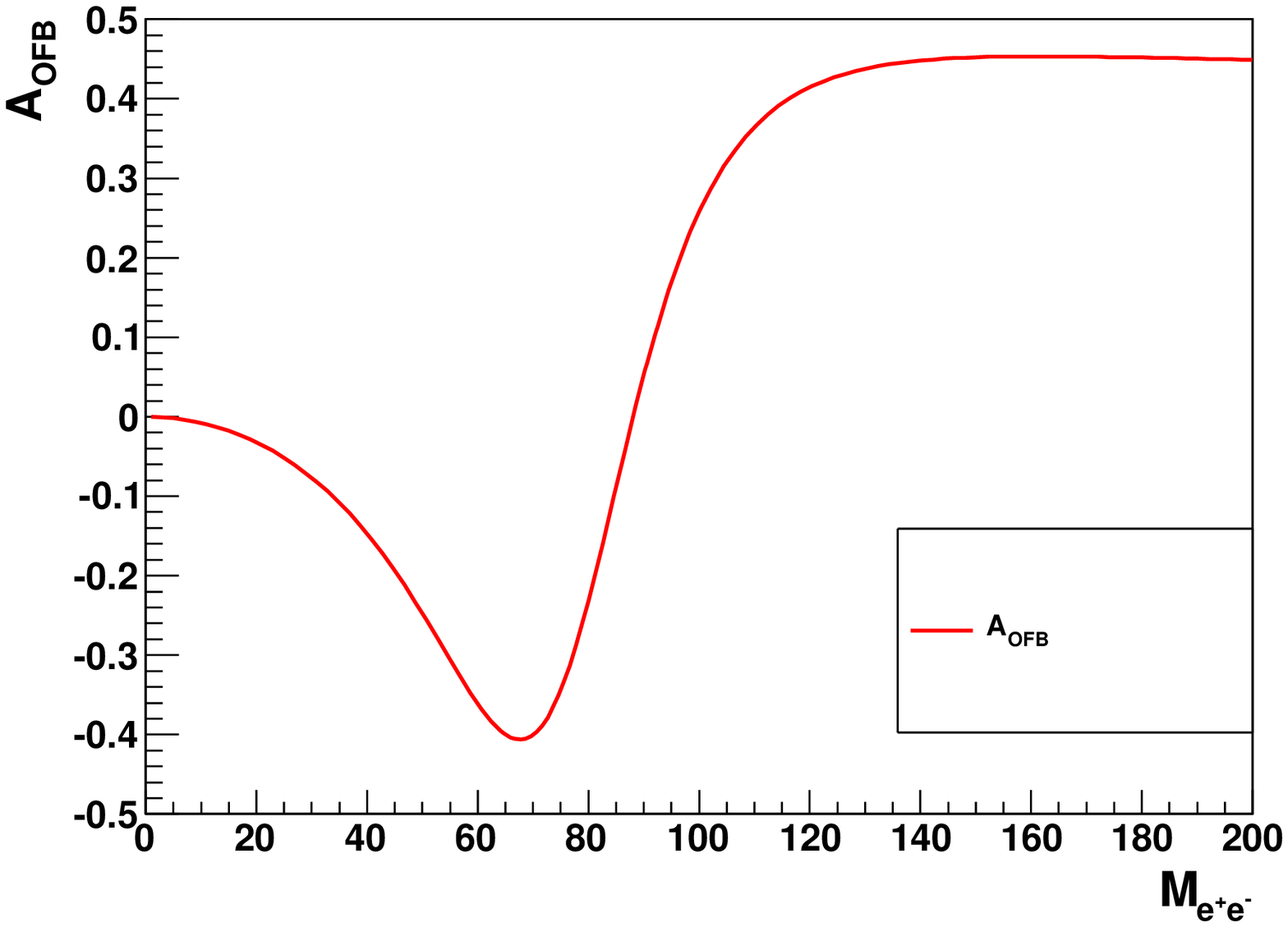}
\includegraphics[height=5cm]
{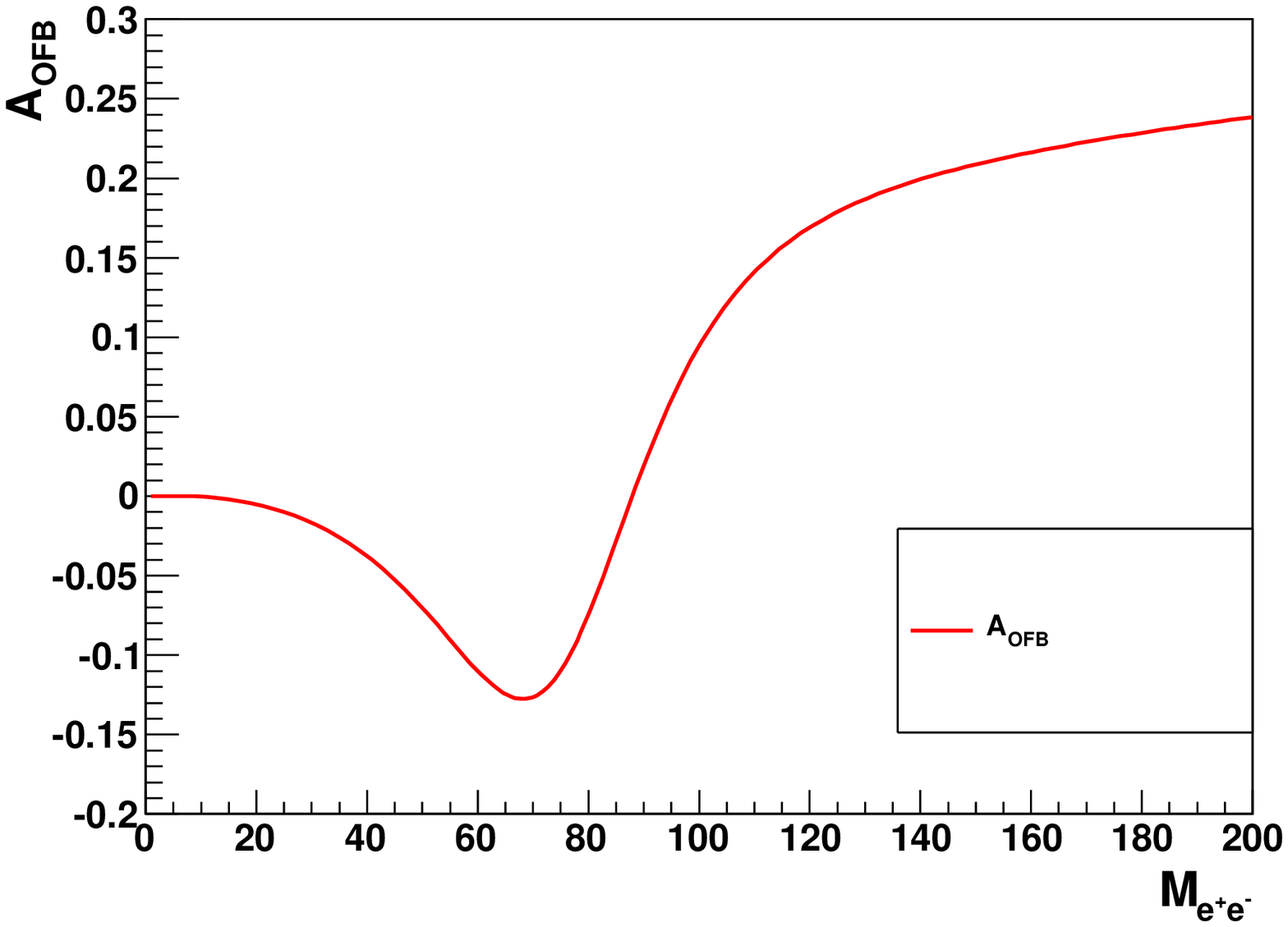}}} \centerline{\hbox{
\includegraphics[height=5cm]
{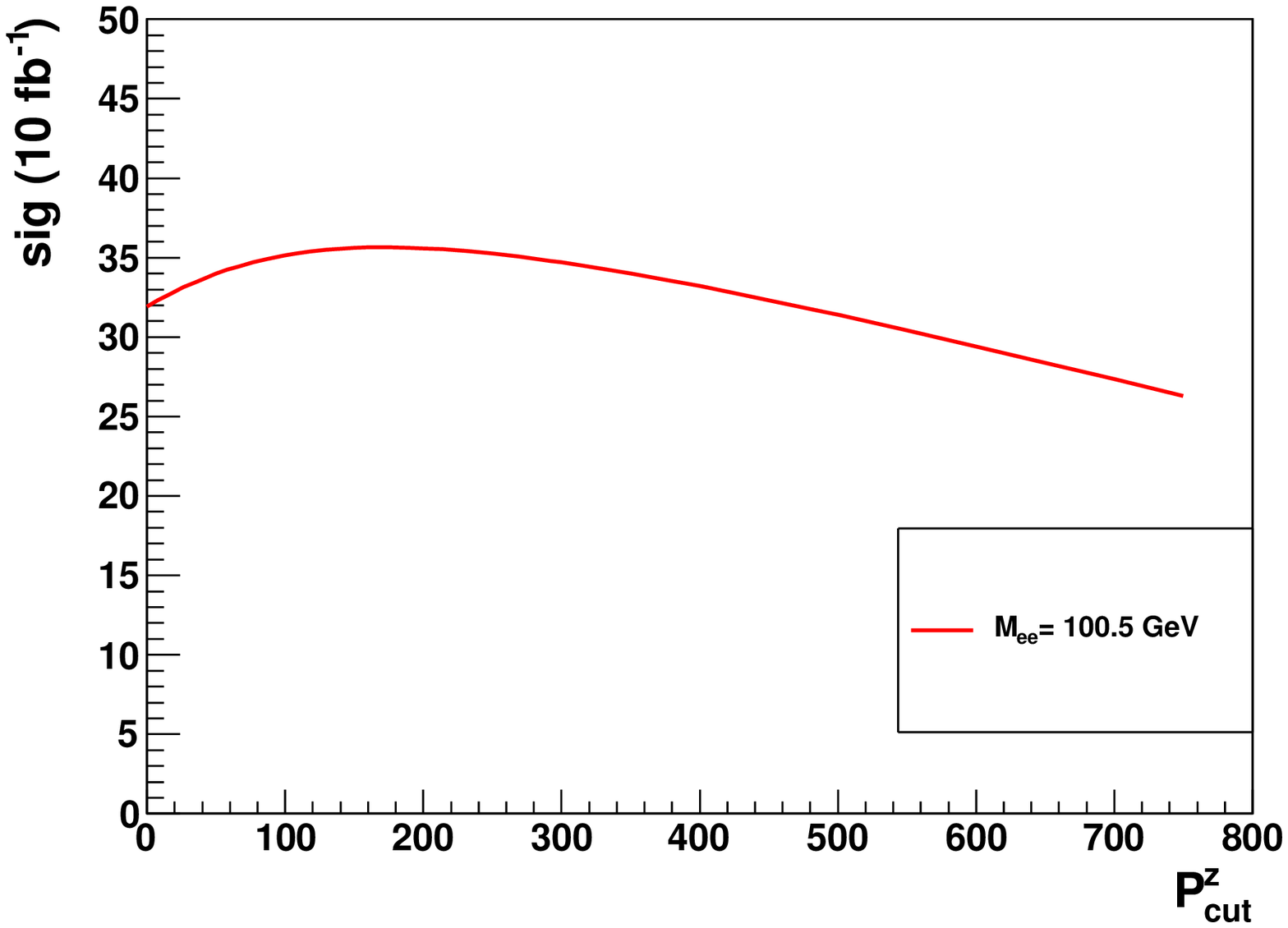}
\includegraphics[height=5cm]
{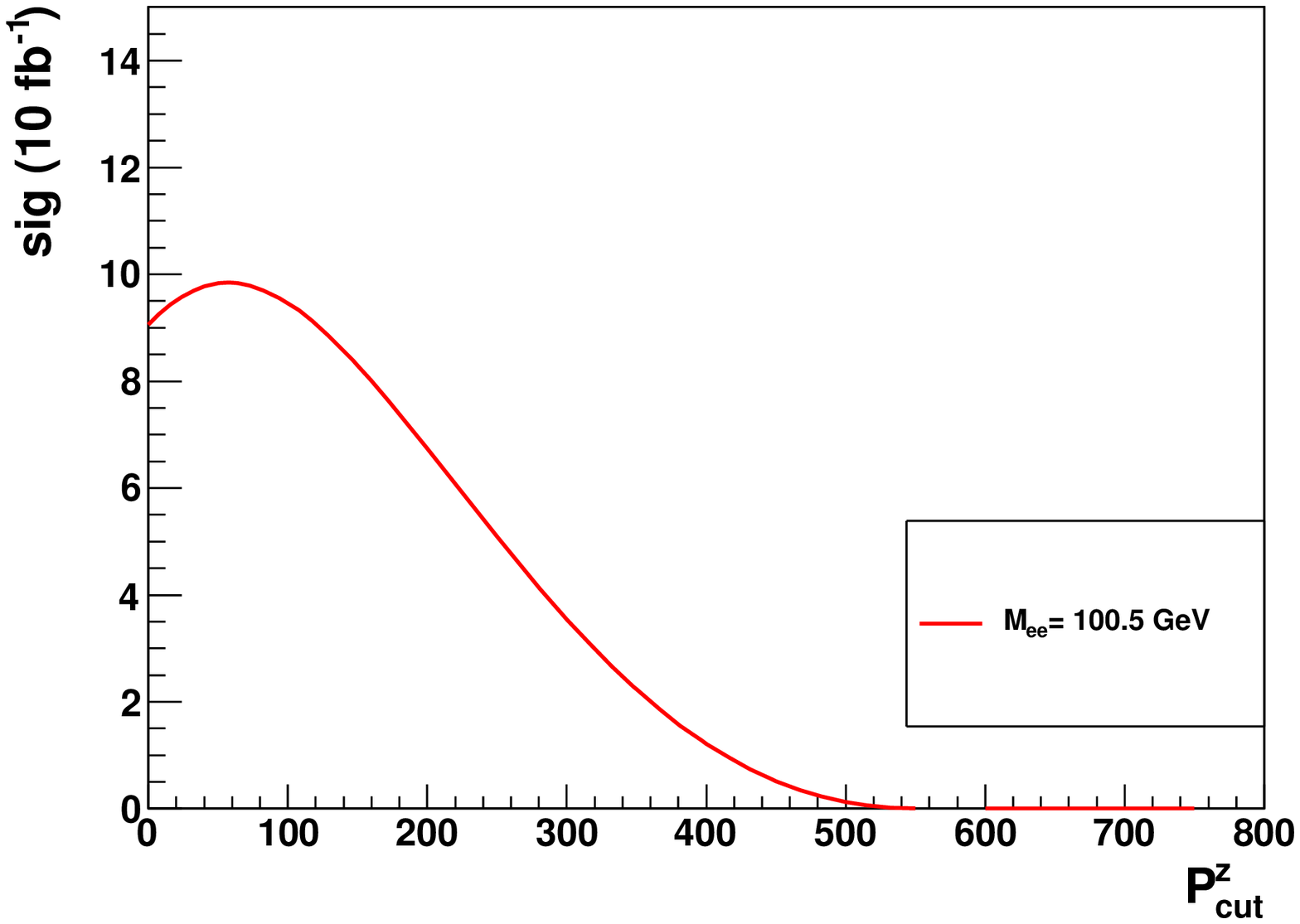}}} \caption{\label{7TeVll}
$d\sigma^A/d M_{e^+e-}$, $d\sigma/d M_{e^+e-}$, $A_{\rm OFB}$ as a
function of $M_{e^+e-}$ and ${\rm sig}$ as a function of $P^z_{\rm
cut}$ at the LHC for $\sqrt{s}=7$ TeV. The left plots have no $\eta$
cut, and the right plots have $|\eta|<2.4$ cut. Optimal $P_{\rm
cut}^z$ are determined by the ${\rm sig}$ plots. The left upper
three plots have $P_{\rm cut}^z=150 \mbox{GeV}$ and the right upper
three plots have $P_{\rm cut}^z=50 \mbox{GeV}$.}
\end{figure}

\begin{figure}[htbp]
\centerline{\hbox{
\includegraphics[height=5cm]
{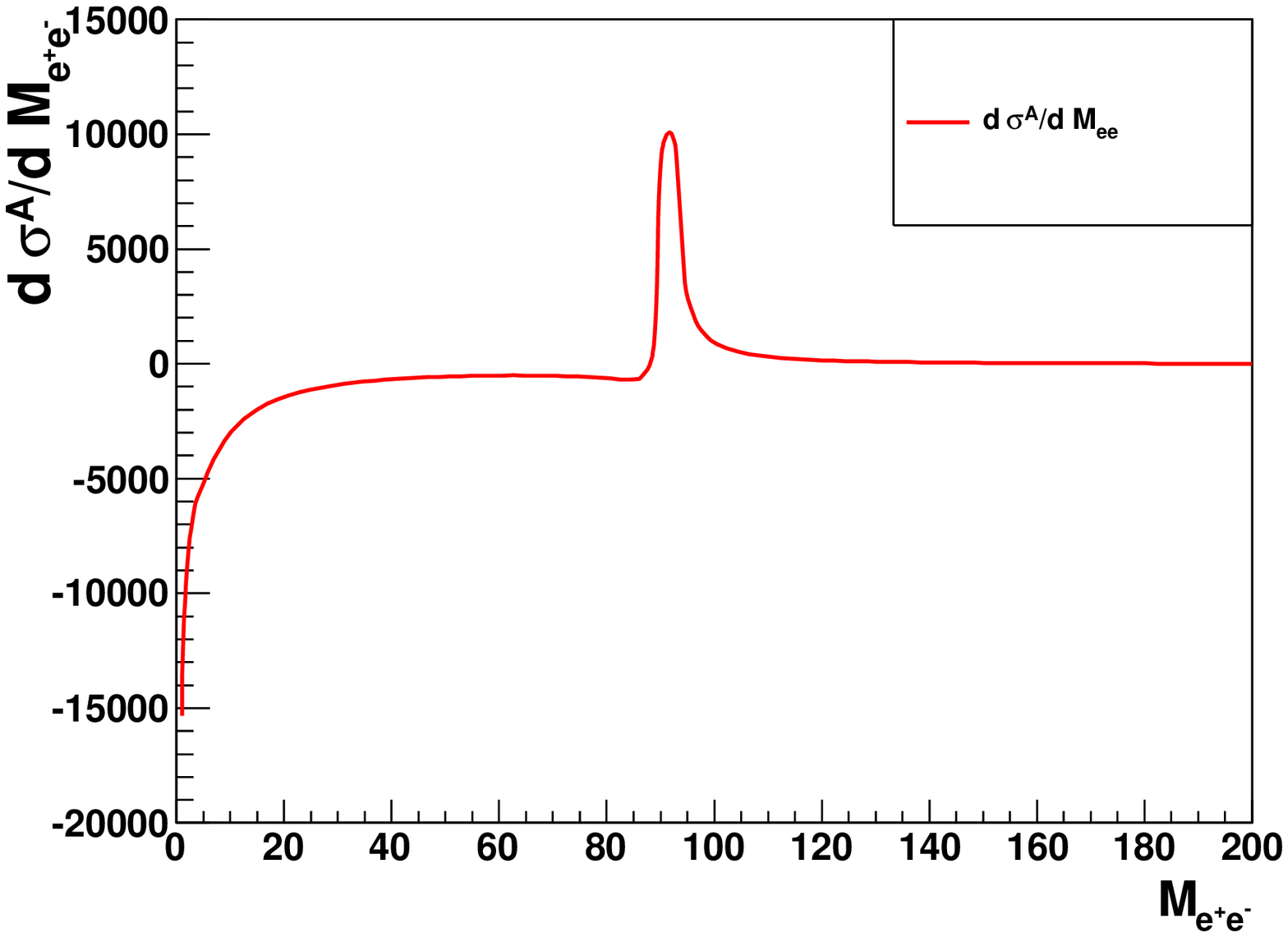}
\includegraphics[height=5cm]
{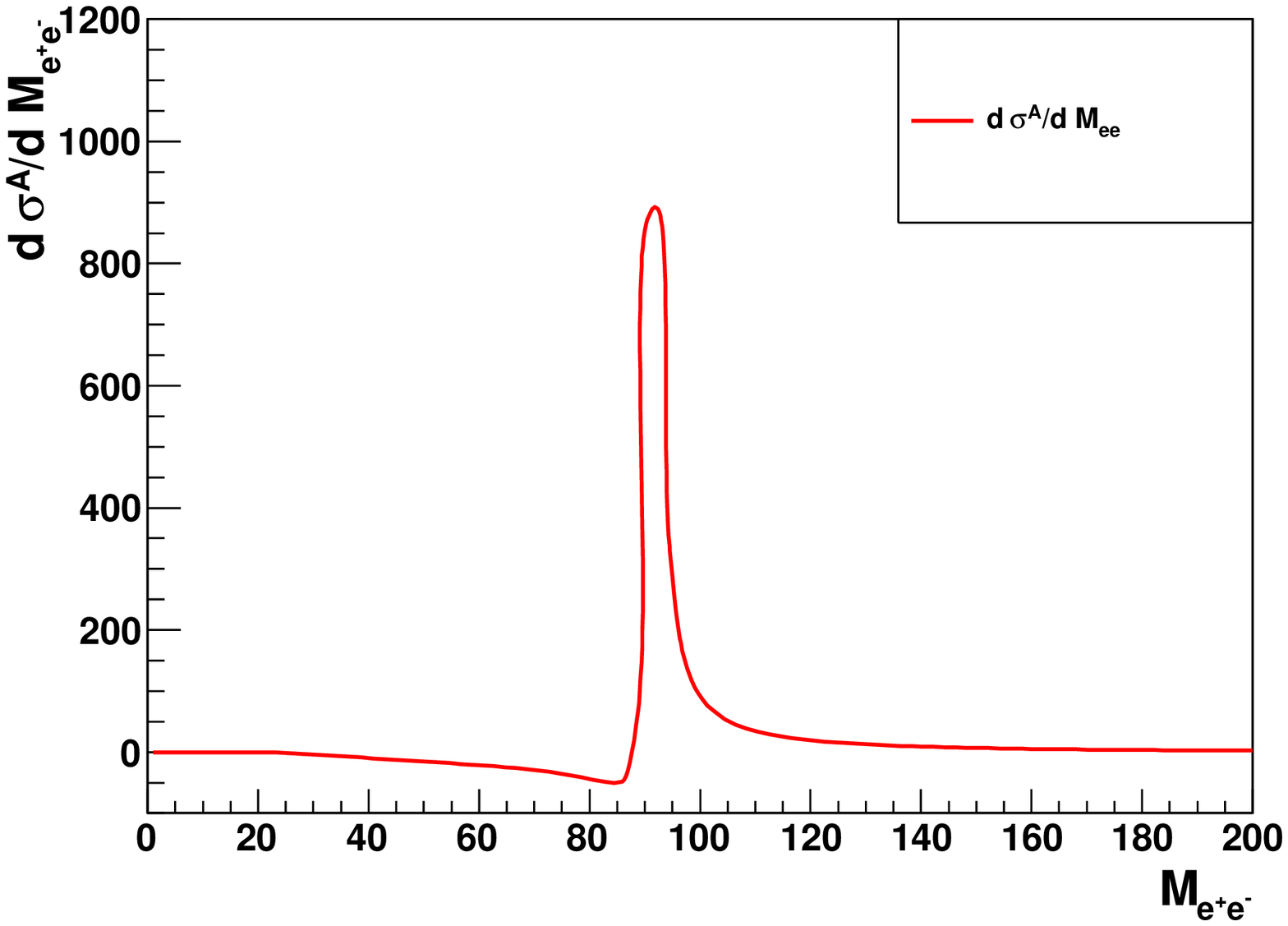}}} \centerline{\hbox{
\includegraphics[height=5cm]
{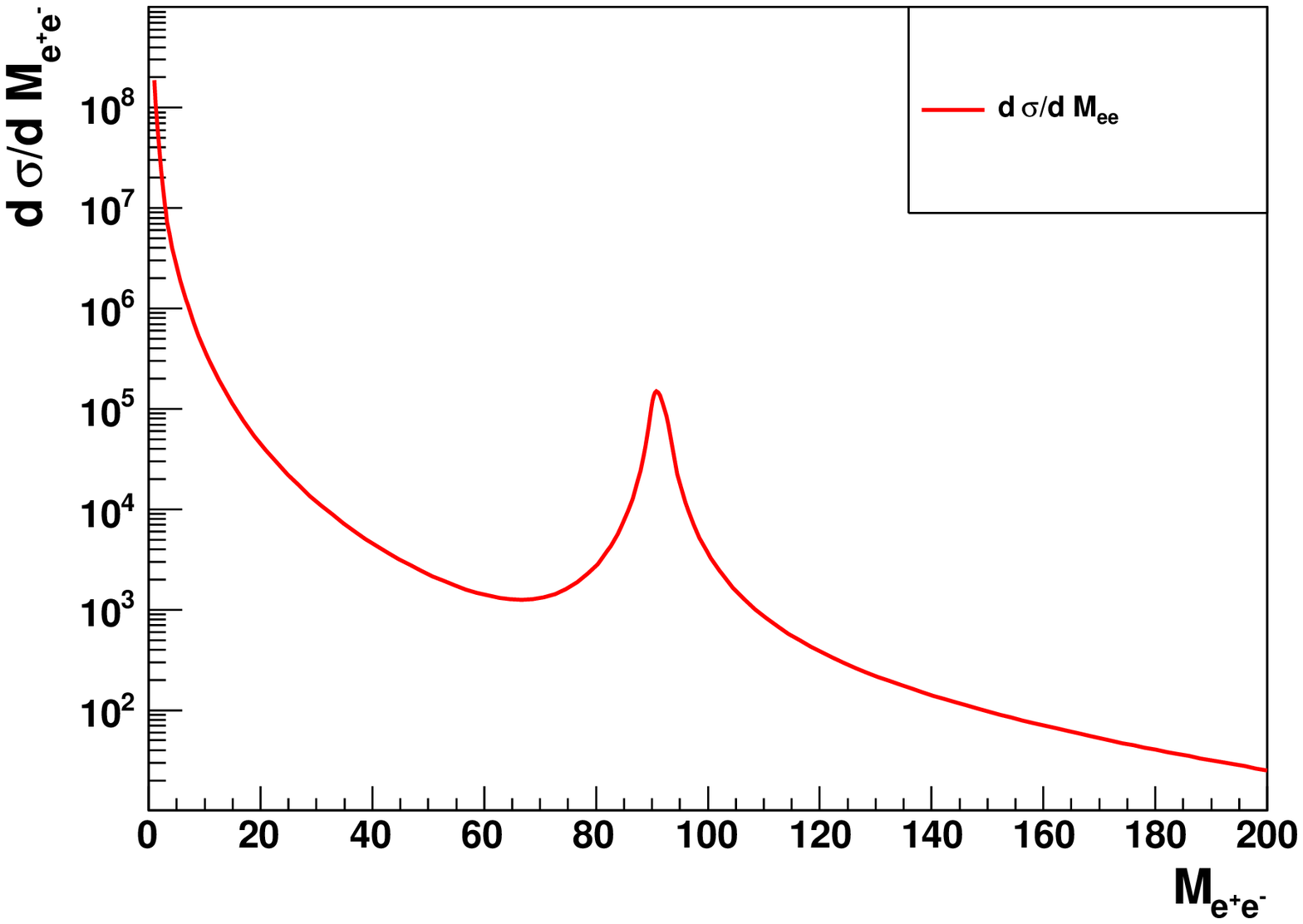}
\includegraphics[height=5cm]
{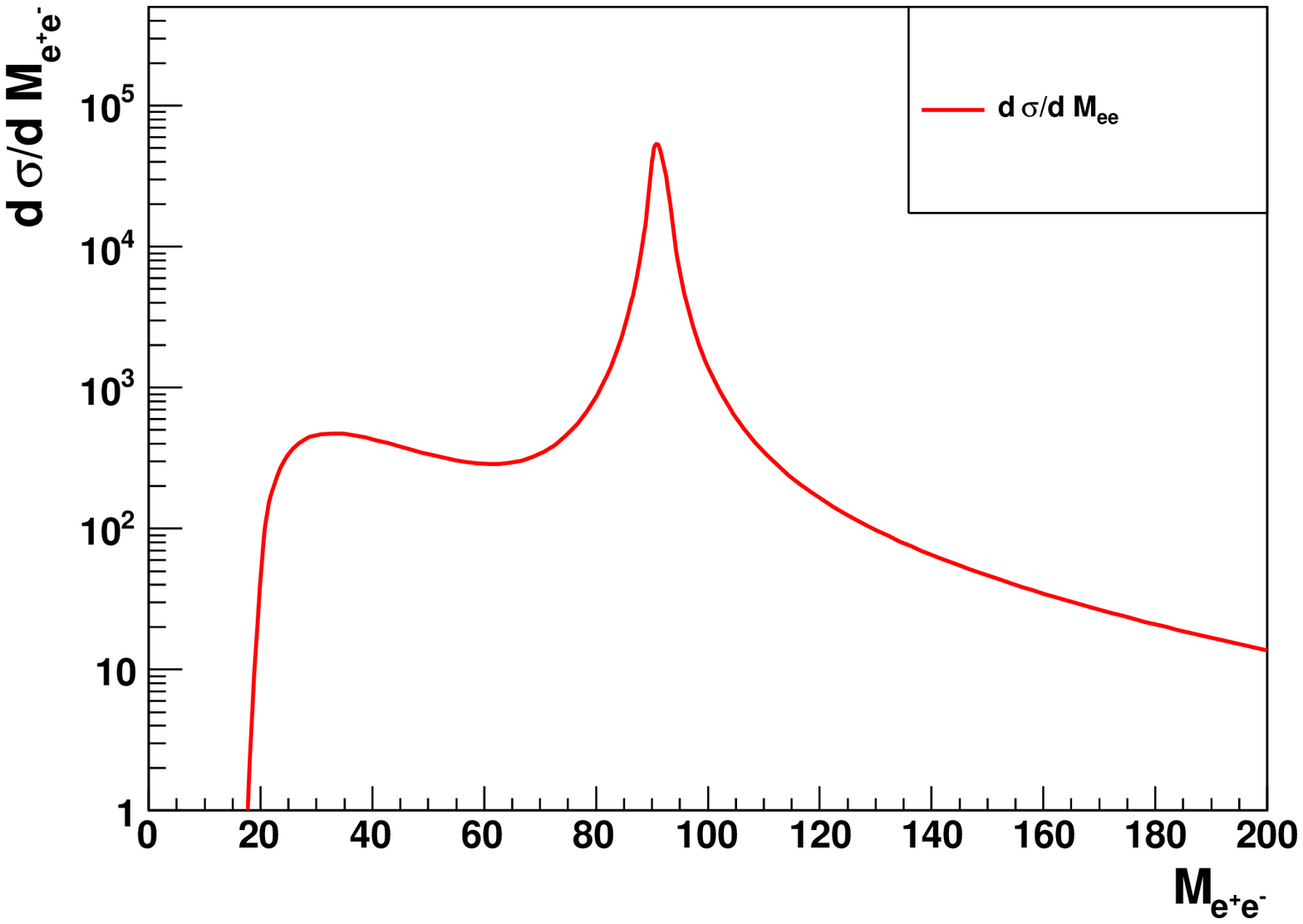}}} \centerline{\hbox{
\includegraphics[height=5cm]
{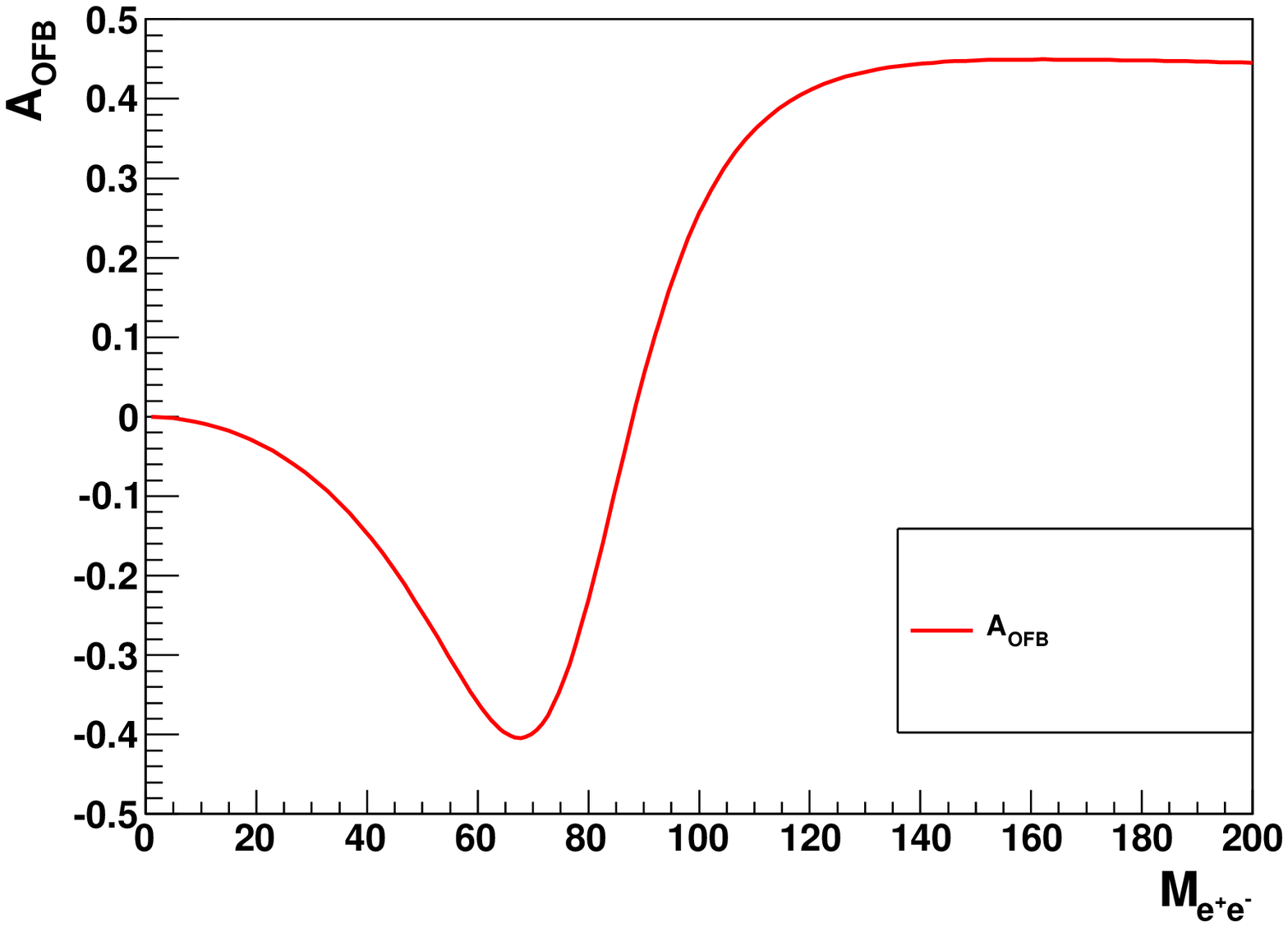}
\includegraphics[height=5cm]
{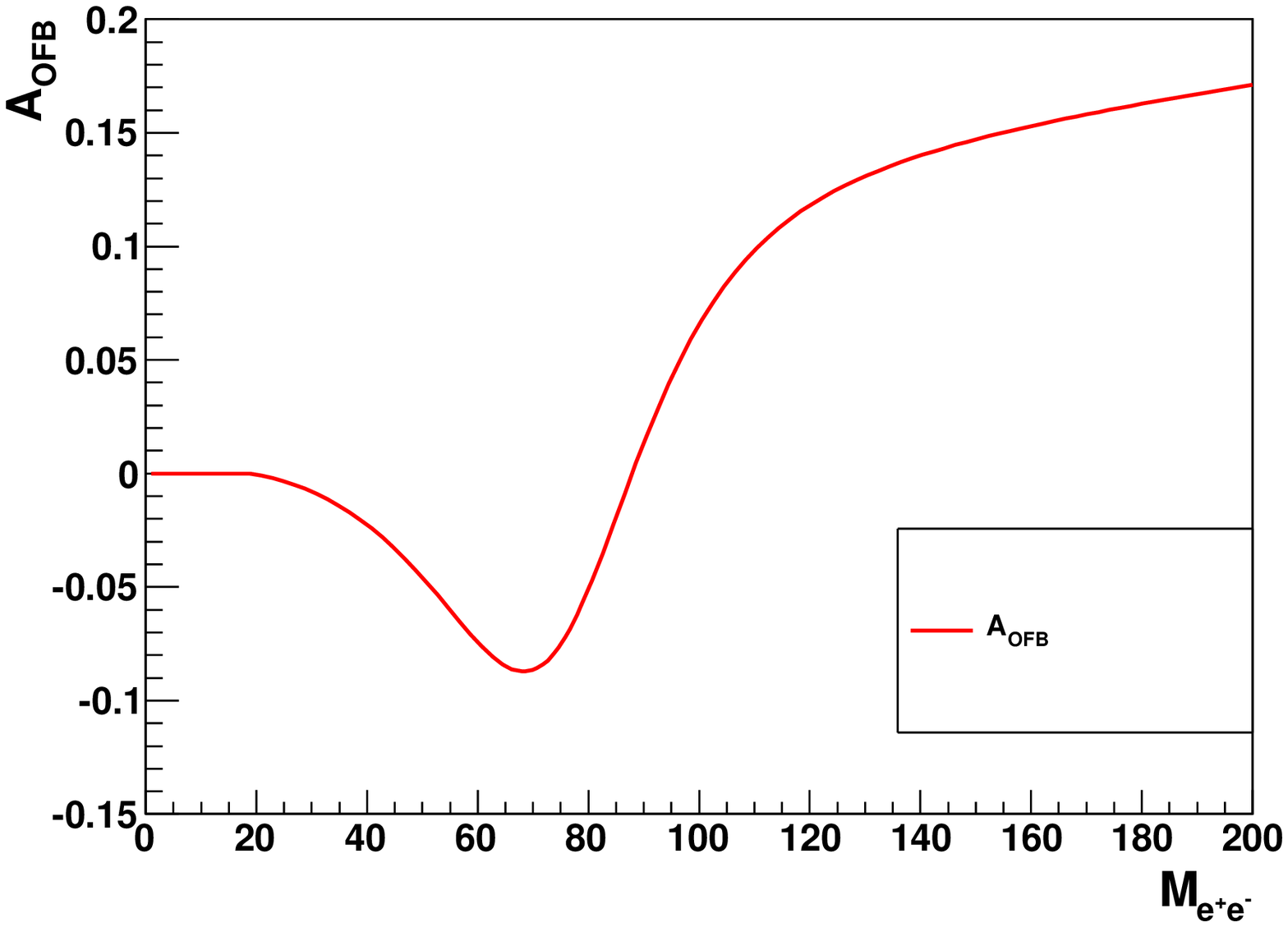}}} \centerline{\hbox{
\includegraphics[height=5cm]
{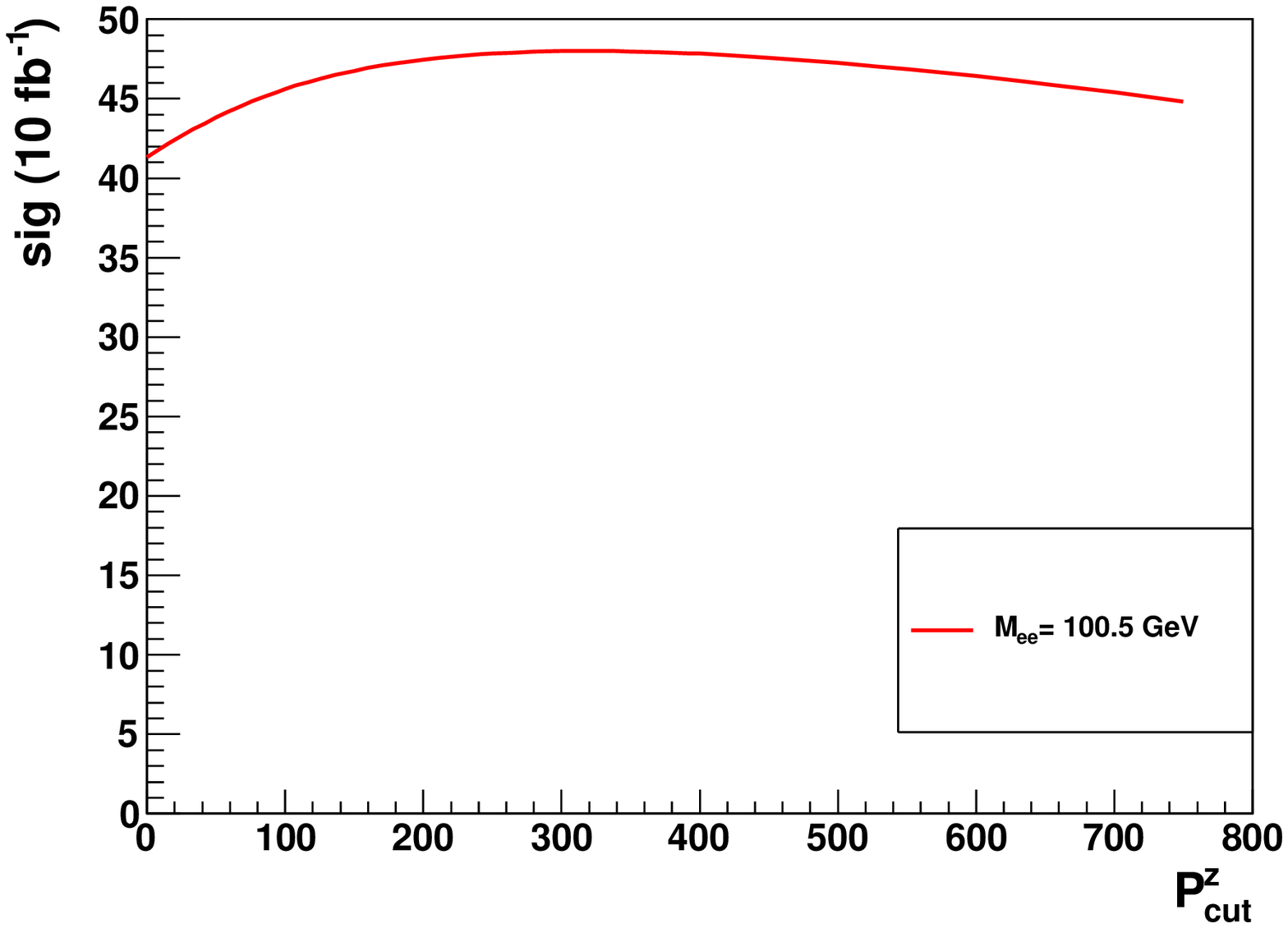}
\includegraphics[height=5cm]
{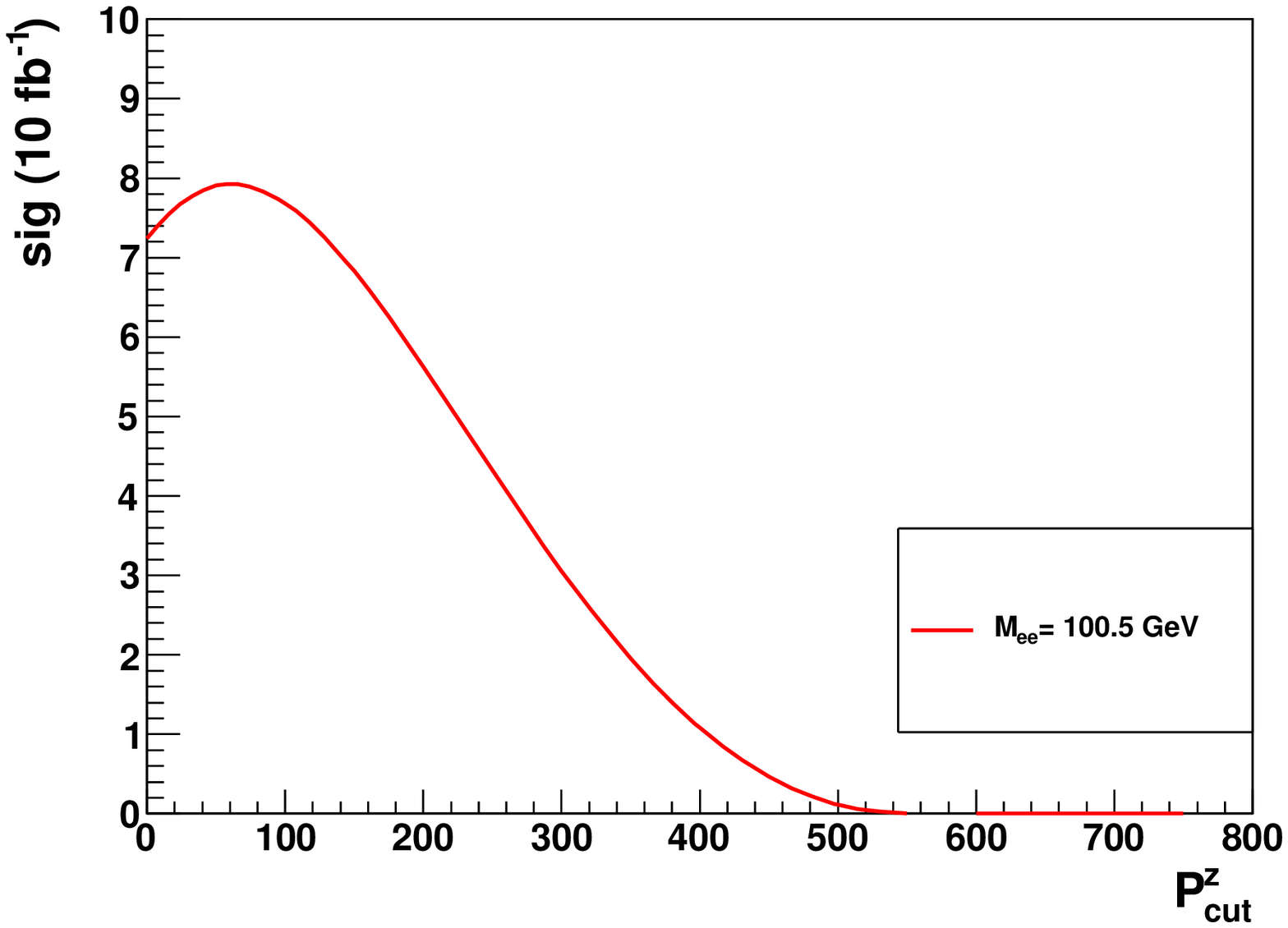}}} \caption{\label{14TeVll} Same
as Fig. \ref{7TeVll} except $\sqrt{s}=14 \mbox{TeV}$. Optimal
$P_{\rm cut}^z=300 \mbox{GeV}$ for left upper three plots and
optimal $P_{\rm cut}^z=100 \mbox{GeV}$ for right upper three plots.}
\end{figure}

Figures \ref{7TeVll} and \ref{14TeVll} show the differential
spectrum of the asymmetric cross section $\sigma^A$, total cross
section $\sigma$, $A_{\rm OFB}$ as a function of the lepton pair
invariant mass $M_{e^+e^-}$ (taking electron as the example), and
the significance ${\rm sig} =\sqrt{\cal{L}} \sigma^A/\sqrt{\sigma} $
( with ${\cal{L}}=10 fb^{-1}$) as a function of $P_{\rm cut}^z$ at
the LHC for $\sqrt{s}=7~\mbox{TeV}$ and 14 TeV respectively. No
$\eta$ cuts are applied for the left column plots, and $|\eta|<2.4$
is applied for the right column plots. The significance plots are
used to select the optimal $P_{\rm cut}^z$. In these plots we take
the typical value $M_{e^+e^-}=100.5 \mbox{GeV}$ as an example. The
optimal $P_{\rm cut}^z$ are not sensitive to $M_{e^+e^-}$. Thus, we
take $P_{\rm cut}^z=150 \mbox{GeV}$ in the left upper three plots
and $P_{\rm cut}^z=50 \mbox{GeV}$ in the right upper three plots.
From the curves with and without $\eta$ cut we  can see that the
asymmetric cross section is sensitive to $\eta$. This behavior
indicates that asymmetric cross section is mostly located in large
$\eta$ region due to the large boost along the longitudinal
direction. From figures we can also see clearly the resonance around
Z in $d\sigma/d M_{e^+ e^-}$ and $d\sigma^A/d M_{e^+ e^-}$.
Moreover, the $A_{\rm OFB}$ varies with $M_{e^+ e^-}$ and the
distribution is similar to that of usual $A_{\rm FB}$ at $e^+e^-$
and $p \bar p$ colliders. The reason is simply because both $A_{\rm
OFB}$  and $A_{\rm FB}$ arise mainly from the same subprocesses
$u\bar{u}\to e^+ e^-$ and $d\bar{d}\to e^+ e^-$.

For our purpose we only study the $A_{\rm OFB}$ at leading order,
namely, at $\mathscr{O}(\alpha^2)$. In practice \cite{Abazov:2008xq}
higher-order effects must be included. Such higher-order effects,
especially the contributions from the high $P_T$ $l^+l^-$ events
which arise from the extra hard photon radiation, can be treated by
adopting Collins-Soper frame \cite{Acosta:2004wq, Abazov:2008xq,
Collins:1977iv}.  The advantage of adopting the Collins-Soper frame
is that $A_{\rm FB}$ is free from the impact of the $2\to3$ process
with initial $\gamma$ radiation which will cause a nonzero $P_T$ of
the lepton pair. $A_{\rm OFB}$ can also be extended to Collins-Soper
frame with extra cut on z-direction momentum of lepton pair. We will
study this issue in detail elsewhere.

One-side forward-backward asymmetry can be tested in the charged
lepton production processes. Theoretically  $A_{\rm OFB}$ can also
be utilized to study the more complicated bottom quark production at
the LHC, though in practice the channel is not as clean as that of
charged leptons.

\section{Bottom quark one-side forward-backward asymmetry $A_{\rm OFB}^b$ in QCD \label{three}}

Unlike the top quark, the bottom quark life time is longer than the
hadronization scale, which means bottom quark will appear as $b$ jet
in the detector. For simplicity in our analysis, we treat the $b$
quark as $b$ jet in the calculation.

As mentioned above, at the LHC, the bottom quark forward backward
asymmetry arises from two sources, namely, the QCD and EW processes.
%Here we ignore the small impact from such as $B\bar{B}$ mixing.
The dominant bottom quark production processes are $gg (q\bar q)
\rightarrow b\bar b$ via strong interaction. However at the leading
order in QCD i.e. $\mathscr{O}(\alpha_S^2)$, the forward backward
asymmetry is zero. The QCD induced asymmetric cross section starts
from $O(\alpha_s^3)$. Same as top pair production, the contributions
can be classified into three categories: (1) Interference among
diagrams for the initial and final state radiation processes
$q\bar{q}\to b\bar{b} g$; (2) Interference among the born diagrams
and virtual box diagrams for the process $q\bar{q}\to b\bar{b}$; (3)
Contribution from diagrams of the real processes $q g\to b\bar{b}q$.
The calculation has been carried out in Ref.
\cite{Kuhn:1998jr,Kuhn:1998kw}. For the EW interaction contribution,
the leading contribution comes from the born cross section
$q\bar{q}\to b\bar{b}$ via a $Z$ and/or $\gamma^*$ boson, similar
with the case of charged lepton. At the LHC the EW contribution is
mostly from the vicinity of the $Z$ pole, while the QCD contribution
extends in the wider energy regime. Moreover except at $Z$ pole the
QCD contribution is much larger than that of the EW one. In this
section we will focus on the QCD contribution to $A_{\rm OFB}^b$.

At the Tevatron the theoretical calculation of heavy quark forward
backward asymmetry arising from the QCD contributions  has been
studied in previous literature \cite{Kuhn:1998jr,Kuhn:1998kw}. Even
at the LHC, the so-called central charge asymmetry $A_C$ has been
constructed to study the forward backward asymmetry of the top quark
\cite{Kuhn:1998jr,Kuhn:1998kw,Antunano:2007da,
Rodrigo:2008qe,Ferrario:2008wm, Ferrario:2010hm}. Some comparison
have been made between the central charge asymmetry and the one-side
forward backward asymmetry in Ref. \cite{Wang:2010du}. At the LHC
$A_{\rm OFB}$ is much larger than $A_C$ because $P_{\rm cut}^z$ can
suppress the huge symmetric $gg$ fusion efficiently.

One-side forward-backward asymmetry for $b$ quark at the LHC can be
defined in the $pp$ rest frame as in Eq. \ref{AOFB},
\begin{equation} F_\pm= \left. \left(\sigma( \Delta Y>0)\pm \sigma(\Delta
Y<0)\right)\right|_{P_{b\bar{b}}^z>P_{\rm cut}^z,M_{b\bar{b}}>M_{\rm
cut}}
\end{equation}
\begin{equation} B_\pm= \left. \left(\sigma(\Delta
Y<0)\pm \sigma(\Delta Y>0)\right)\right|_{P_{b\bar{b}}^z<-P_{\rm
cut}^z, M_{b\bar{b}}>M_{\rm cut}}. \end{equation}
Here we only
consider QCD contributions and ignore the electroweak contributions.
The purpose to apply constraints on $P_{t\bar{t}}^z$ and
$M_{t\bar{t}}$ is to suppress the symmetric $gg\to t\bar{t}$ events,
which will be illustrated in the following figures.

To measure $A_{\rm OFB}$ at the LHC, the charge of the $b$ jet
should be identified to distinguish the bottom or anti-bottom jet.
So one bottom/antibottom quark is required to decay into a charged
lepton, and the other antibottom/bottom can decay hadronically. For
the b tagging, there are two selecting criteria \cite{Lowette}:
$P_T>40\mbox{GeV}$ and $|\eta|<1.5$ without second vertex
reconstruction, and $P_T>10~\mbox{GeV}$ and $|\eta|<2.4$ with second
vertex reconstruction. We find that our signal $b$ jets locate
mostly in large $\eta$ regions due to the high longitudinal boost.
So we take the second cut criteria in the following analysis.

Note that the definition in Eq.(\ref{AOFB}) is based on the CMS or
ATLAS detector at the LHC. For the LHCb, one can take real
``one-side'' definition, namely
\begin{equation}
A_{\rm OFB}=\frac{F_- }{F_+}. \label{AFBLHCb}
\end{equation}
In this case $b$ tagging requirements should also be adjusted
accordingly.
 The obvious difference is that there will be a lower bound on $\eta$,
e.g. $2.0<\eta<5.5$ \cite{:2001hf}. As most of $b\bar{b}$ events are
boosted in the $z$-direction, LHCb has the unique advantage to
collect more bottom events to reach higher precision measurement of
forward-backward asymmetry.

\begin{figure}[htbp]
\centerline{\hbox{
\includegraphics[height=5cm]
{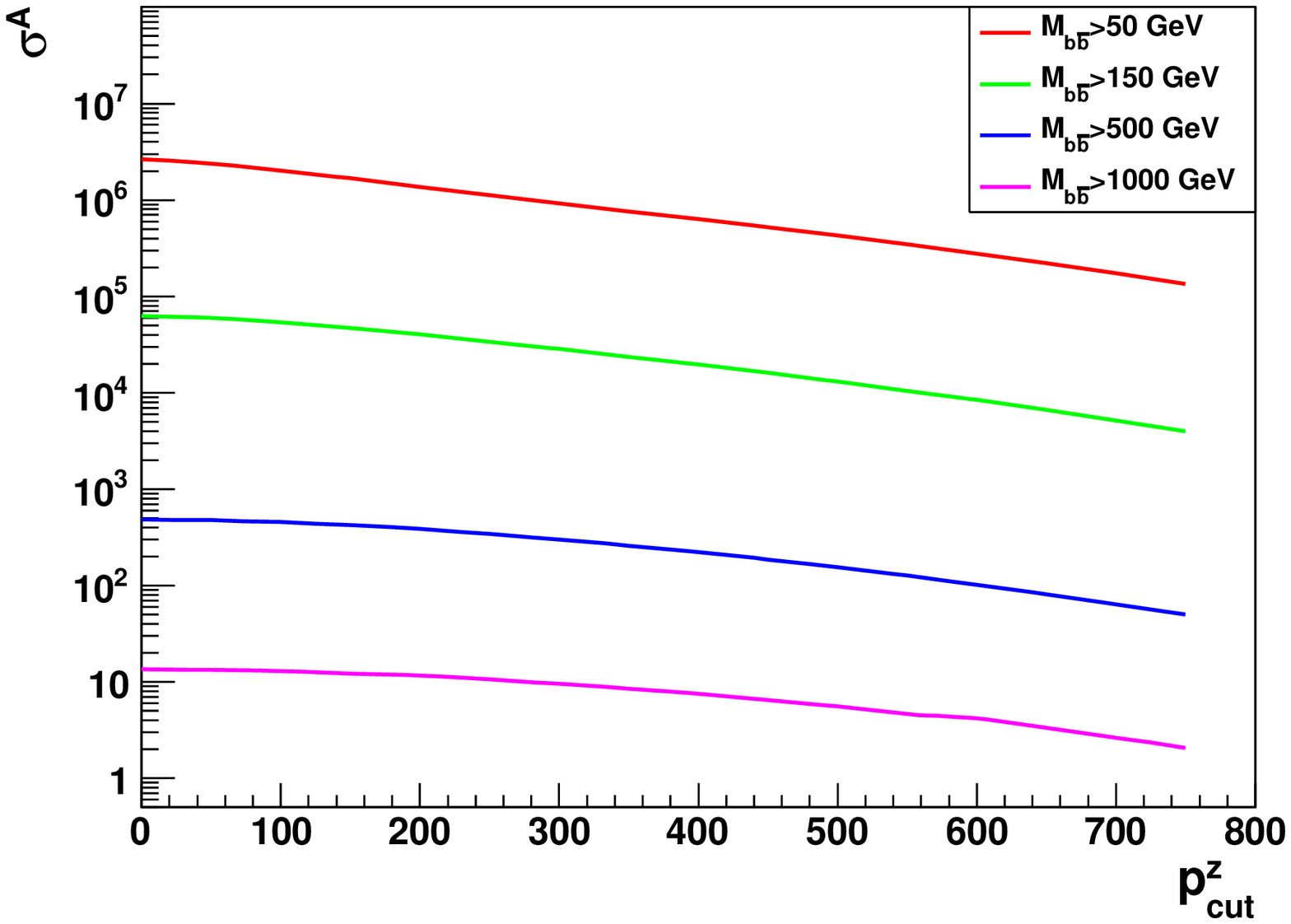}
\includegraphics[height=5cm]
{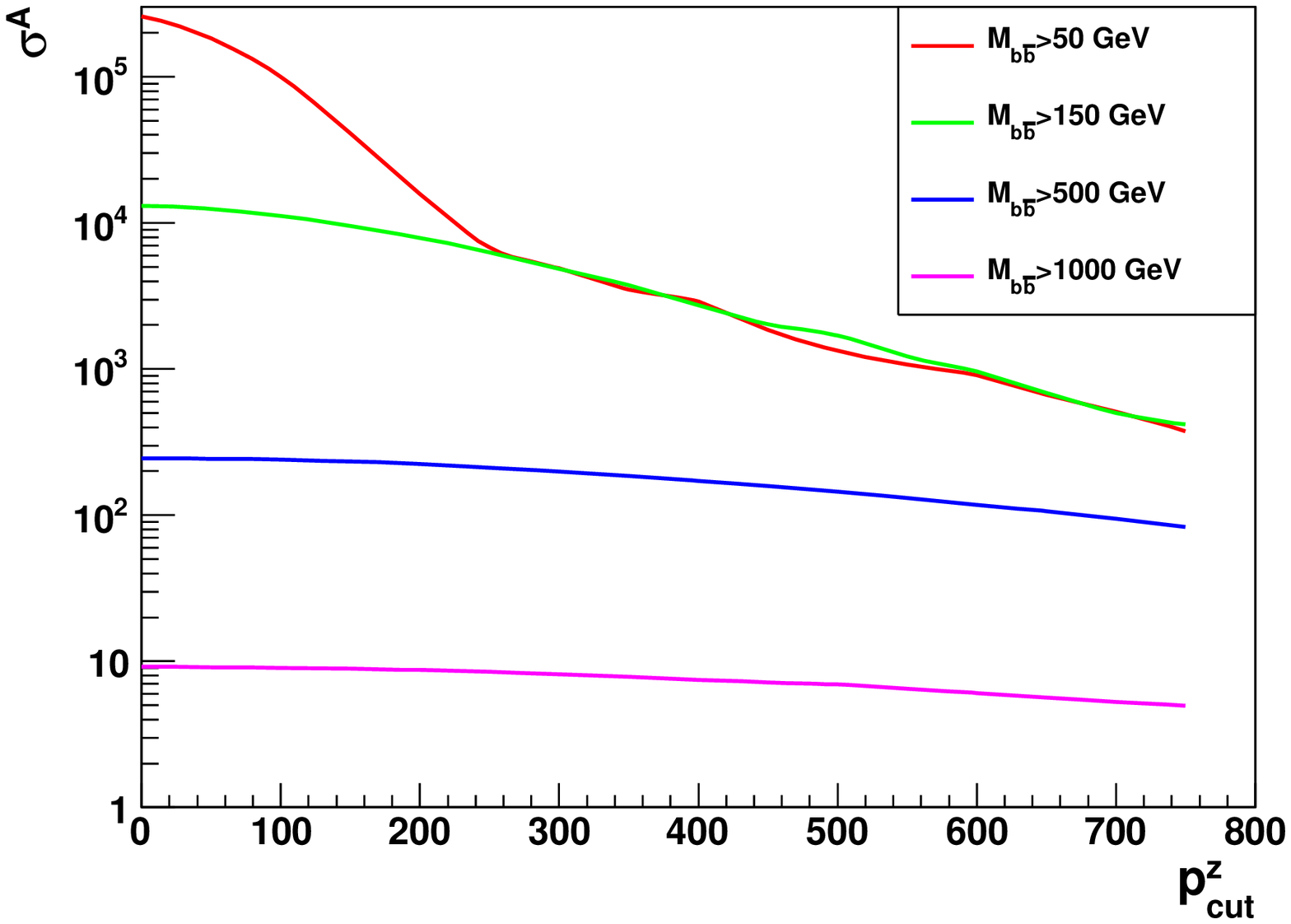}}} \centerline{\hbox{
\includegraphics[height=5cm]
{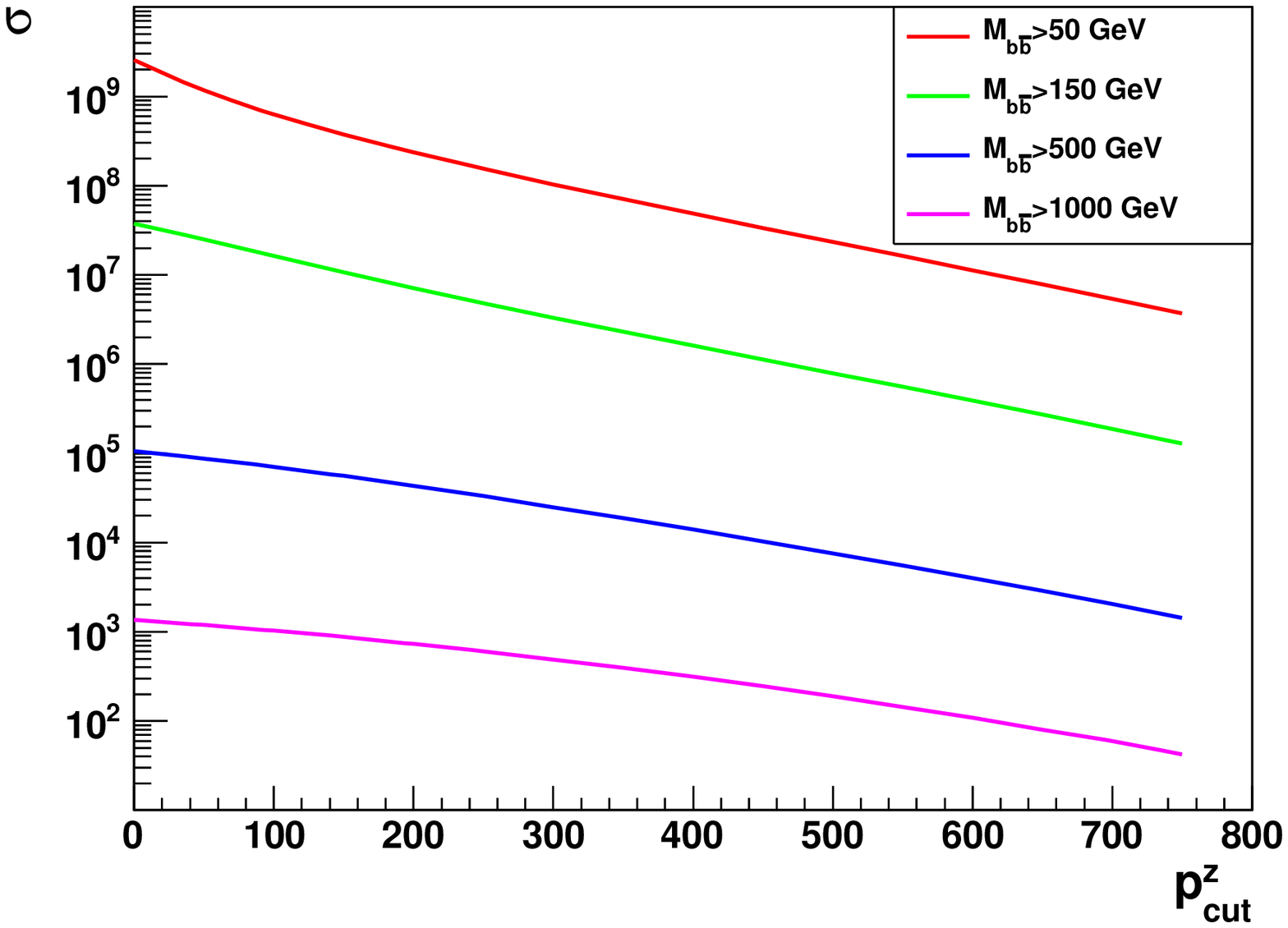}
\includegraphics[height=5cm]
{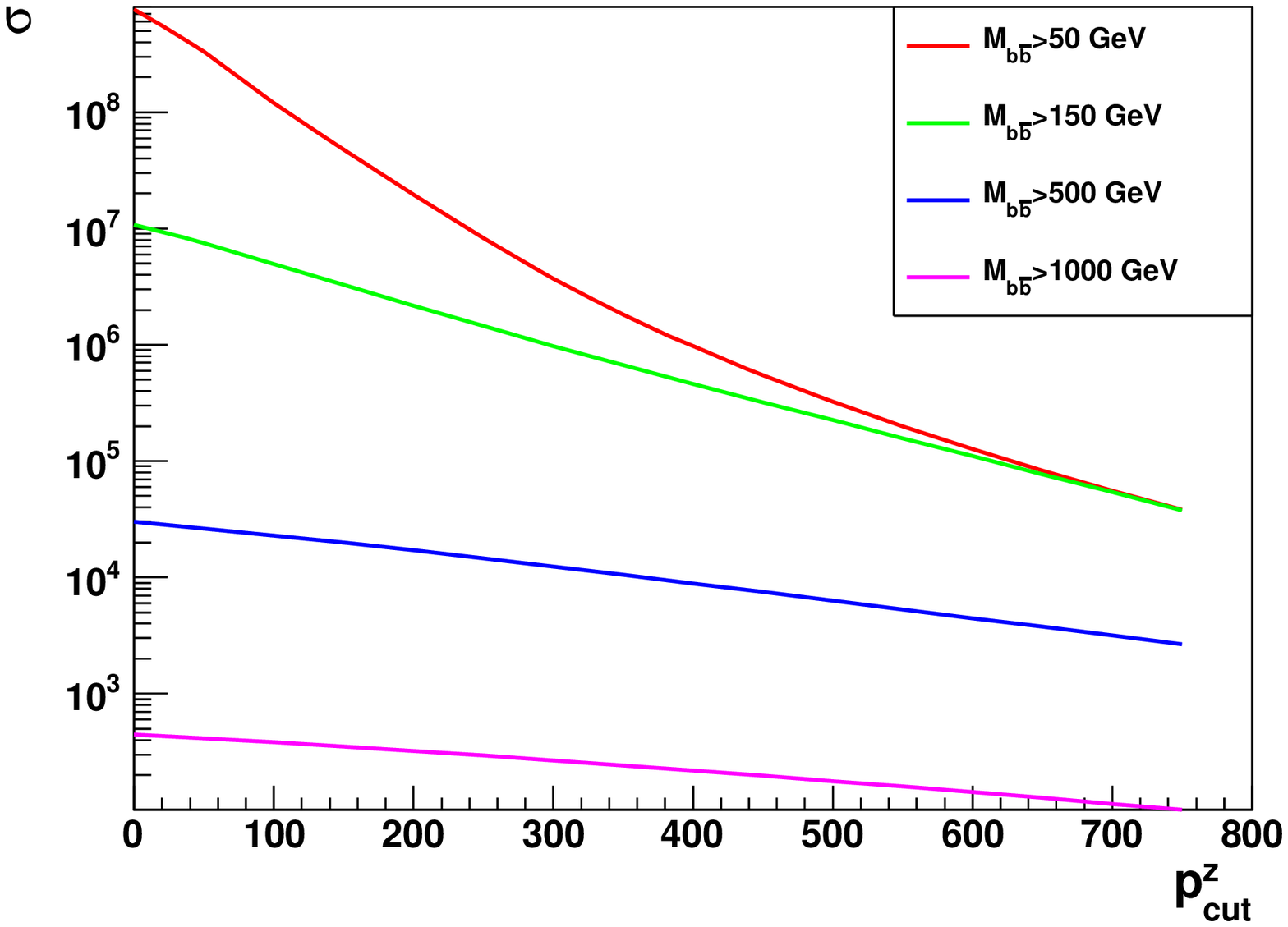}}} \centerline{\hbox{
\includegraphics[height=5cm]
{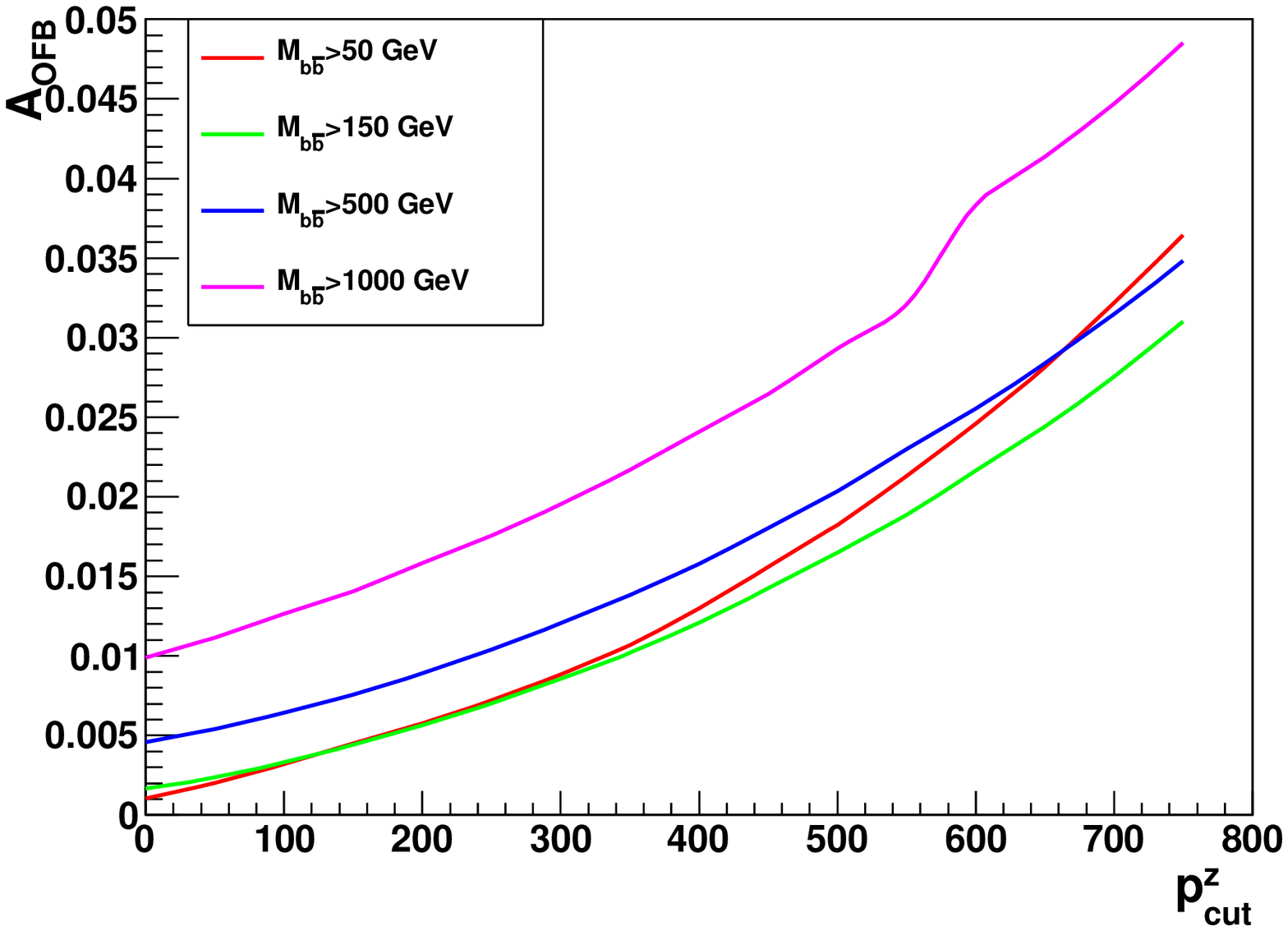}
\includegraphics[height=5cm]
{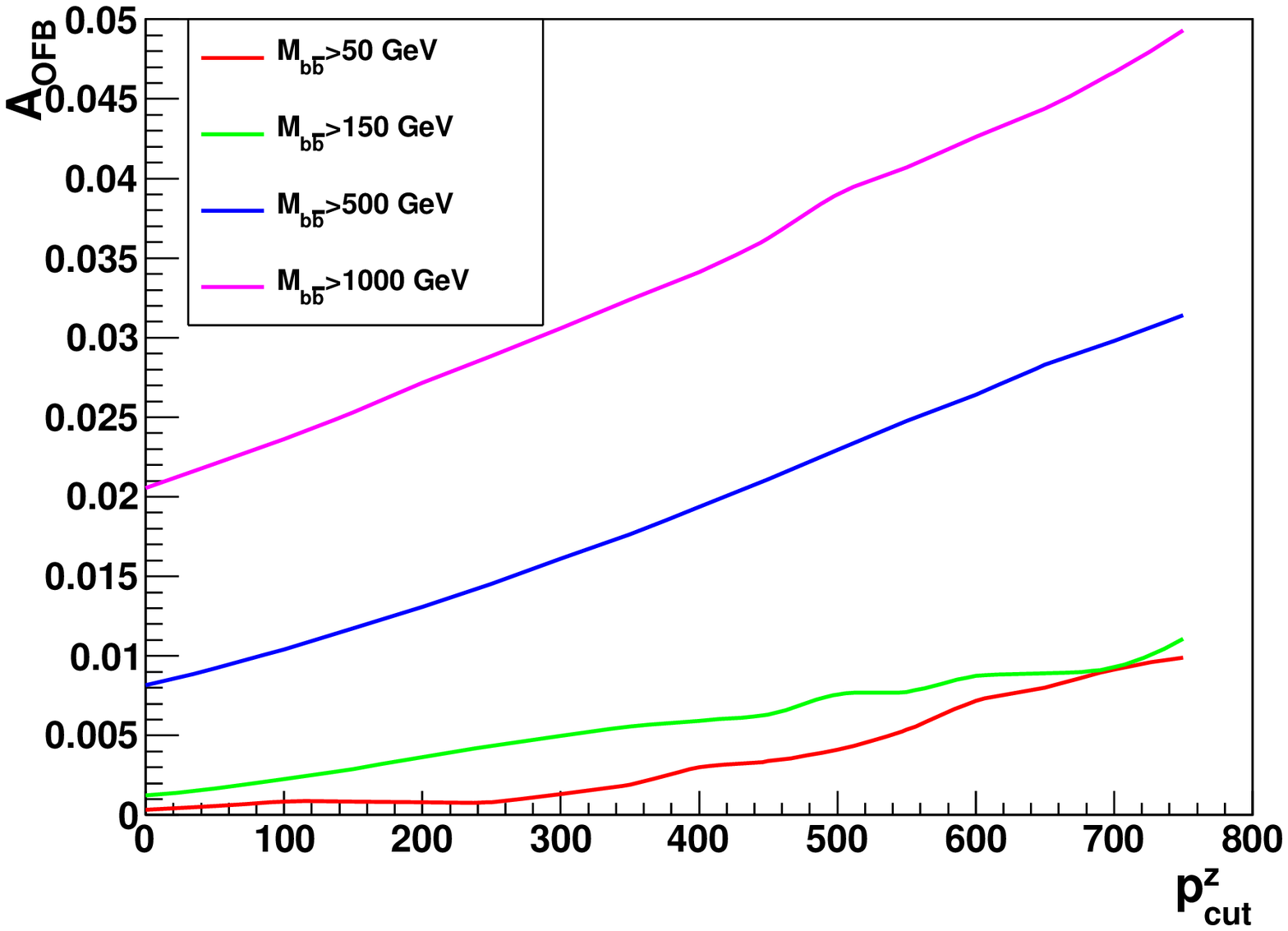}}} \centerline{\hbox{
\includegraphics[height=5cm]
{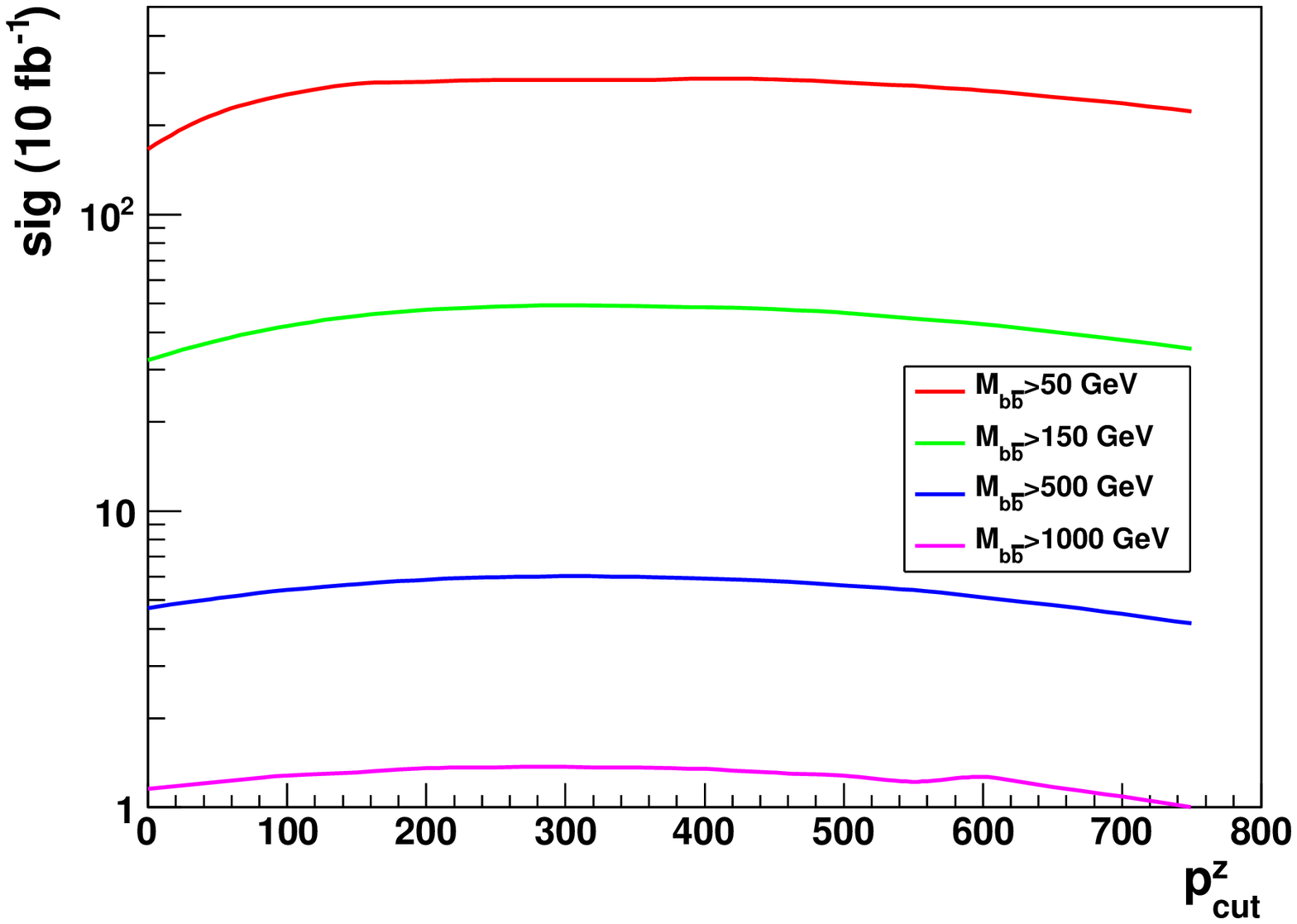}
\includegraphics[height=5cm]
{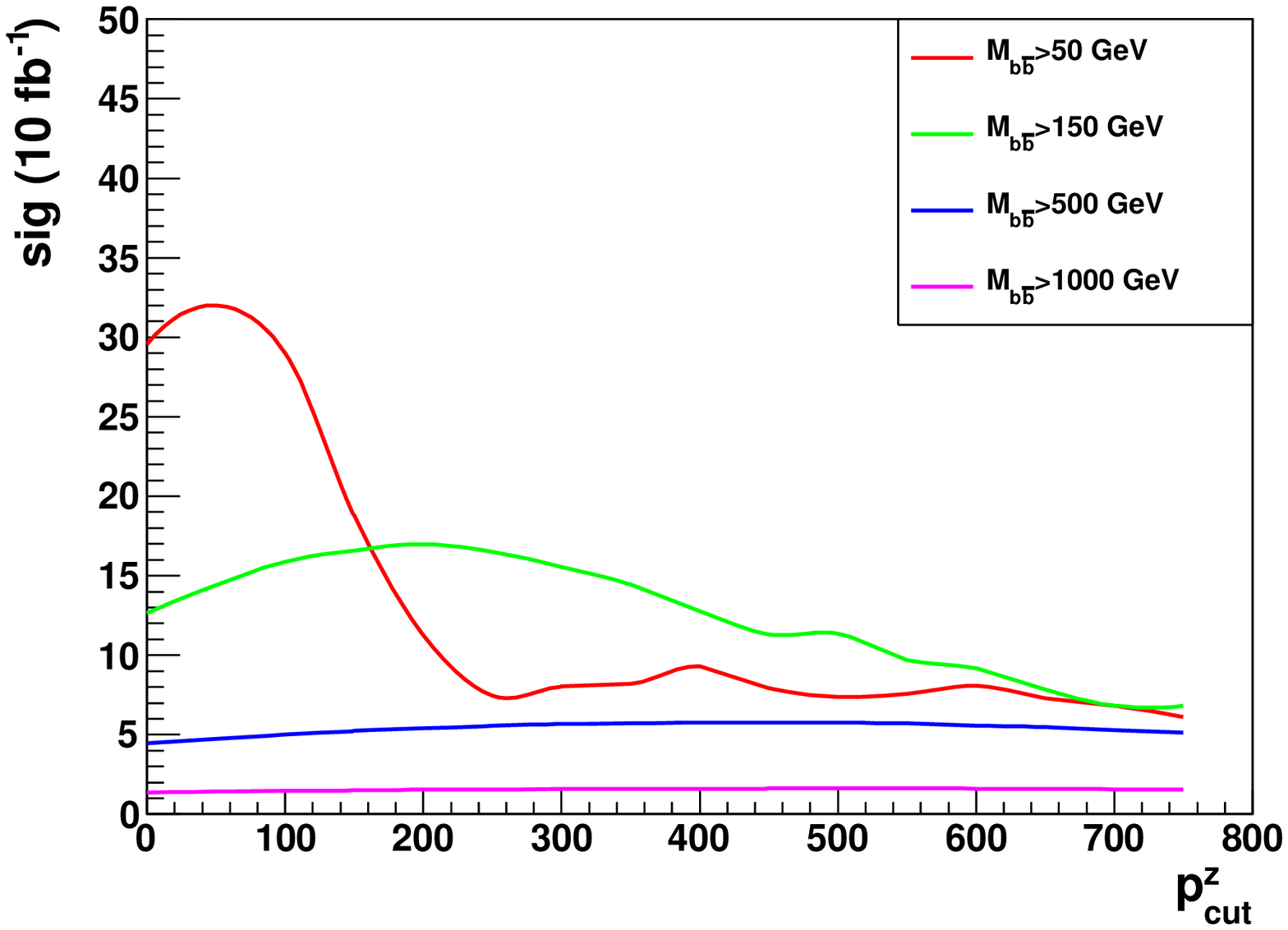}}} \caption{\label{7TeVbb}
$\sigma^A$, $\sigma$, $A_{\rm OFB}$, and ${\rm sig}$ as a function
of $P_{b\bar{b}}^z$ for $\sqrt{s}=7~ \mbox{TeV}$. For the right
(left) column plots $b$ jet cuts are (not) applied.}
\end{figure}

\begin{figure}[htbp]
\centerline{\hbox{
\includegraphics[height=5cm]
{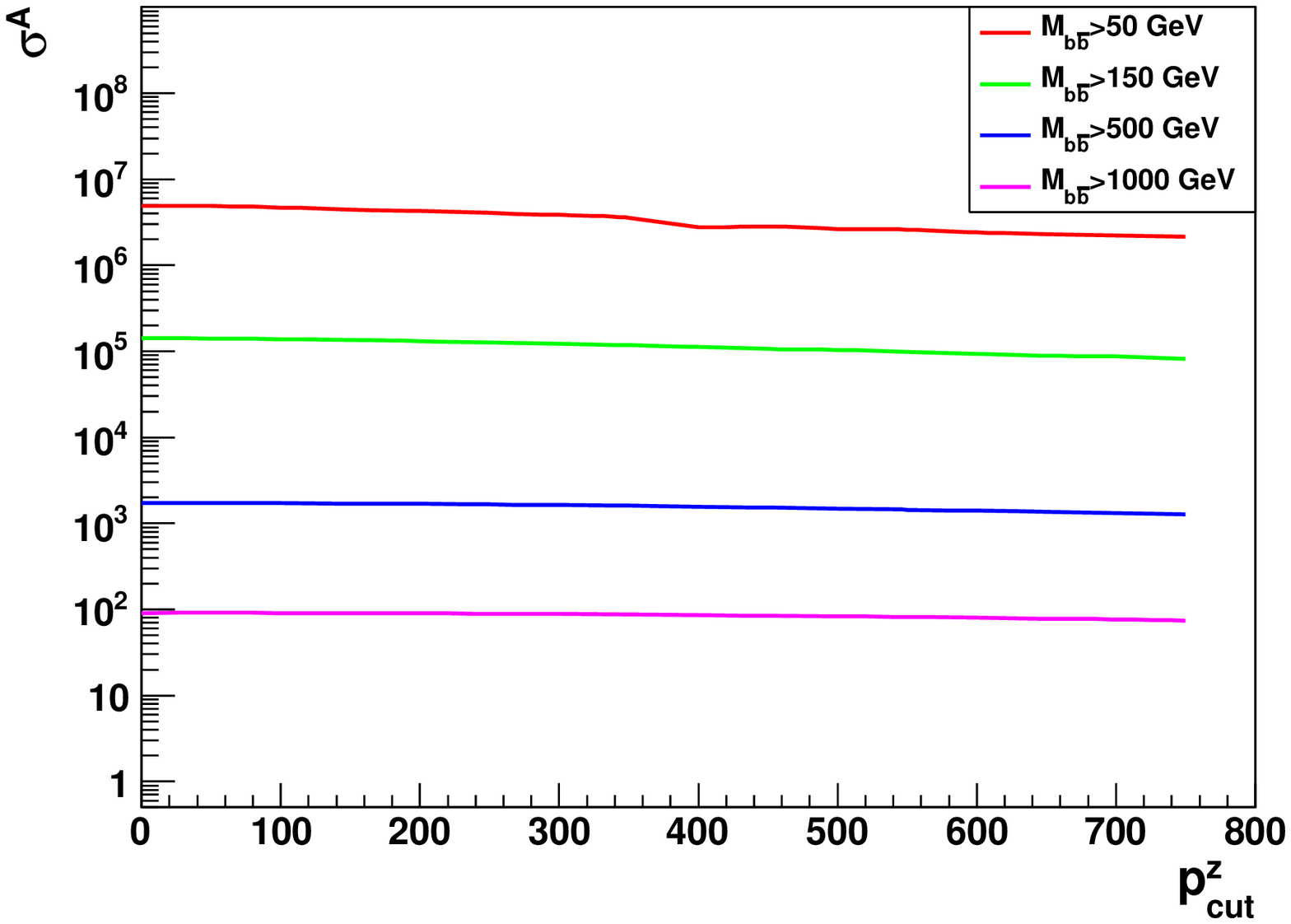}
\includegraphics[height=5cm]
{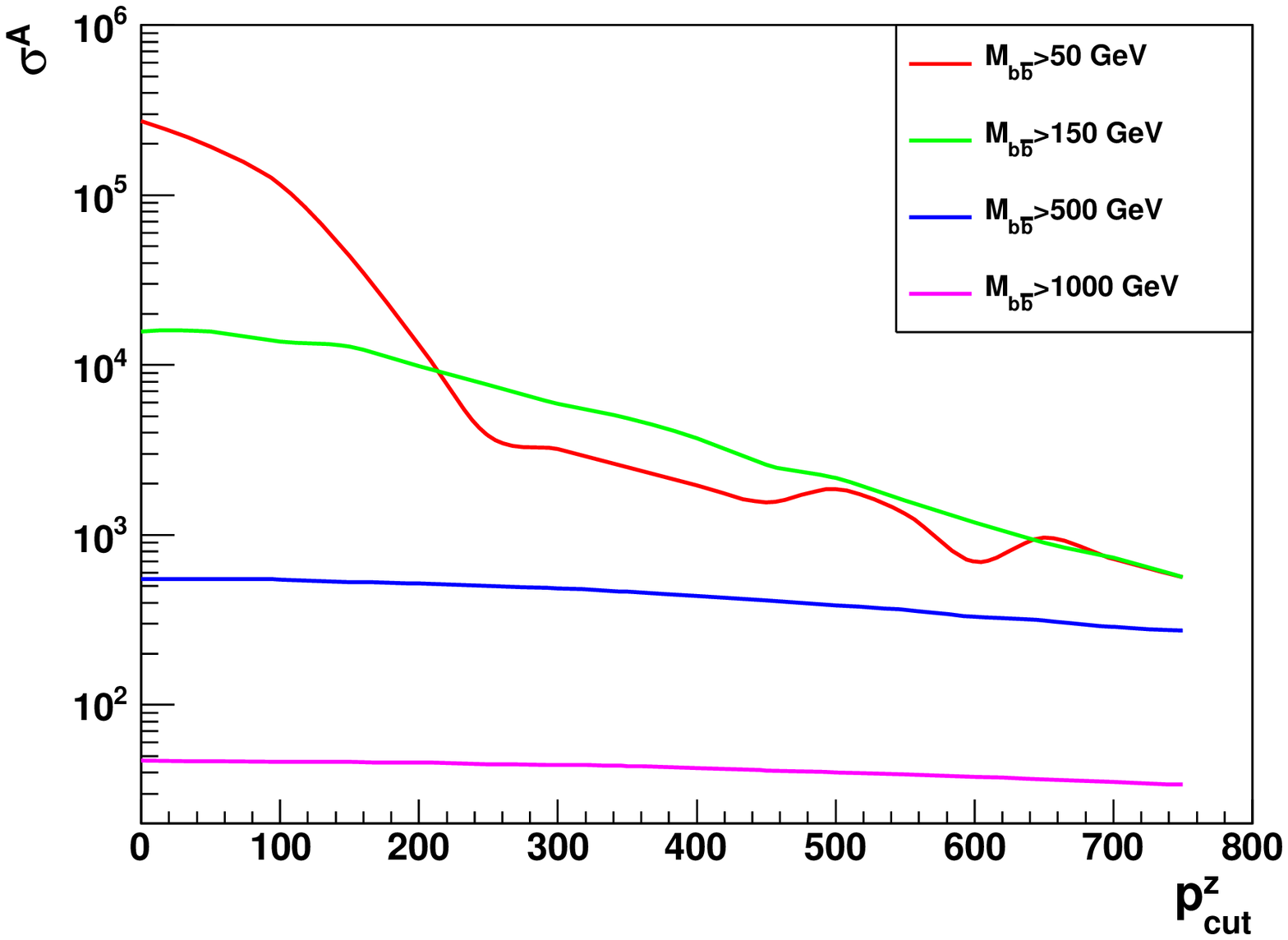}}} \centerline{\hbox{
\includegraphics[height=5cm]
{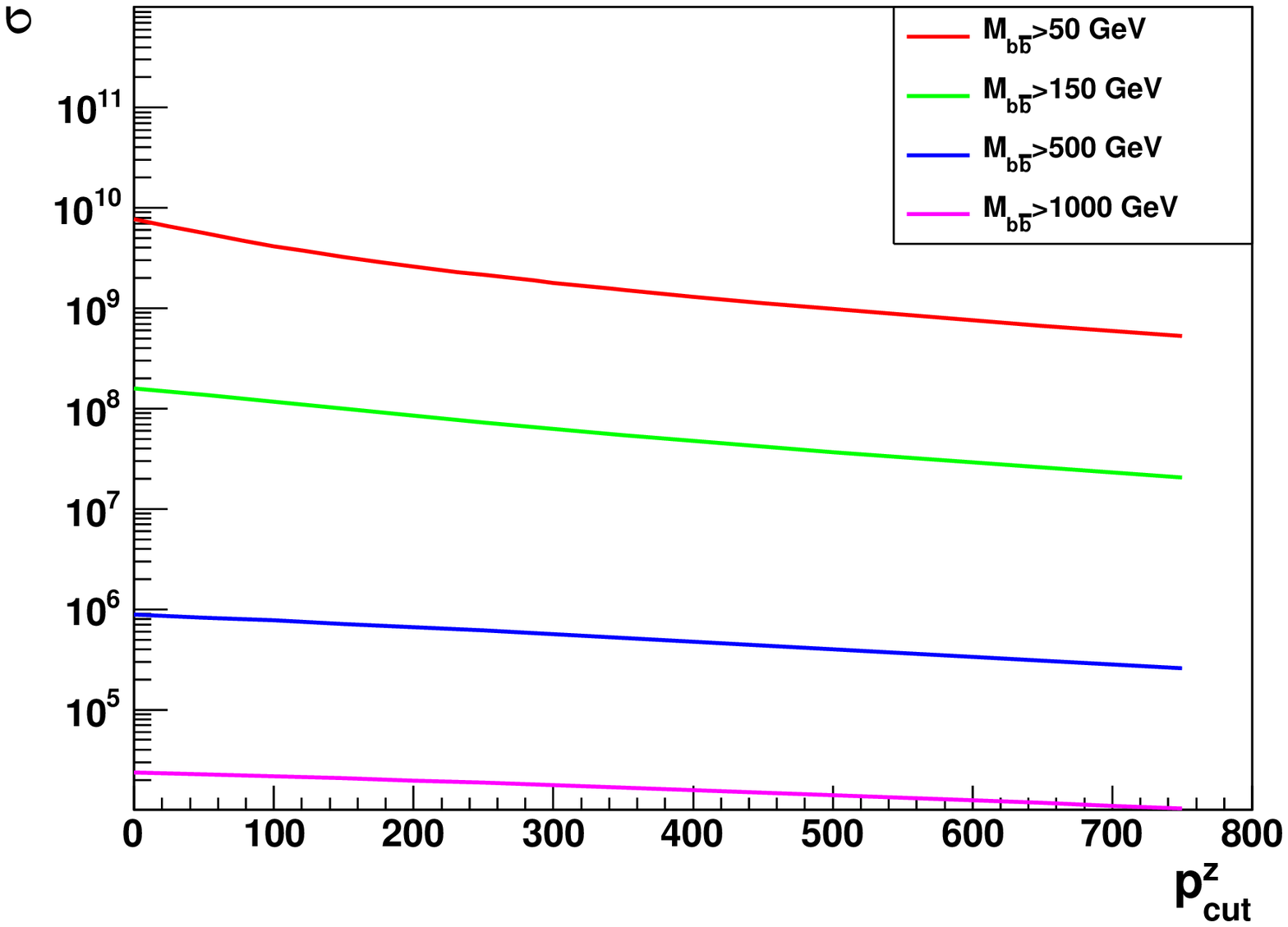}
\includegraphics[height=5cm]
{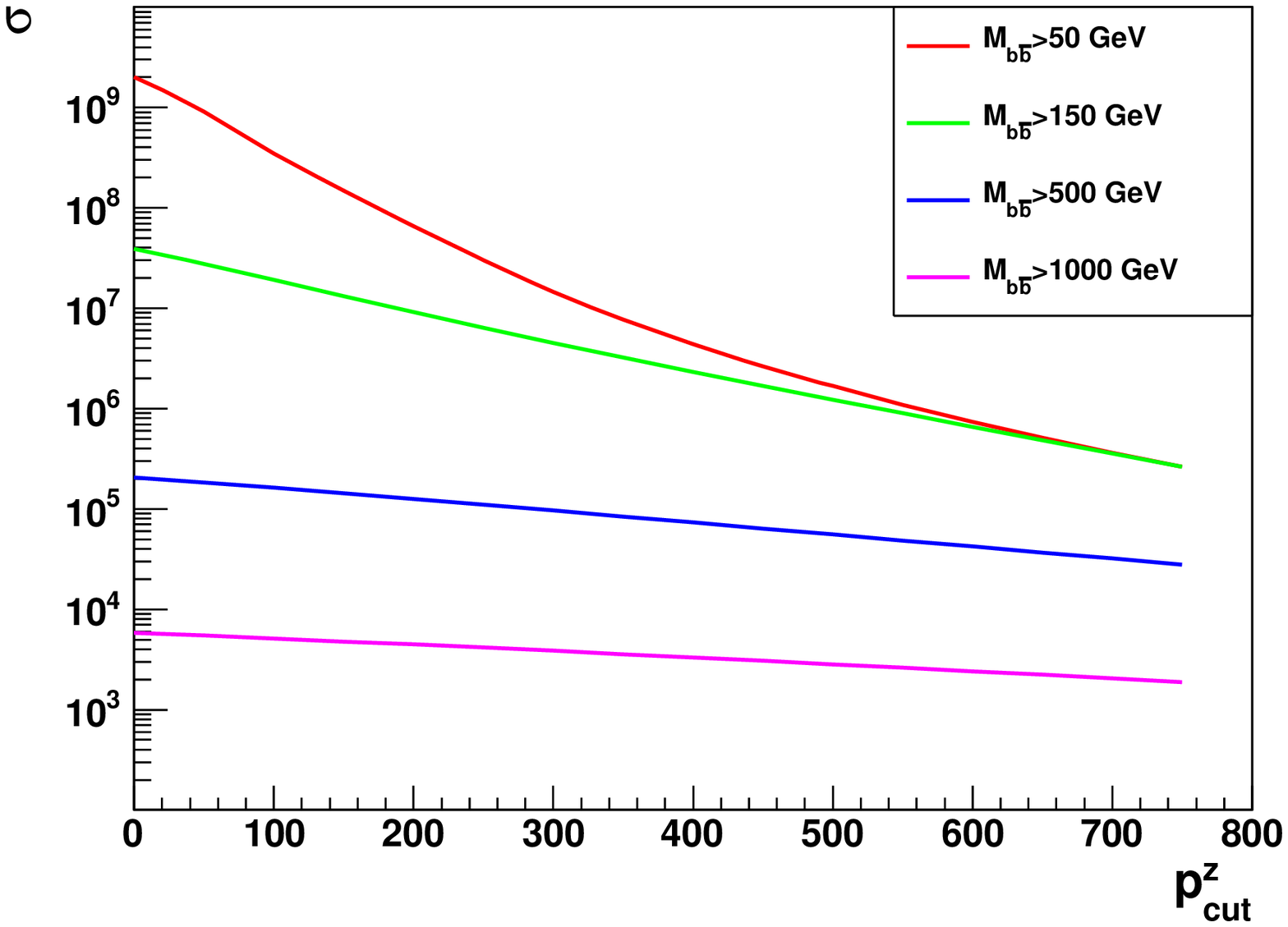}}} \centerline{\hbox{
\includegraphics[height=5cm]
{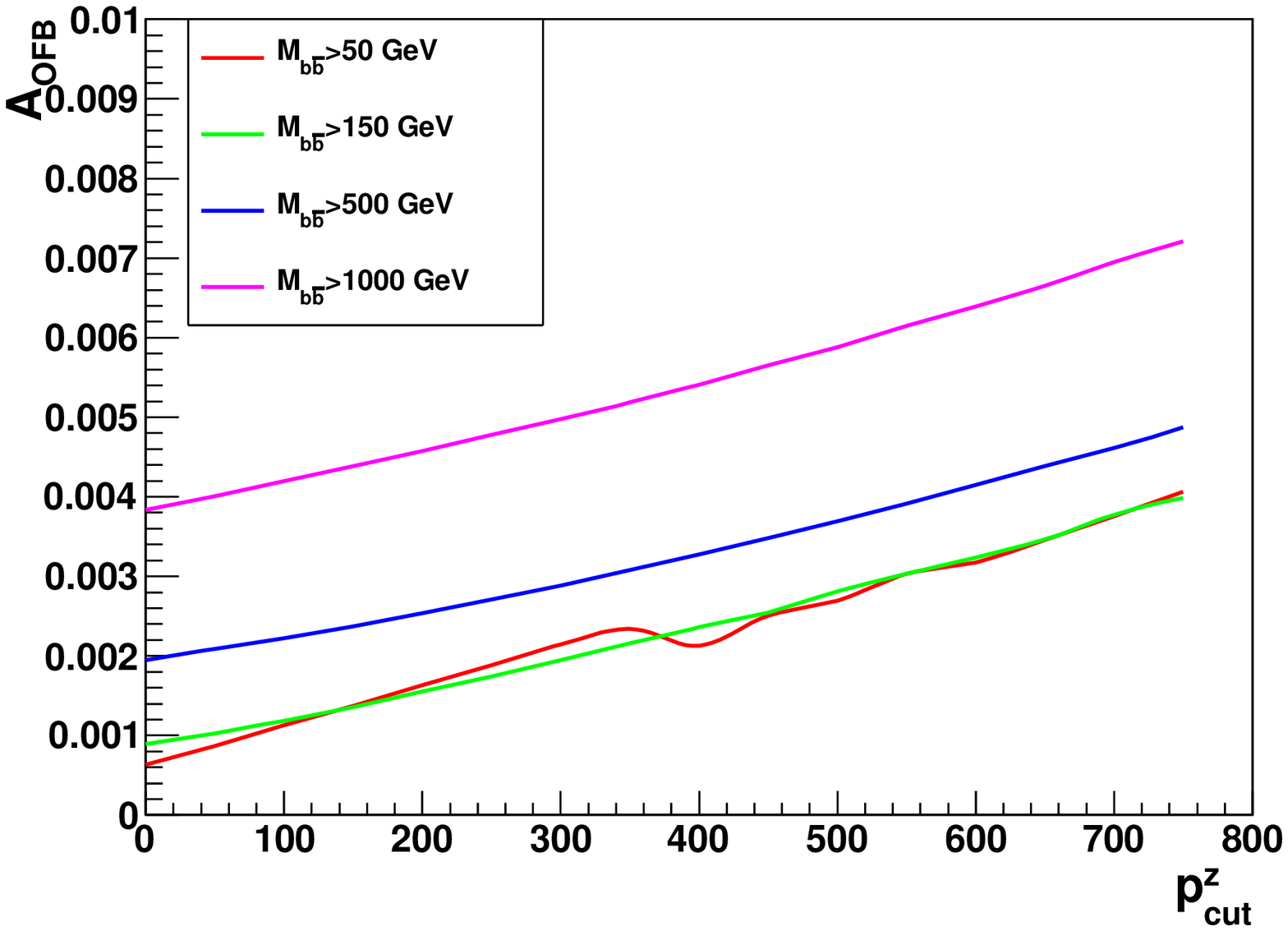}
\includegraphics[height=5cm]
{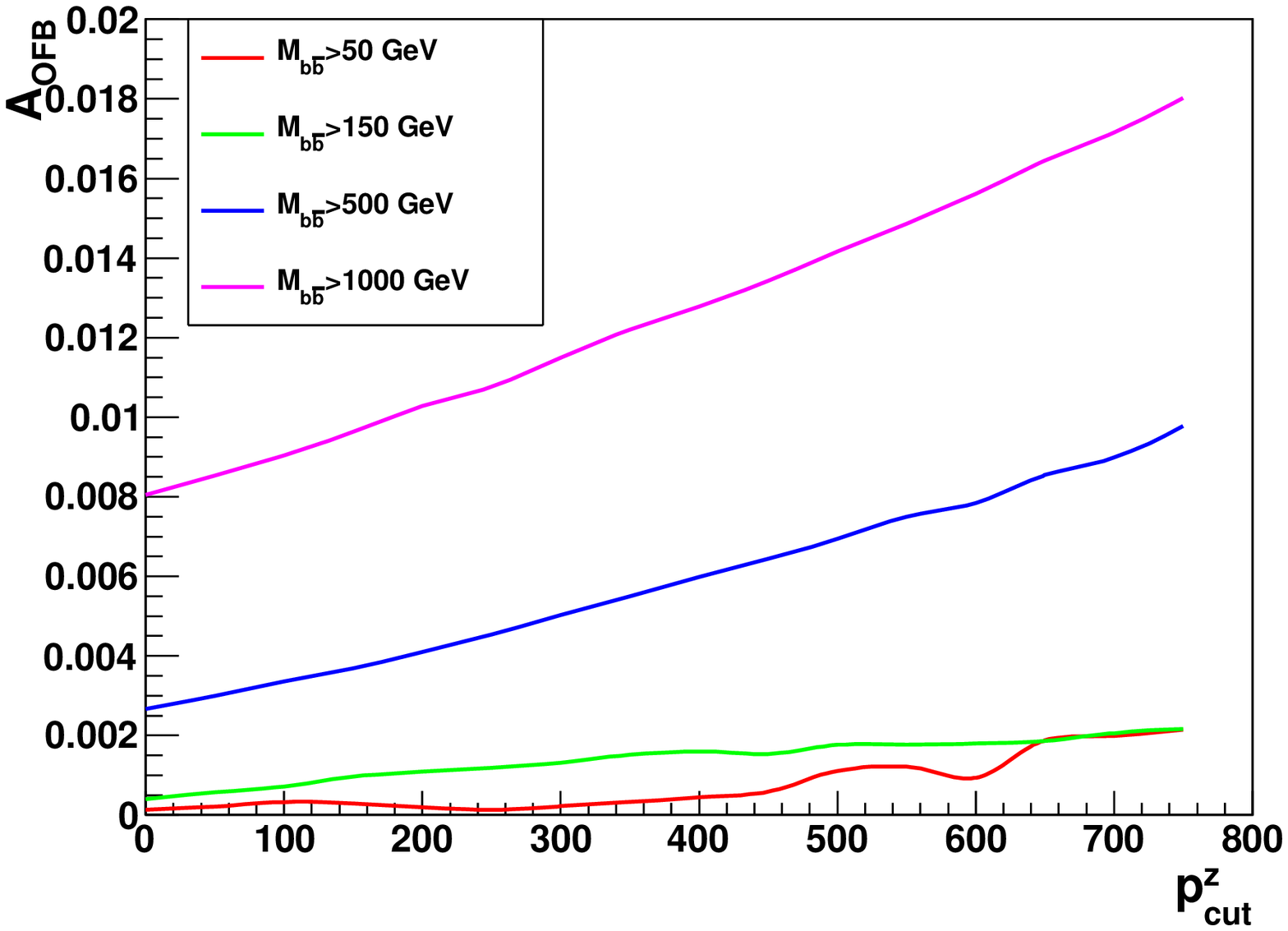}}} \centerline{\hbox{
\includegraphics[height=5cm]
{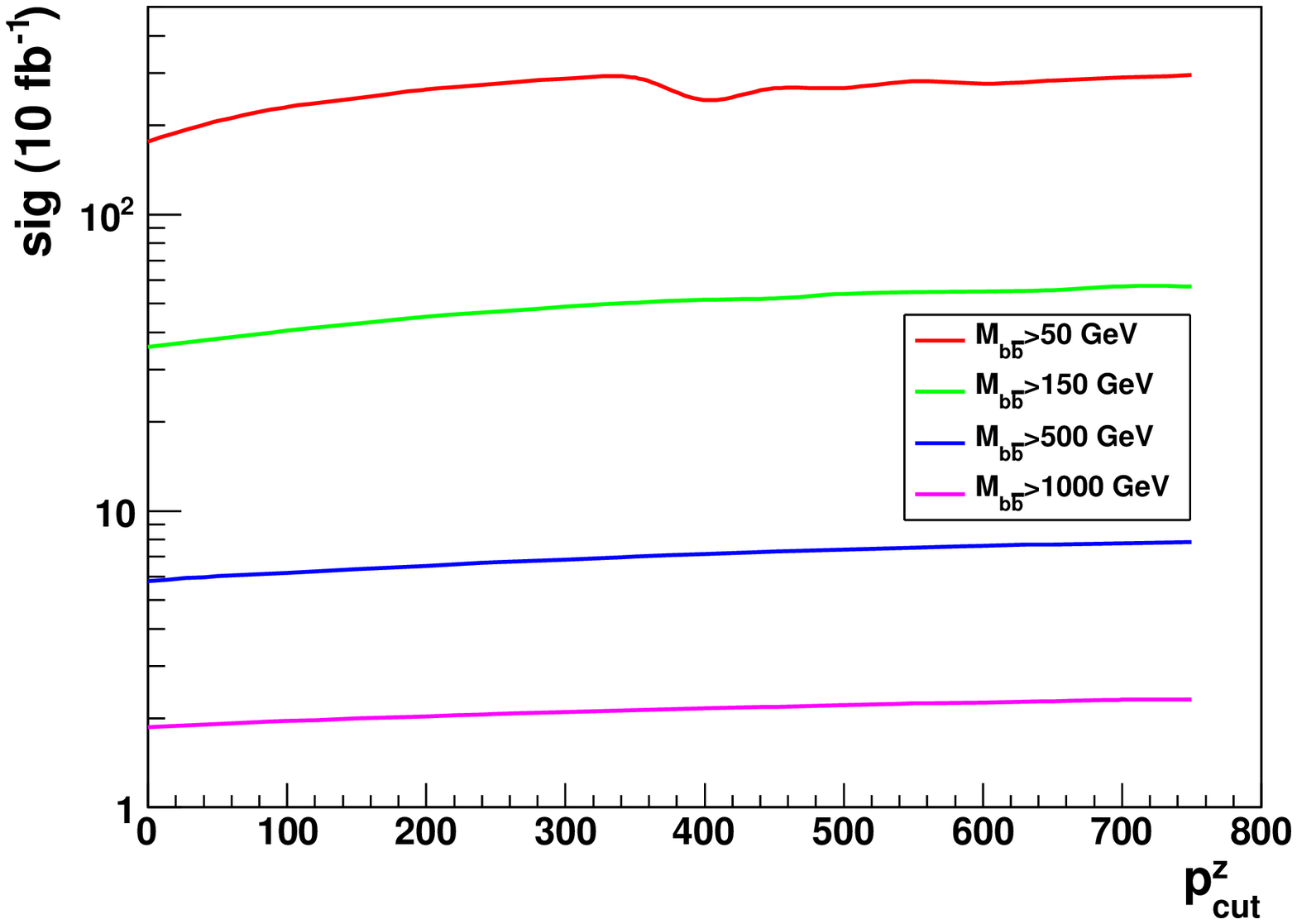}
\includegraphics[height=5cm]
{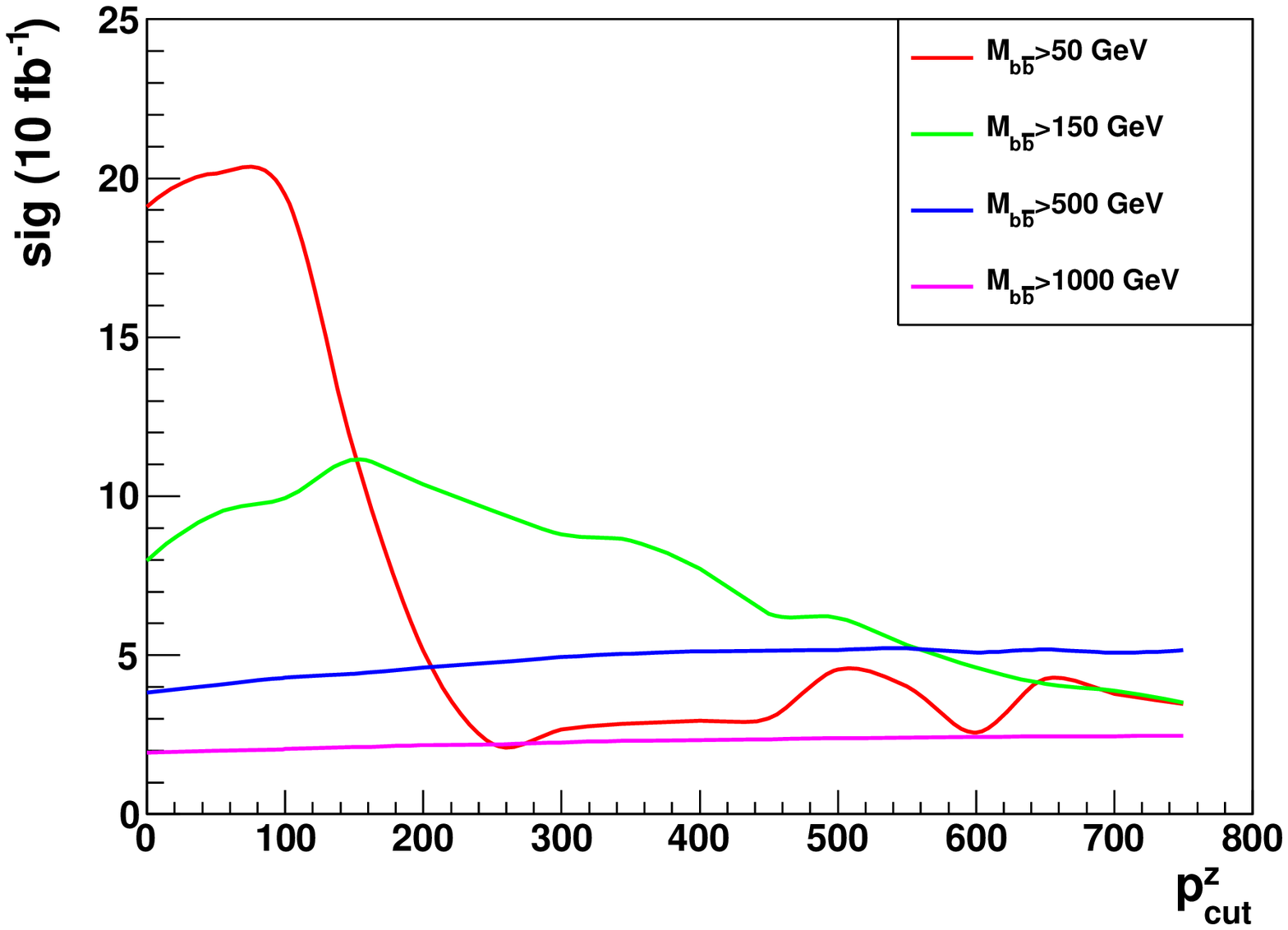}}} \caption{\label{14TeVbb}
Same as Fig. \ref{7TeVbb} except $\sqrt{s}=14~\mbox{TeV}$. }
\end{figure}

Figure \ref{7TeVbb} shows the asymmetric cross section $\sigma^A$,
symmetric cross section $\sigma$, $A_{\rm OFB}$ and significance
${\rm sig}$ ( with $\cal{L}=10 {\rm fb}^{-1}$) as a function of
$P_{\rm cut}^z$ without and with b jet cut for
$\sqrt{s}=7~\mbox{TeV}$. From the left column plots, we see that
both $\sigma^A$ and $\sigma$ drop with the increase of $P_{\rm
cut}^z$. $\sigma$ decreases even faster so $A_{\rm OFB}$ rises with
the increase of $P_{\rm cut}^z$. This is due to two reasons. First,
as mentioned in above sections, $A_{\rm OFB}$ will be polluted by
the negative contributions to asymmetric cross section in the case
that the sea quark's momentum is larger than the valence quark's
momentum. These events locates mostly in small $P_{b\bar{b}}^z$
region. A larger cut of $P_{b\bar{b}}^z$ can increase the portion of
positive sign asymmetric cross section. Second, due to the
properties of the parton distribution function, the symmetric $gg\to
b\bar{b}$ events are mostly distributed in the small
$P_{b\bar{b}}^z$ region and the asymmetric events are likely to be
highly boosted along the z direction. The $P_{\rm cut}^z$ can remove
more symmetric backgrounds.

The right column plots indicate that $b$ jet cut can change the
above distributions significantly. Most of the events are lost. Low
$M_{b\bar{b}}$ events are more sensitive to $b$ jet cuts than the
large $M_{b\bar{b}}$ events, which indicates that they tends to have
lager $\eta$ and small $P_T$. The $\sigma^A$ plot shows that
asymmetric events with $50\mbox{GeV}<M_{b\bar{b}}<150\mbox{GeV}$ and
$P_{b\bar{b}}^z>250\mbox{GeV}$ are completely removed by $b$ jet
cut.
%As $A_{\rm OFB}$ and significance curves corresponding to
%$M_{b\bar{b}}>50, 150\mbox{GeV}$ are not so stable, we suggests that
%in real experiment measurements, $M_{b\bar{b}}$ should be greater
%than that although in paid of statistics.
Generally speaking, $b$ jet cut is a very strong constraint on the
forward-backward asymmetry measurements at the LHC. Because of the
high energy at the LHC, most of the highly longitudinal boosted b
quark are difficult to record in the detector. The significance can
drop greatly in the real experimental environment. The situation
becomes even harder for $\sqrt{s}=14~\mbox{TeV}$ as shown in
Fig.\ref{14TeVbb}. Because of the even larger longitudinal boosts,
precision measurements require higher integrated luminosity for
$\sqrt{s}=14~\mbox{TeV}$.

As mentioned above, forward backward asymmetry of the bottom quark
has been measured at the LEP near the $Z$ pole. As an $e^+ e^-$
collider, LEP's measurement can only show the EW contribution to the
asymmetric cross section of the $b$ quark. Although Tevatron has
already measured the top quark forward backward asymmetry, the
forward backward asymmetry of the bottom quark at the hadron
collider, which are mainly contributed from the QCD interference
diagrams, is still not  investigated yet. According to the above
studies, it is still hopeful that this QCD induced asymmetric
signature can be seen at the LHC detectors. Previous measurements
show the top quark forward backward asymmetry has about 2 standard
deviation from the QCD prediction \cite{CDFAfb53, D0AFB43}. Many new
physics beyond the SM have been studied to explain this novel
signature \cite{Frampton:2009rk,Shu:2009xf,Chivukula:2010fk,
Jung:2009jz, Cheung:2009ch,
Cao:2010zb,Djouadi:2009nb,Jung:2009pi,Cao:2009uz,Barger:2010mw,
Arhrib:2009hu, Xiao:2010hm,Bauer:2010iq, Xiao:2010ph}. It will be
very interesting to see whether the bottom quark, which belongs to
the third generation, has similar deviation as the top quark.

\section{Bottom quark one-side forward-backward asymmetry $A_{\rm OFB}^b$ around Z-pole \label{four}}

In Sec. \ref{three} we investigated how to study QCD induced $A_{\rm
OFB}^b$.  In a wide energy regime, electroweak contribution to the
forward-backward asymmetry is much less than the QCD contribution.
By requiring the final $b\bar{b}$ invariant mass near the $Z$-pole,
most of QCD contribution to the asymmetric cross section can be
suppressed and the remaining QCD and the electro-weak induced
asymmetric cross section can be comparable. So, the LHC can also
explore the electroweak induced asymmetric cross section for the
bottom quark. Such measurements are the nearest hope to do the
cross-check to the corresponding measurements at the LEP.

The one-side forward-backward asymmetry near the $Z$ pole can be
defined as in Eq. \ref{AOFB} with
\begin{equation} F_\pm= \left. \left(\sigma( \Delta Y>0)\pm \sigma(\Delta
Y<0)\right)\right|_{P_{b\bar{b}}^z>P_{\rm cut}^z, m_Z-\delta
E/2<M_{b\bar{b}}<m_Z+\delta E/2}
\end{equation}
\begin{equation}
B_\pm= \left. \left(\sigma(\Delta Y<0)\pm \sigma(\Delta
Y>0)\right)\right|_{P_{b\bar{b}}^z<-P_{\rm cut}^z, m_Z-\delta
E/2<M_{b\bar{b}}<m_Z+\delta E/2}
\end{equation}
in which $\delta E$ is the energy window near the $Z$ pole. Both EW
and QCD contribution should be included in the calculation. For the
EW processes, we calculate the contribution at the leading order
 $\mathscr{O} (\alpha^2)$. For the QCD processes, we include the NLO QCD contribution in the numerator
 and LO QCD ones in the denominator \cite{Wang:2010du}.

\begin{figure}[htbp]
\centerline{\hbox{
\includegraphics[height=5cm]
{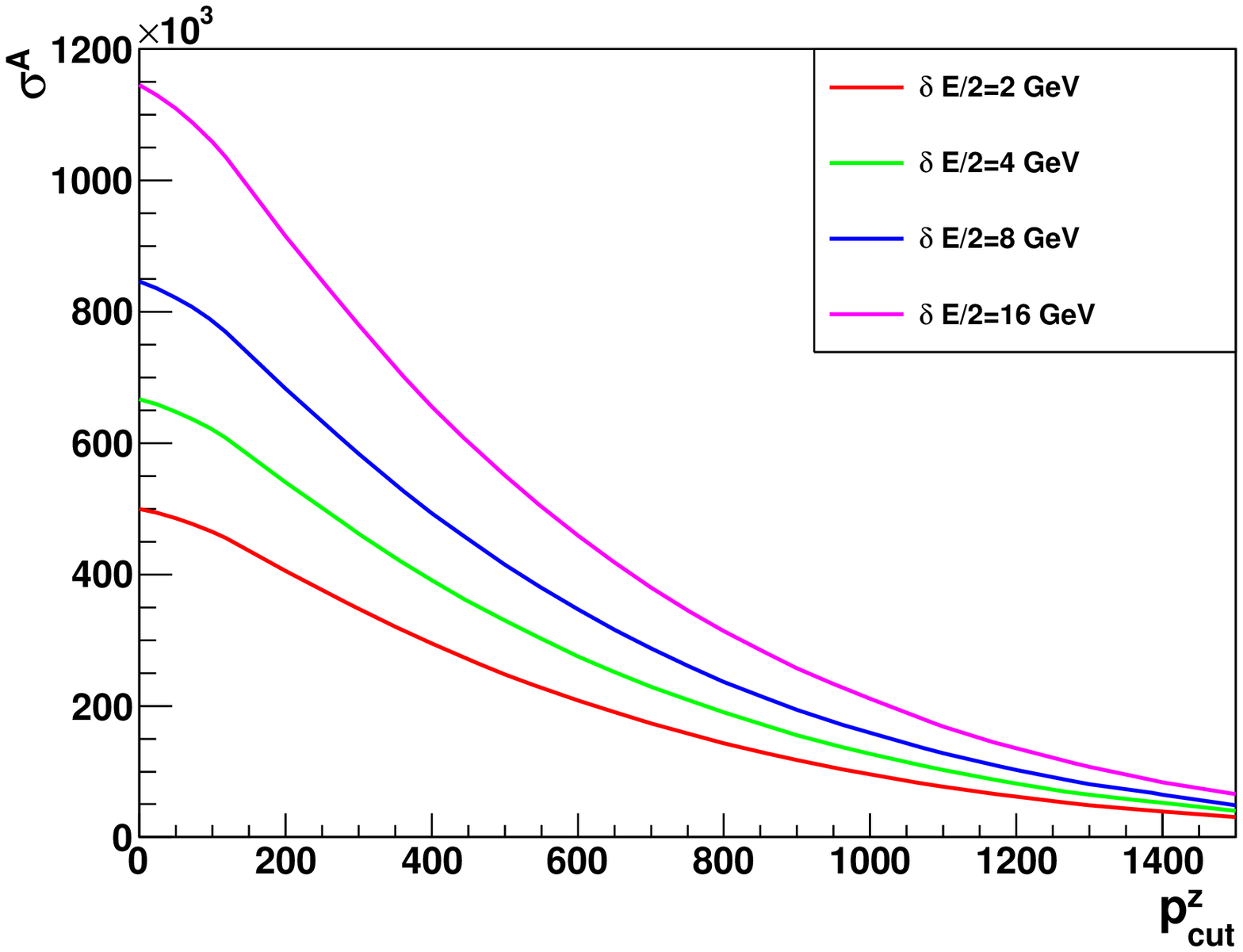}
\includegraphics[height=5cm]
{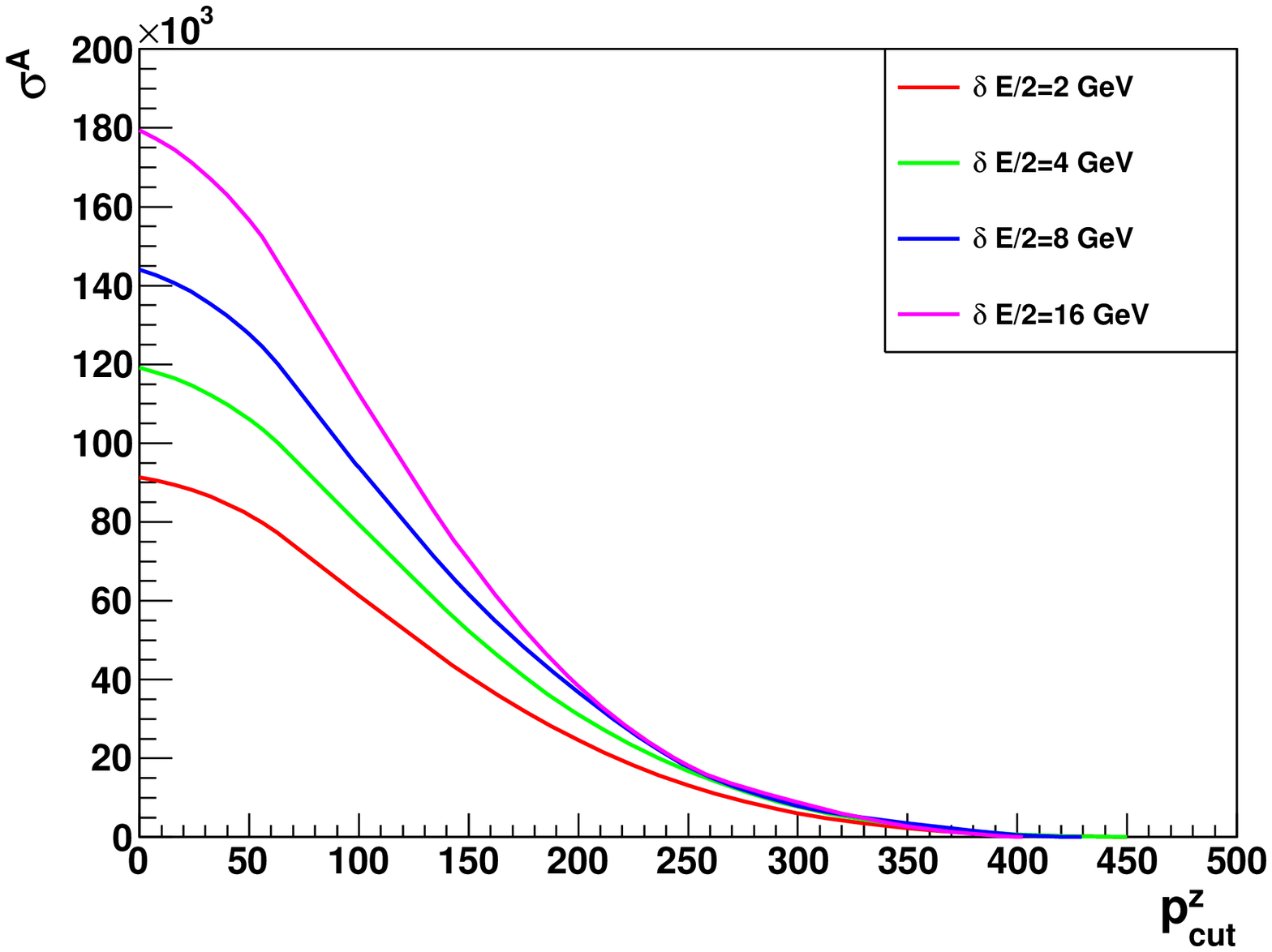}}} \centerline{\hbox{
\includegraphics[height=5cm]
{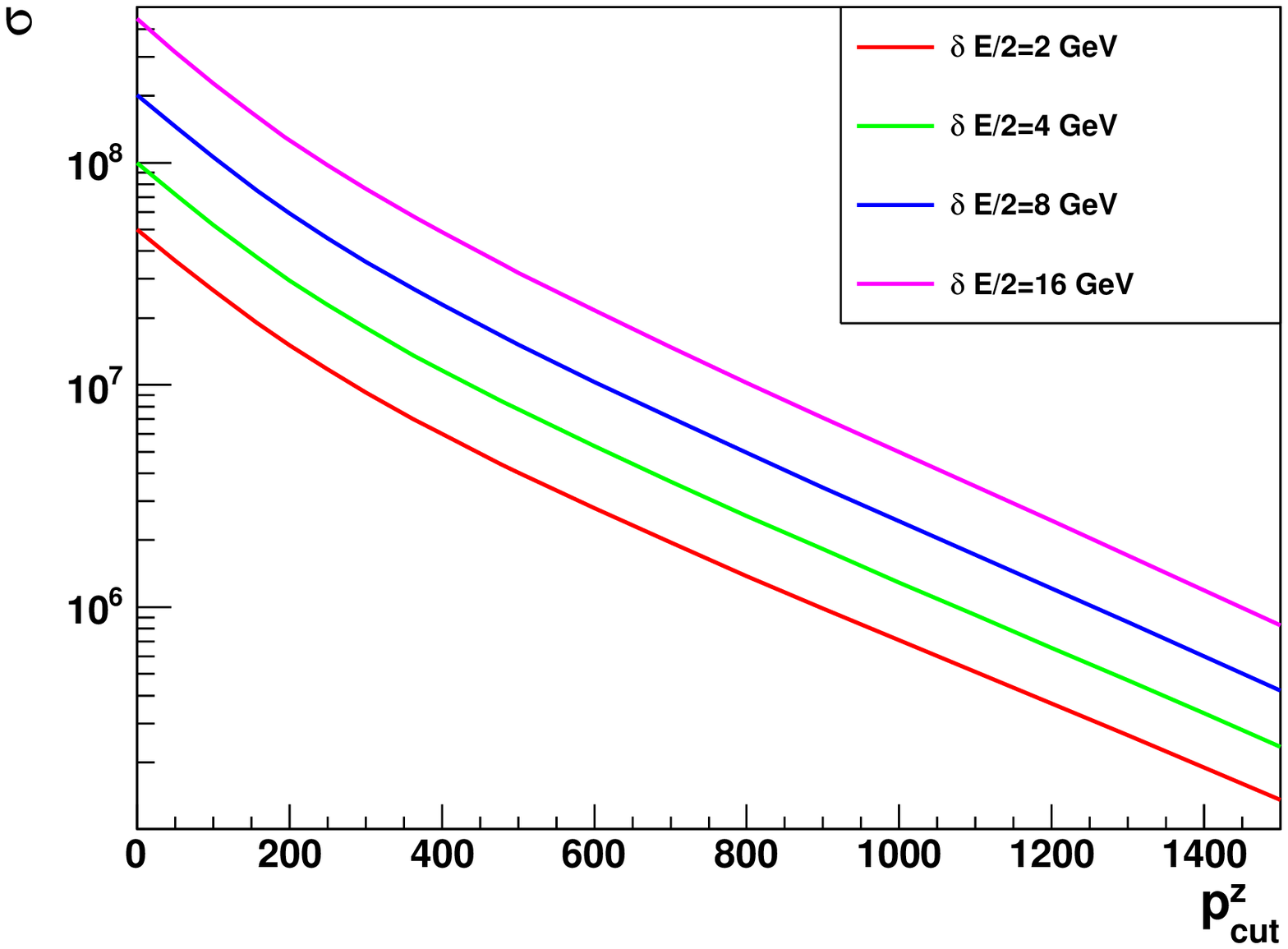}
\includegraphics[height=5cm]
{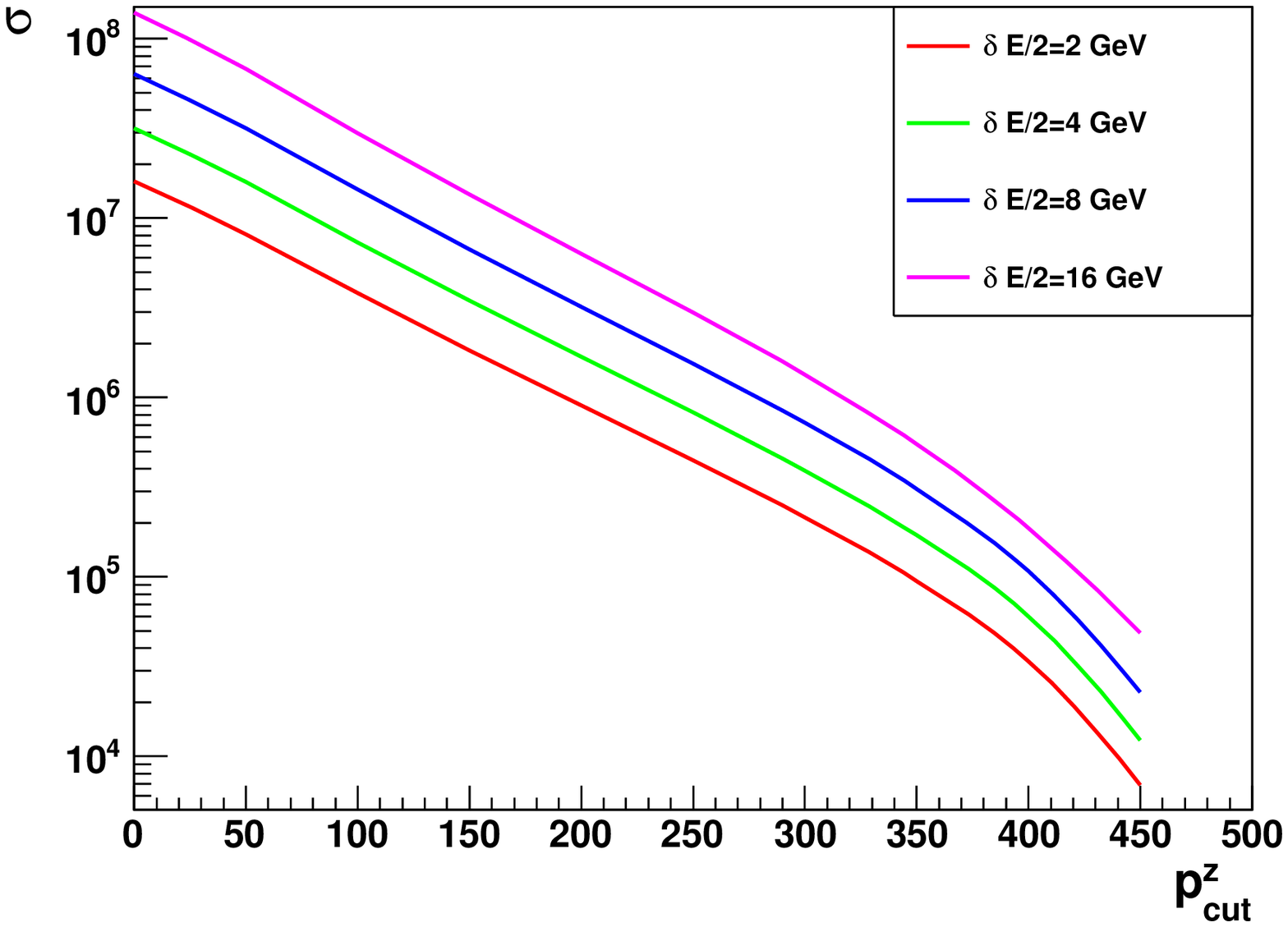}}} \centerline{\hbox{
\includegraphics[height=5cm]
{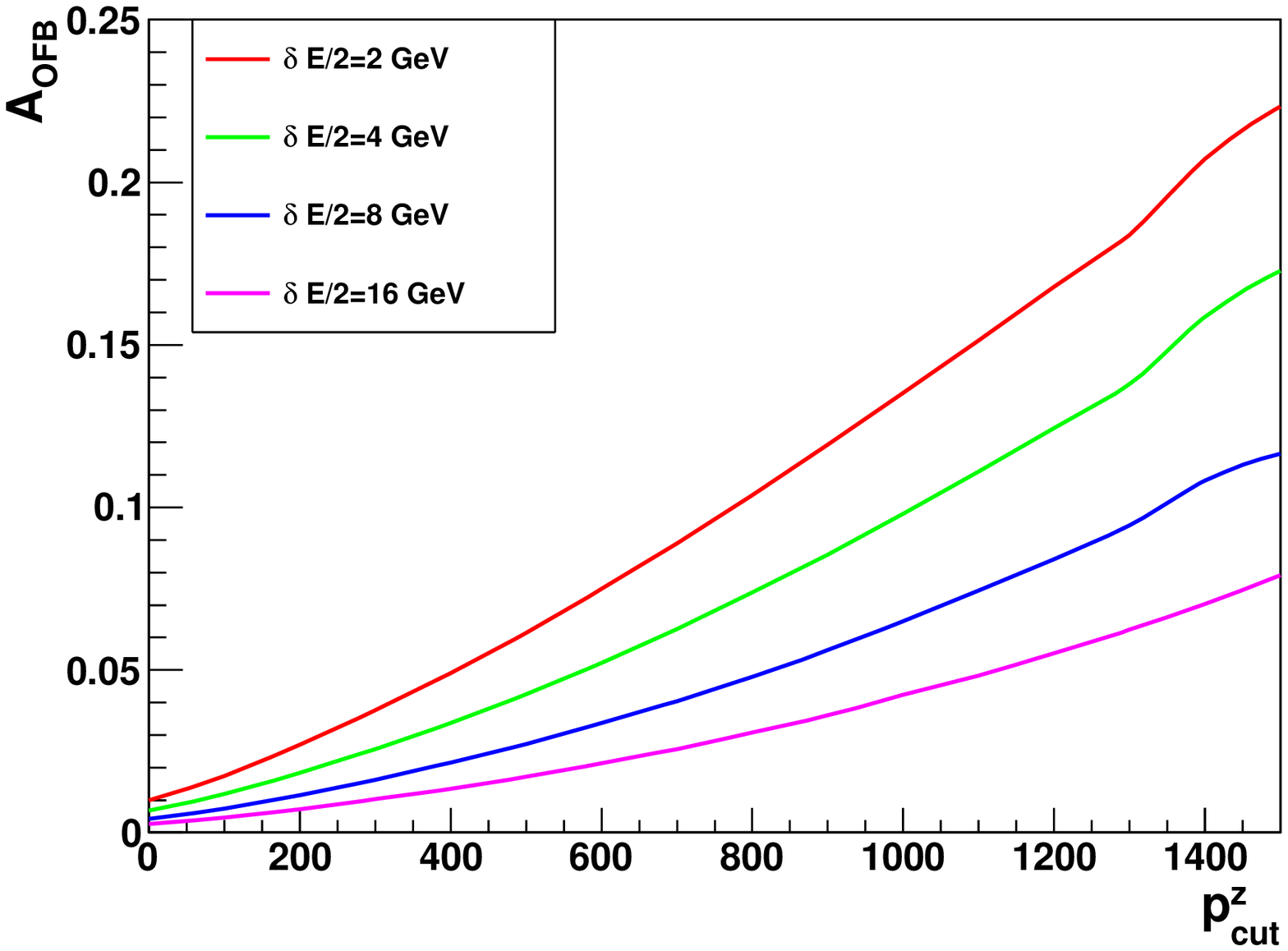}
\includegraphics[height=5cm]
{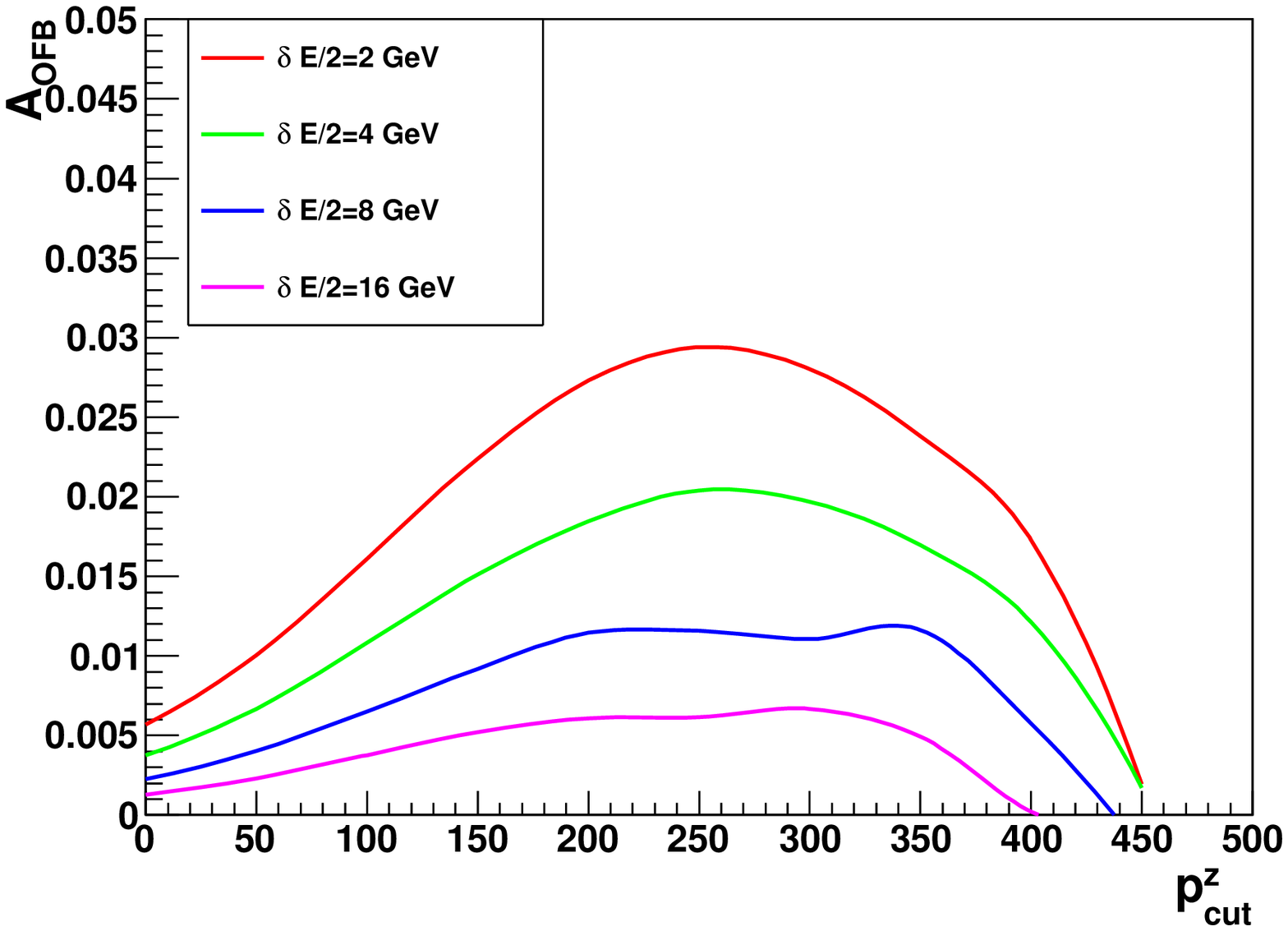}}} \centerline{\hbox{
\includegraphics[height=5cm]
{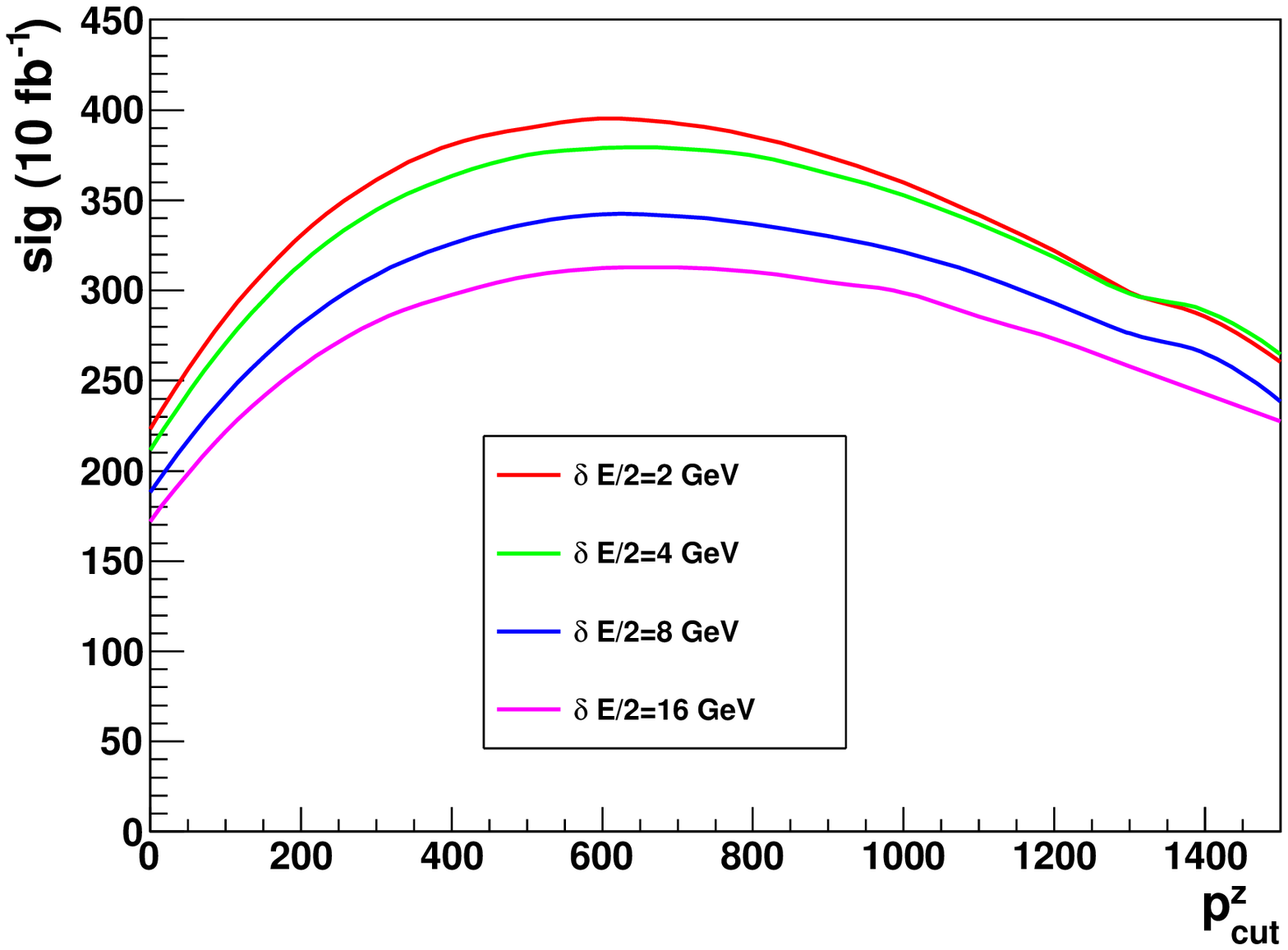}
\includegraphics[height=5cm]
{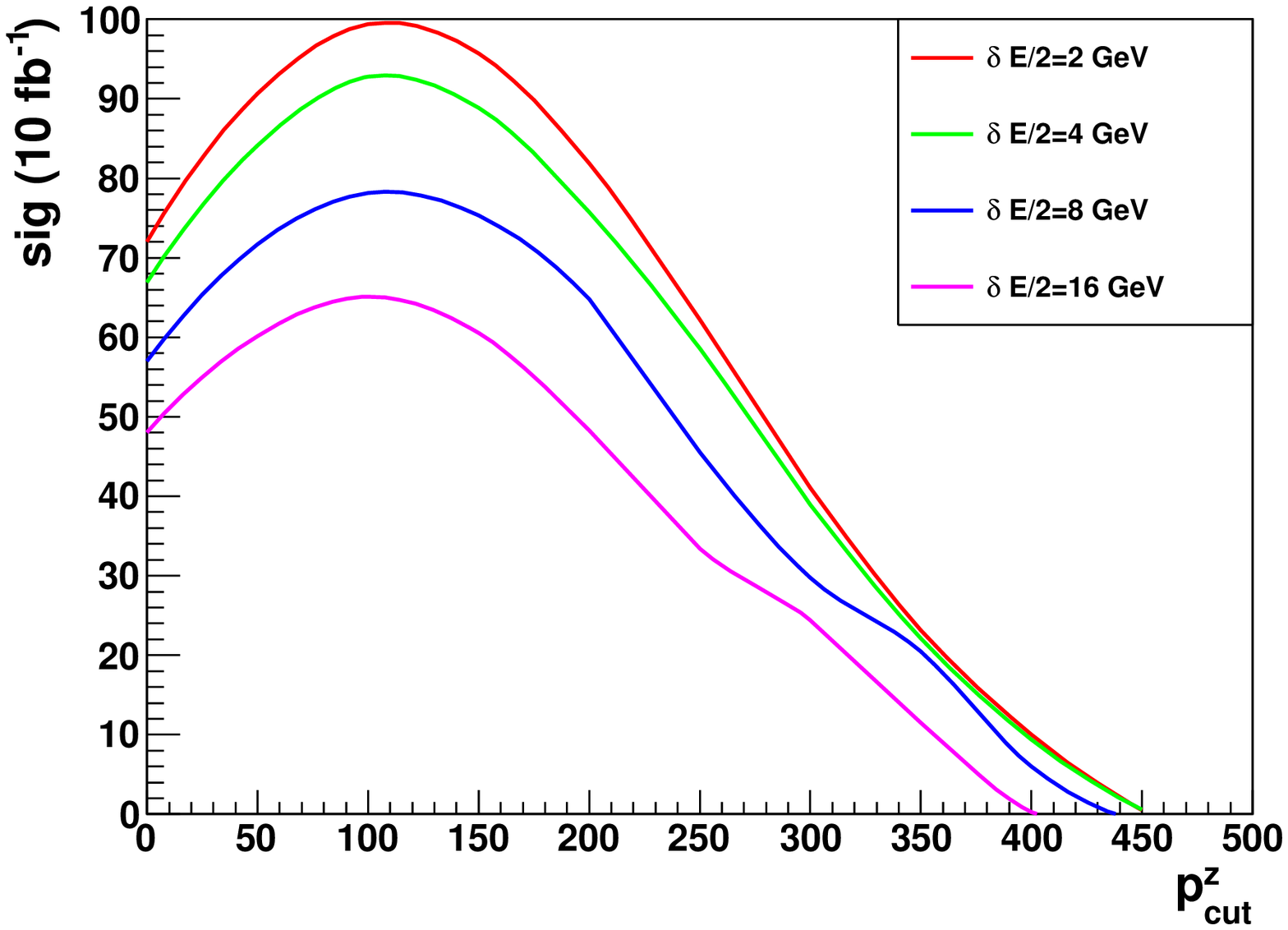}}}
\caption{\label{7TeVbbzpole}$\sigma^A$, $\sigma$, $A_{\rm OFB}$ and
${\rm sig}$ as a function of $P_{\rm cut}^z$ at LHC with
$\sqrt{s}=7~$TeV. For the right (left) column plots $b$ jet cuts are
(not) applied. }
\end{figure}

\begin{figure}[htbp]
\centerline{\hbox{
\includegraphics[height=5cm]
{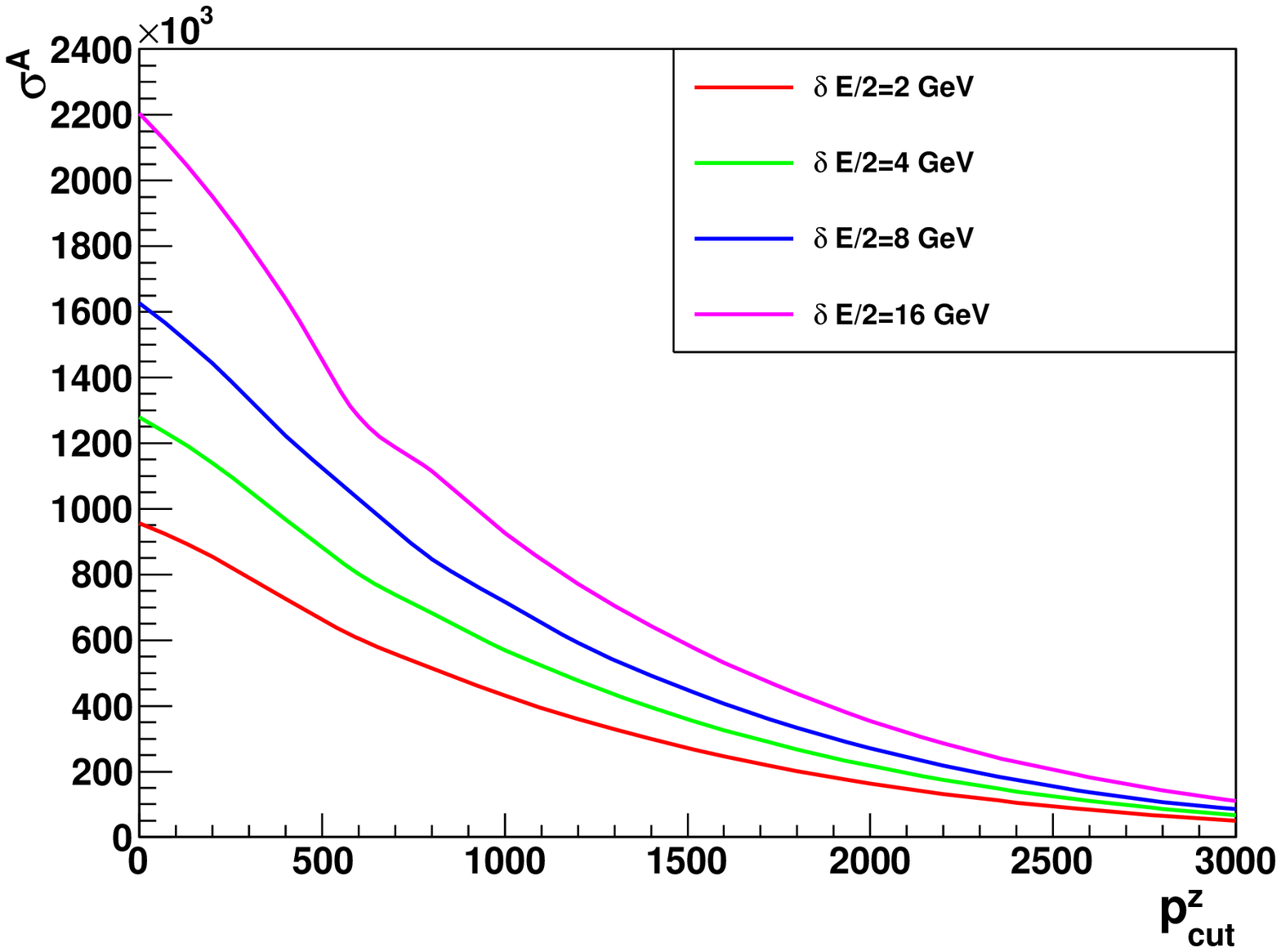}
\includegraphics[height=5cm]
{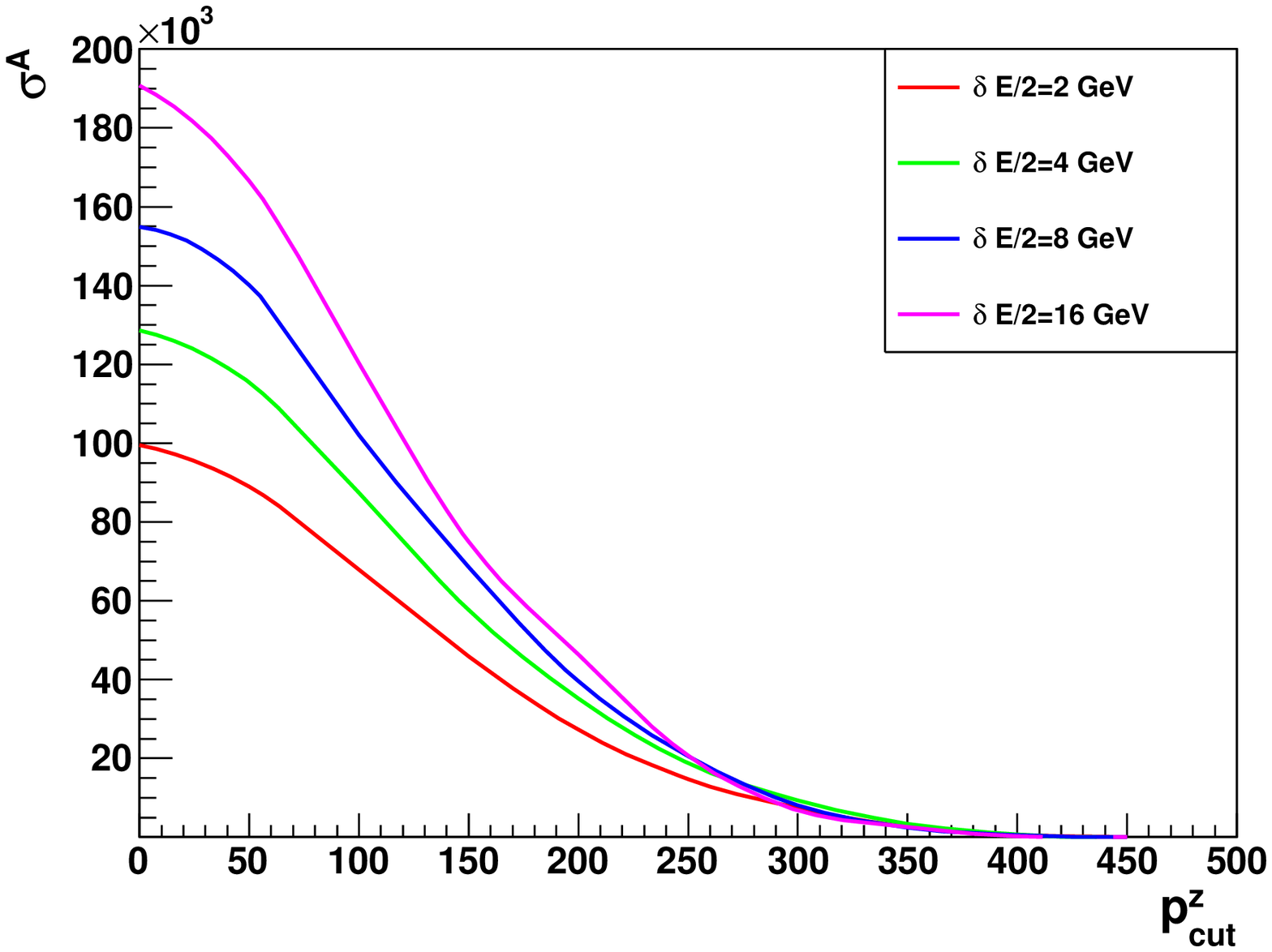}}} \centerline{\hbox{
\includegraphics[height=5cm]
{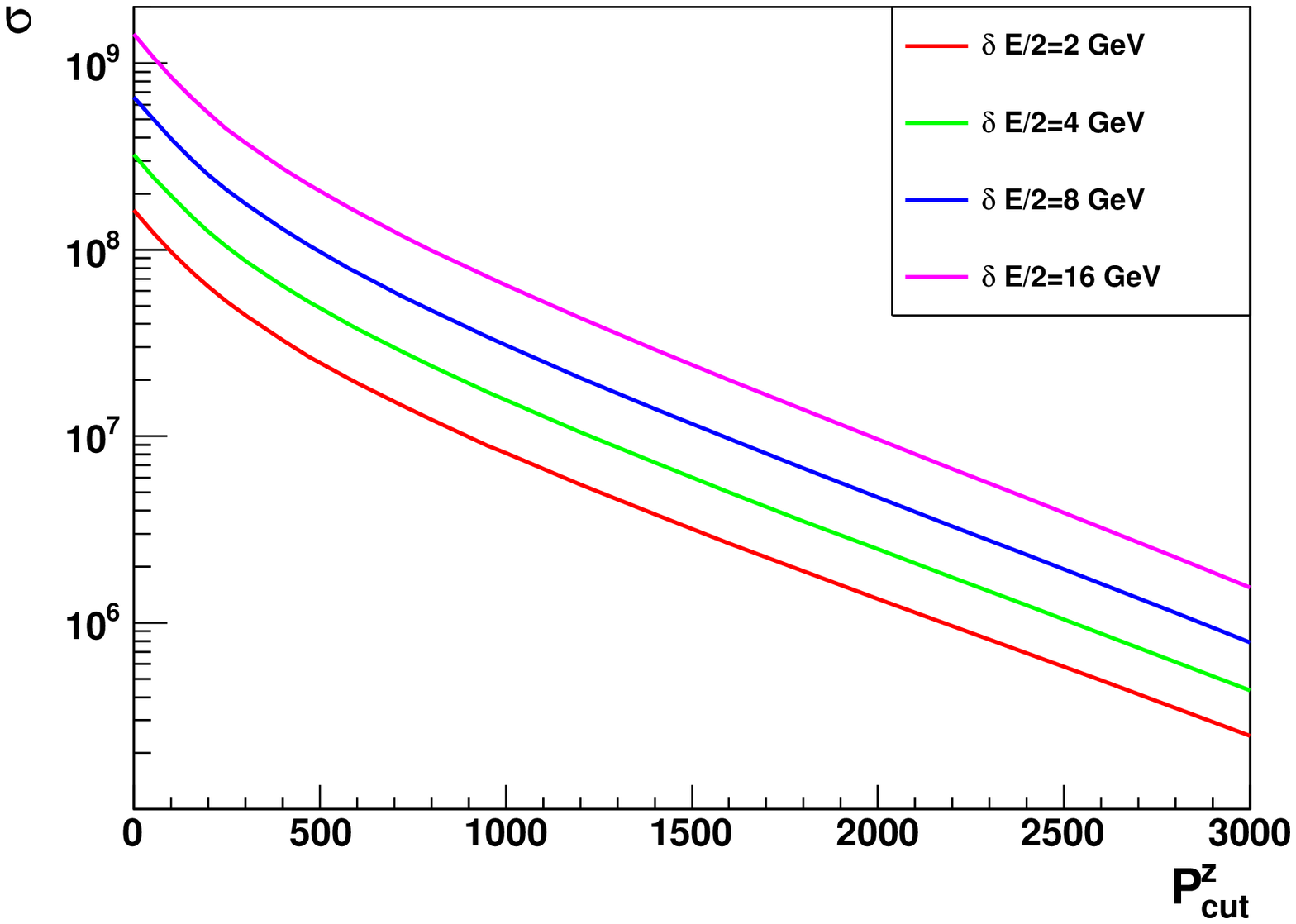}
\includegraphics[height=5cm]
{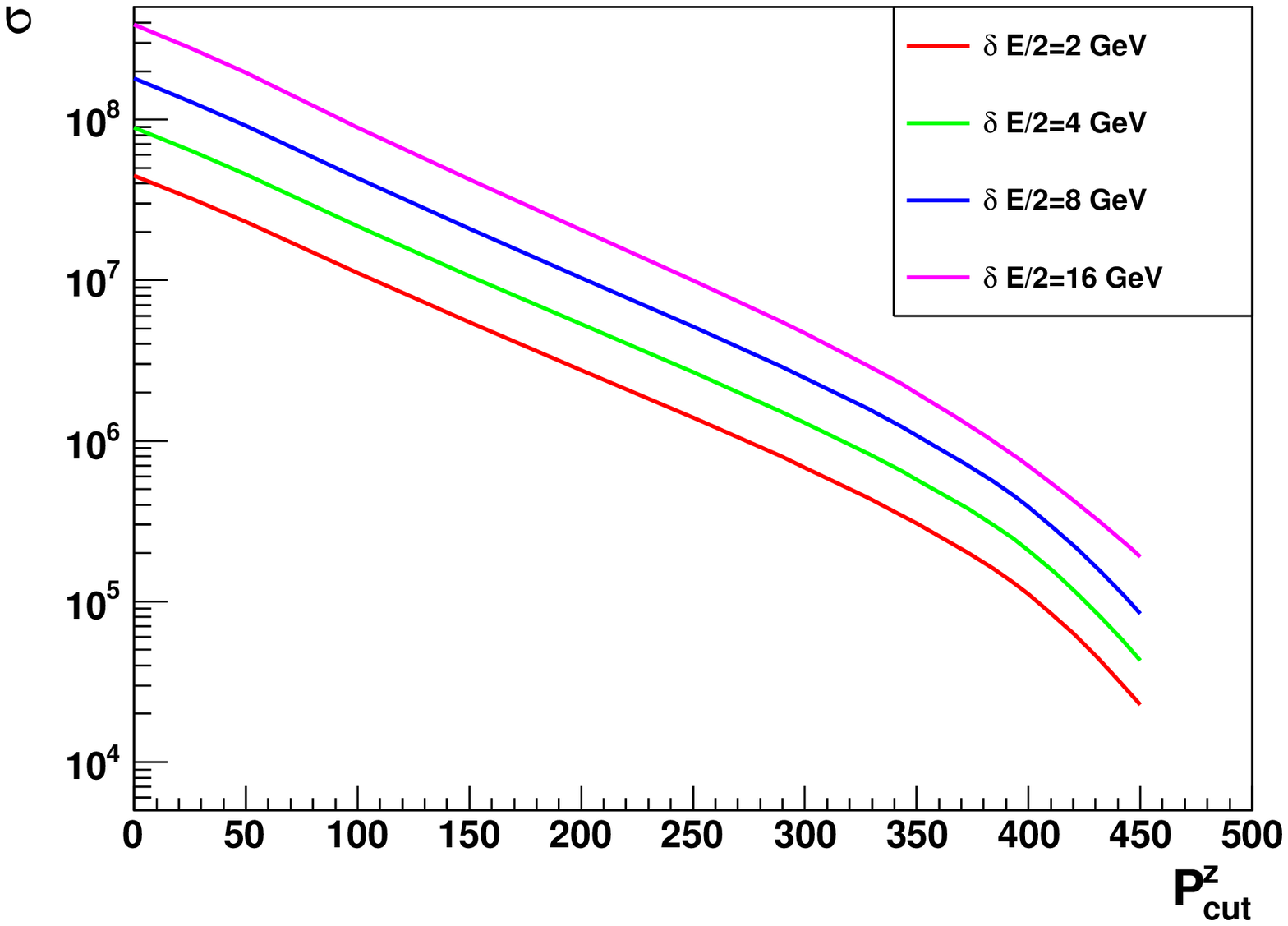}}} \centerline{\hbox{
\includegraphics[height=5cm]
{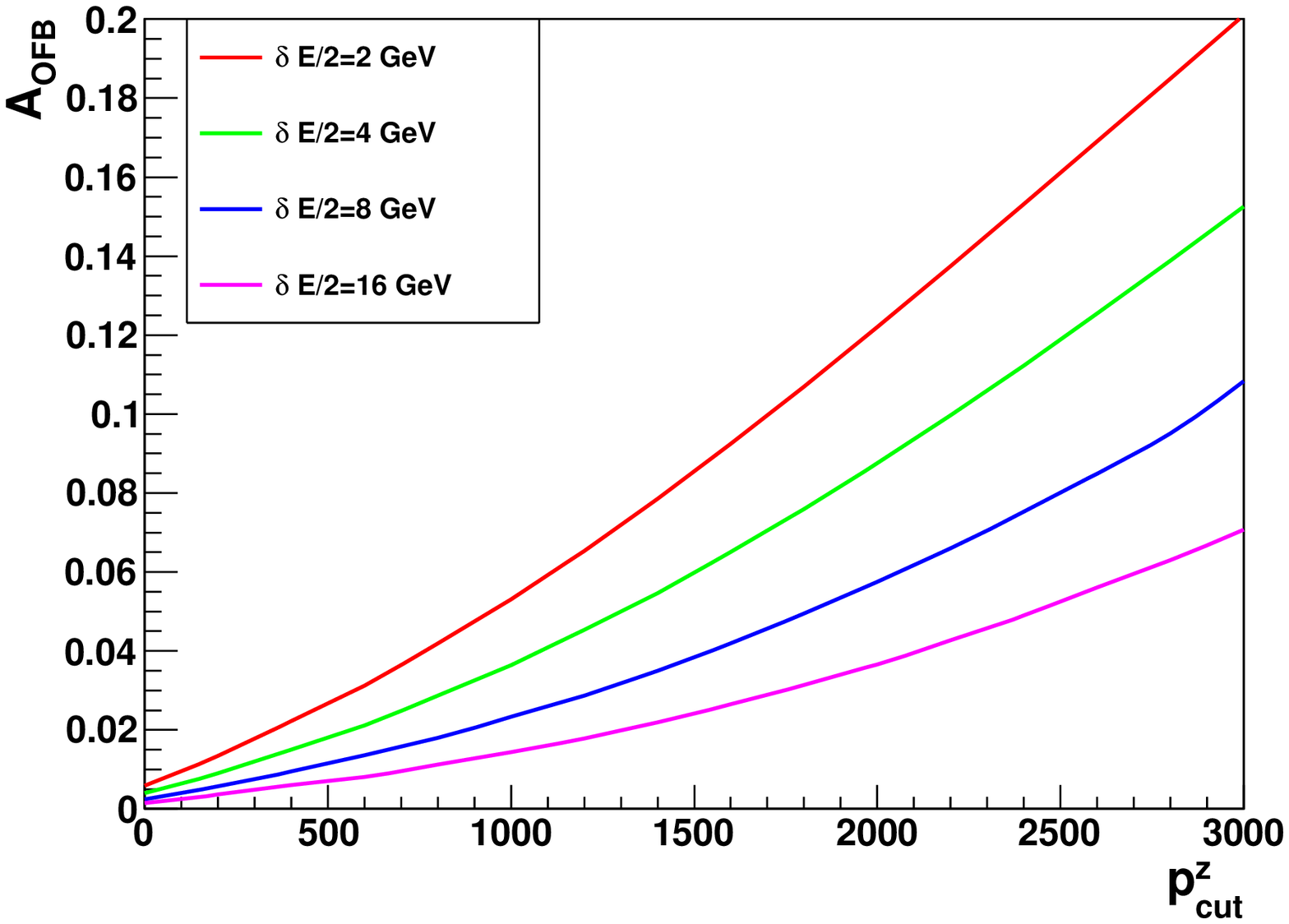}
\includegraphics[height=5cm]
{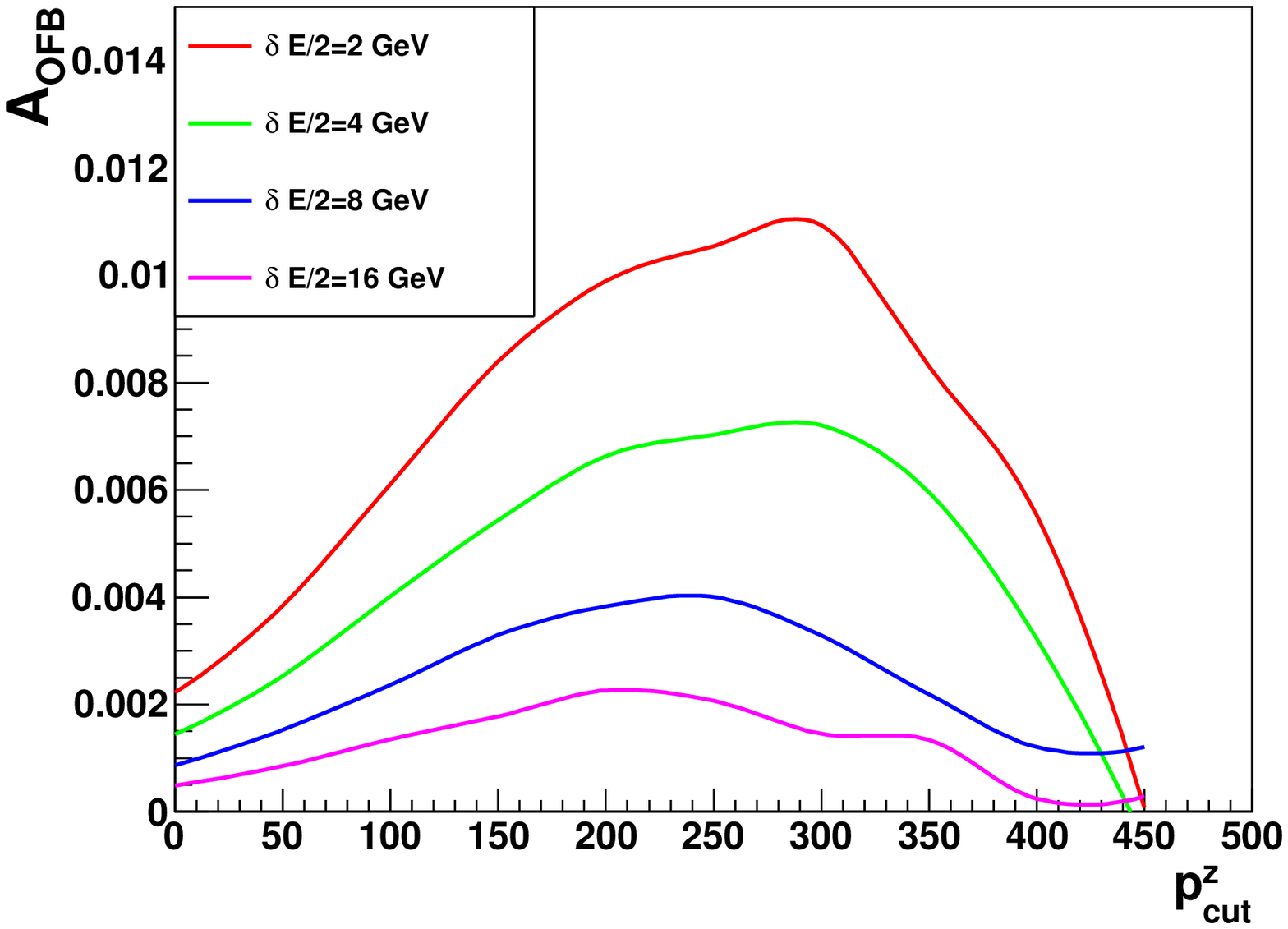}}} \centerline{\hbox{
\includegraphics[height=5cm]
{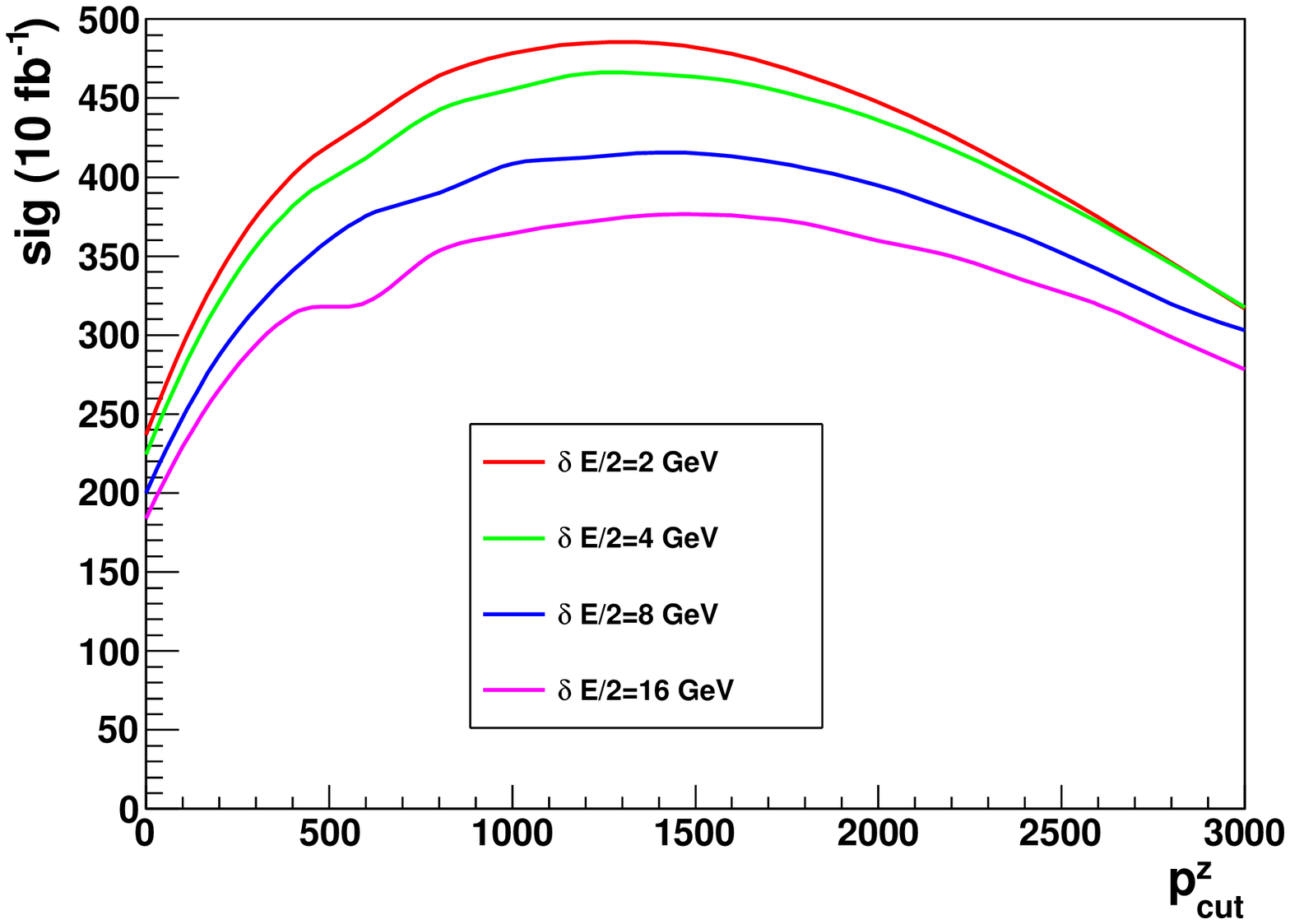}
\includegraphics[height=5cm]
{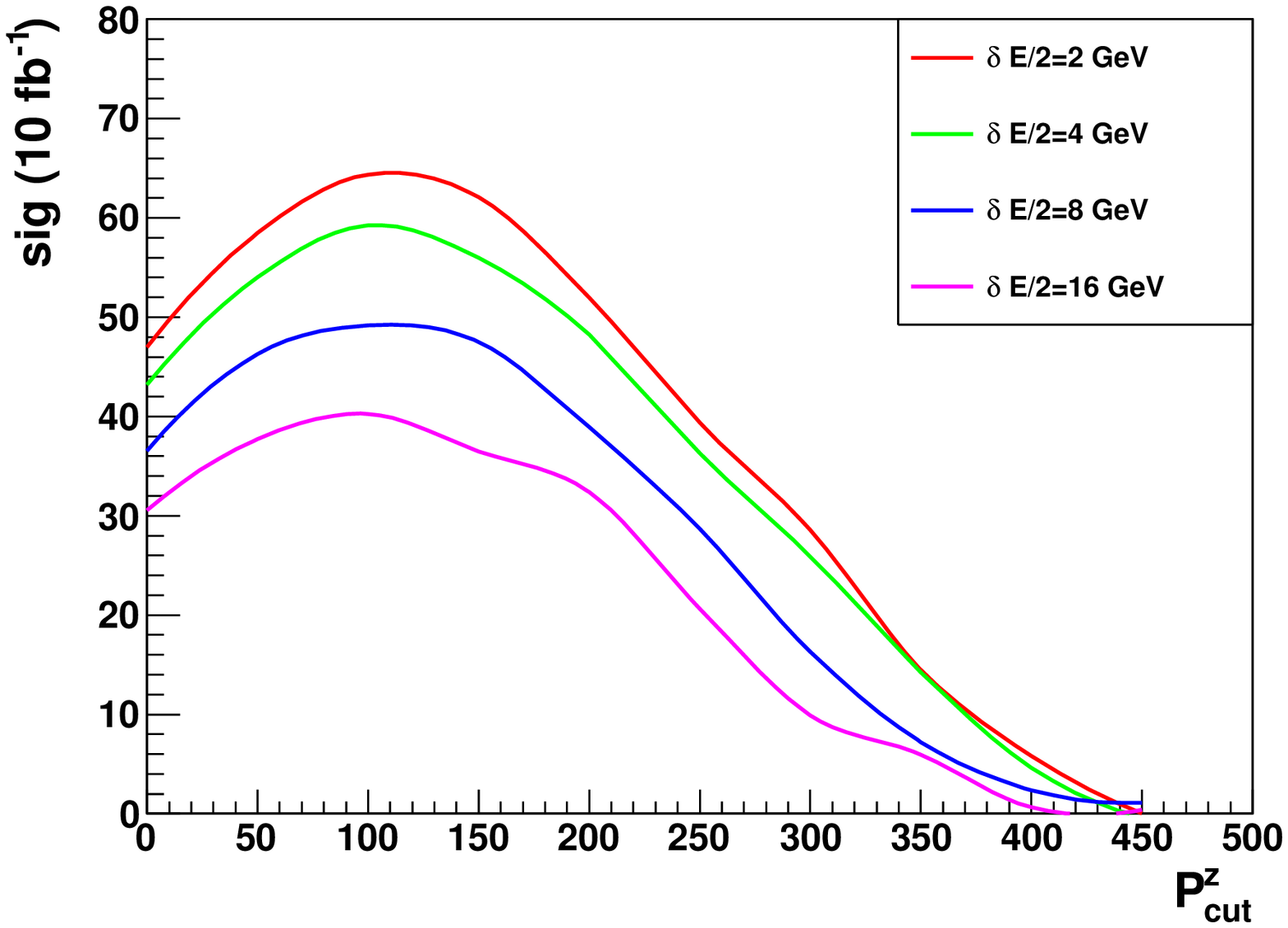}}}
\caption{\label{14TeVbbzpole}Same as Fig. \ref{7TeVbbzpole} except
$\sqrt{s}=14~\mbox{TeV}$.}
\end{figure}

Figure \ref{7TeVbbzpole} shows  $\sigma^A$, $\sigma$, $A_{\rm OFB}$
and $sig$ of the bottom quark as a function of $P_{\rm cut}^z$ at
the LHC with $\sqrt{s}=7~\mbox{TeV}$ with and without $b$ jet cut.
Without b-jet cuts, $A_{\rm OFB}$ will rise with the increase of
$P_{b\bar{b}}^z$, and the behavior is the same with the case in
section \ref{three}. Cuts on b jet will greatly change the
distribution of $A_{\rm OFB}$ and ${\rm sig}$. The figure indicates
that large $P_{b\bar{b}}^z$ events are highly boosted in the
$z$-direction, which means they have small $P_T$ and large $\eta$.
All $\sigma^A$ events with $P_{b\bar{b}}^z>450\mbox{GeV}$ will be
cut off by requiring the $b$ jet $|\eta|<2.4$ and
$P_T>10\mbox{GeV}$. Theoretically, $A_{\rm OFB}$ can be very large
in high $P_{b\bar{b}}^z$ region. However, limited by the coverage of
the real detector, identifying these events is quite challenging.

In Fig. \ref{7TeVbbzpole}, we can see that $\sigma$ is approximately
proportion to $\delta E$ while $\sigma^A$ raise with the increase of
$\delta E$ more slowly. This is because $\sigma$ mainly comes from
QCD contribution, which distributes evenly around the $Z$-pole,
while $\sigma_A$ mainly comes from EW contribution, which
distributes sharply around $Z$-pole. So $A_{\rm OFB}$ decrease with
the increase of $\delta E$. $A_{\rm OFB}$ is harder to be measured
for $\sqrt{s}=14 \mbox{TeV}$ due to the larger longitudinal boost as
shown in Fig. \ref{14TeVbbzpole}.

\begin{figure}[htbp]
\centerline{\hbox{
\includegraphics[height=5cm]
{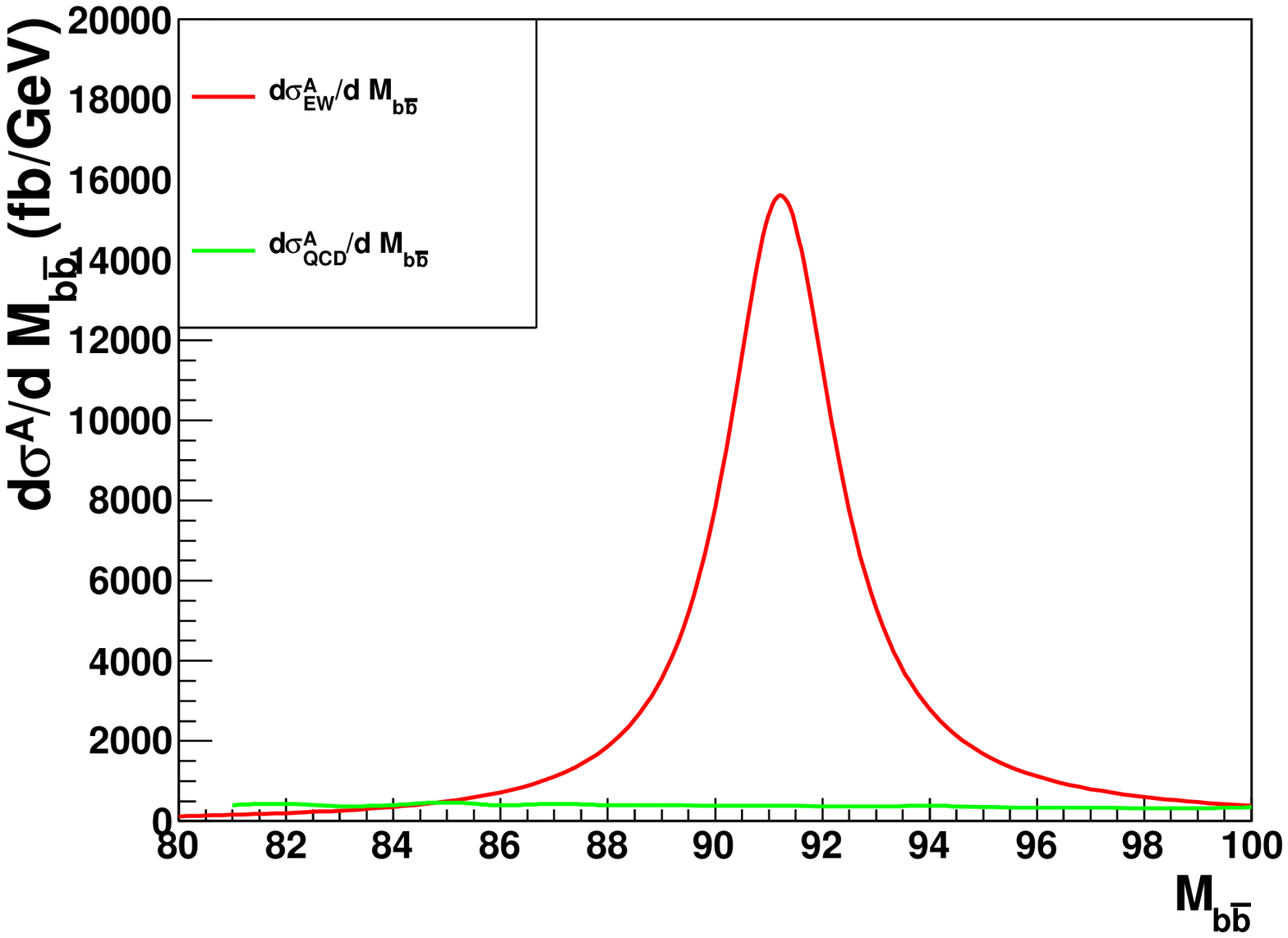}
\includegraphics[height=5cm]
{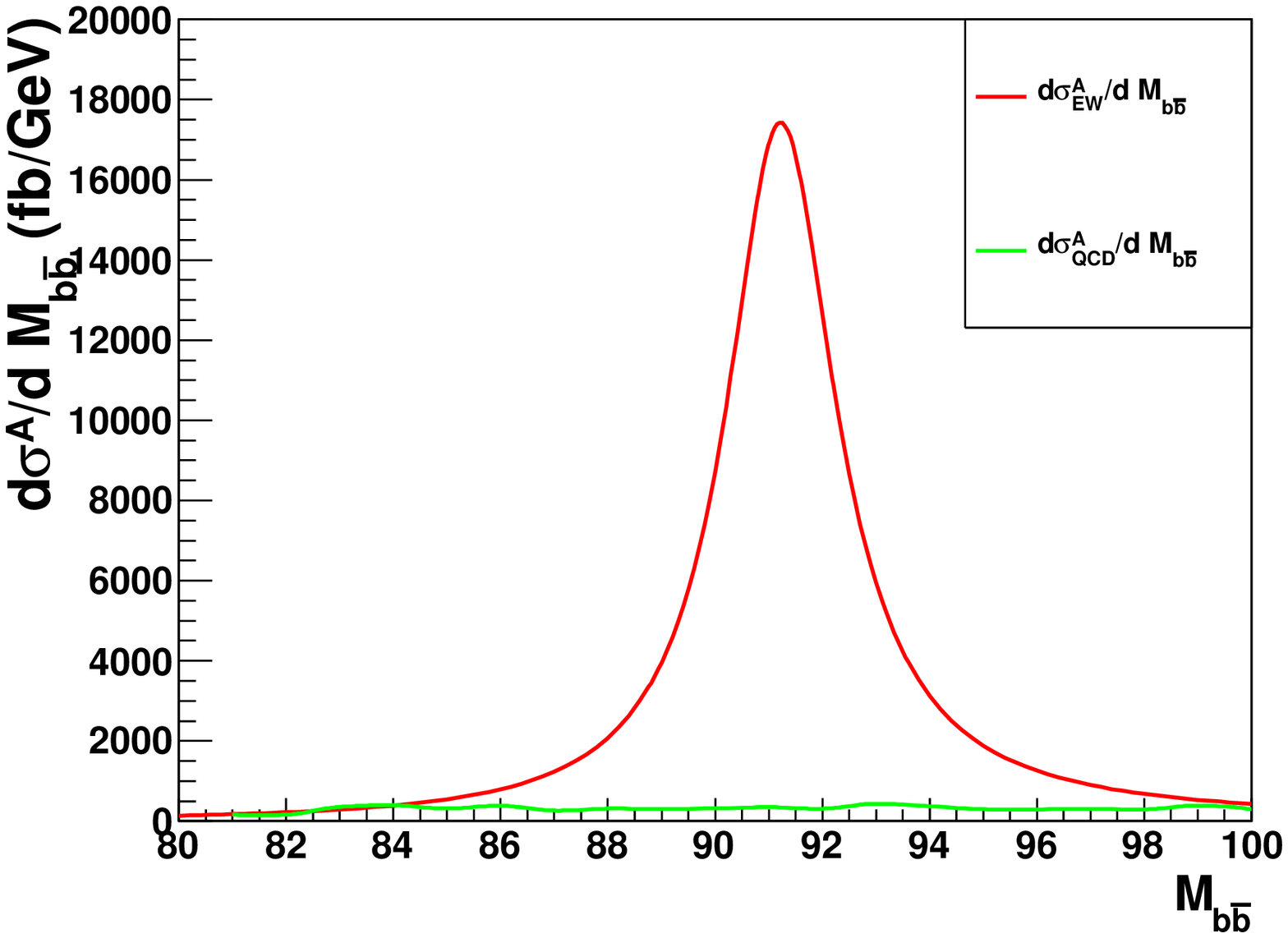}}}
\centerline{\hbox{
\includegraphics[height=5cm]
{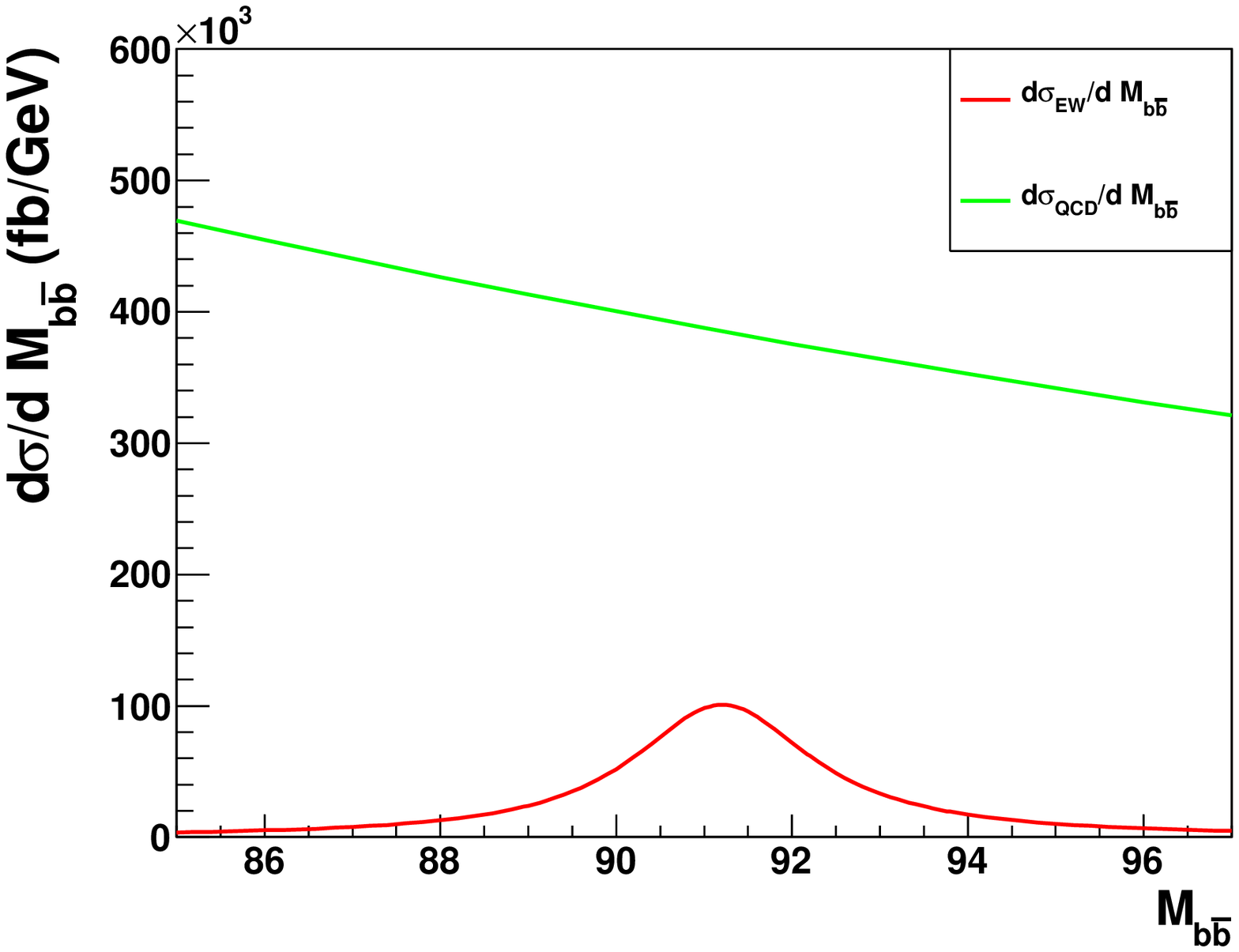}
\includegraphics[height=5cm]
{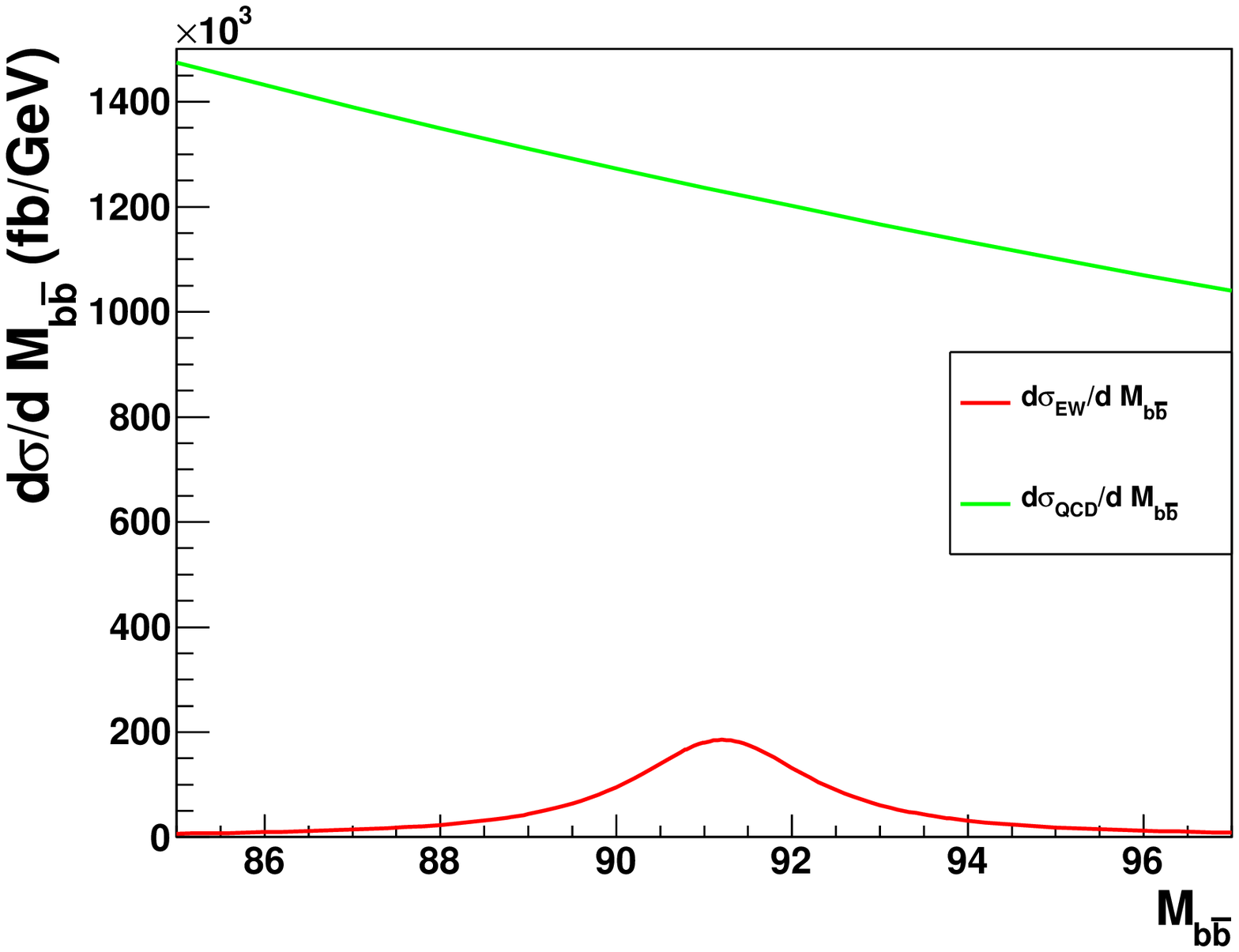}}}
\centerline{\hbox{
\includegraphics[height=5cm]
{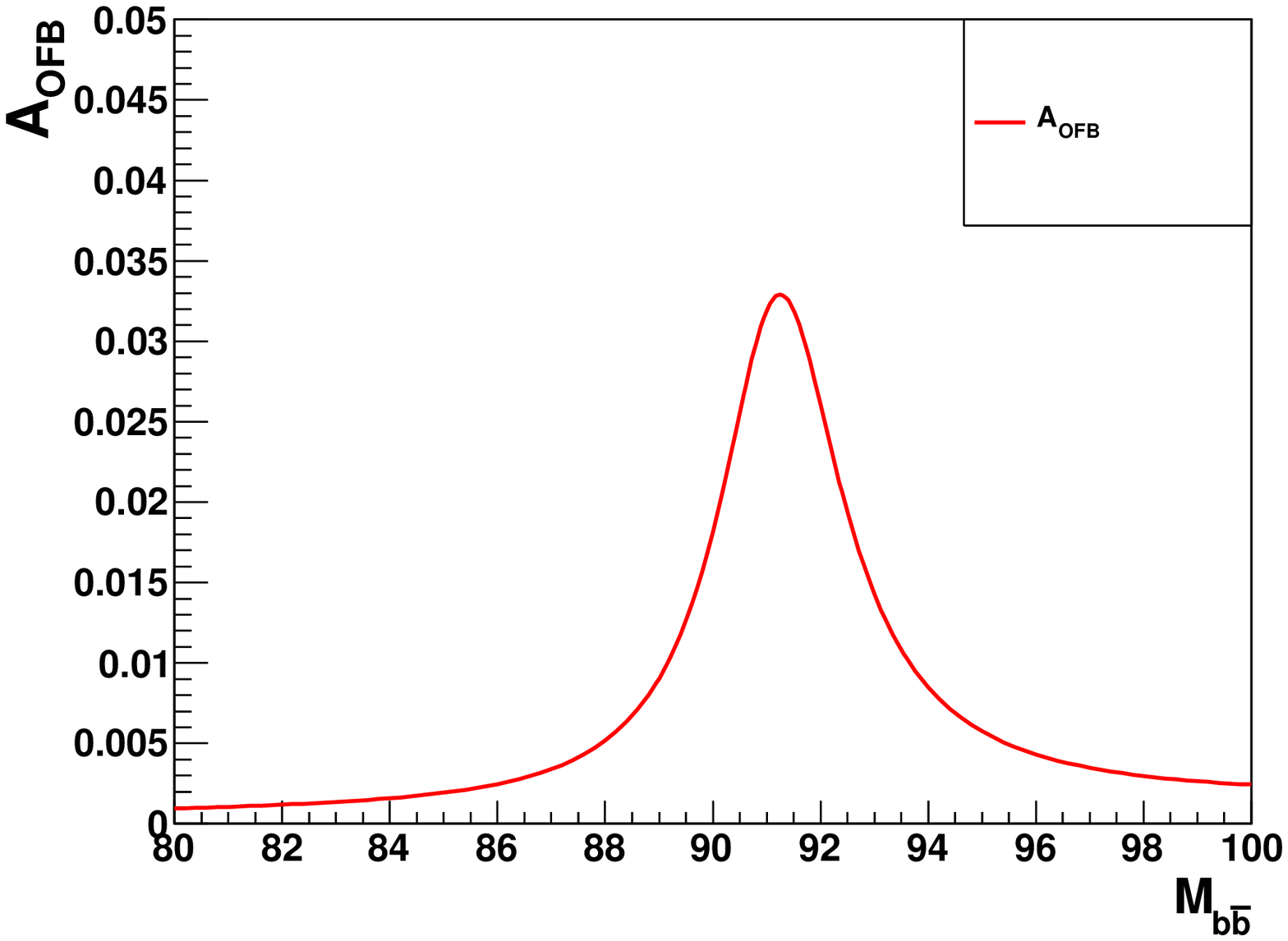}
\includegraphics[height=5cm]
{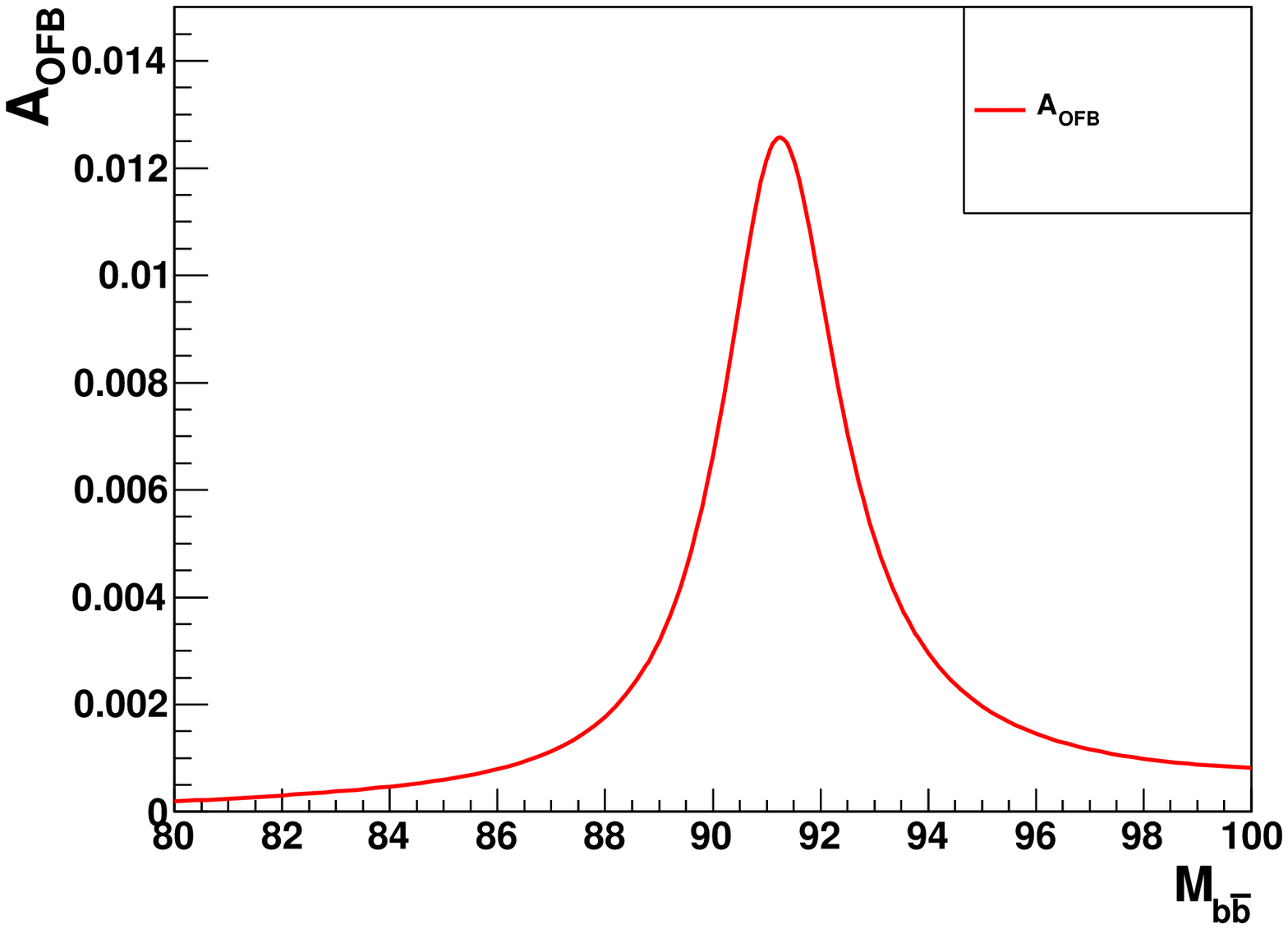}}}
\caption{\label{7TeV14TeVdMbb} Differential distributions of
$\sigma^A, \sigma$ and $A_{\rm OFB}$ with
$\sqrt{s}=7~\mbox{TeV}$(left column) and
$\sqrt{s}=14\mbox{TeV}$(right column). Here $b$ jet cut is applied.
$P_{b\bar{b}}^z=150~\mbox{GeV}$. Here $A_{\rm OFB}$ contains
contribution from both EW and QCD.}
\end{figure}

Figure \ref{7TeV14TeVdMbb} show the differential $\sigma^A$,
$\sigma$ and $A_{\rm OFB}$ as a function of $M_{b\bar{b}}$ at the
LHC with $\sqrt{s}=7$~TeV  and $14$~TeV respectively. Here the $b$
jet cut $|\eta|<2.4$ and $P_T>10\mbox{GeV}$ are applied.
$P_{b\bar{b}}^z=150~\mbox{GeV}$ is the optimal cut which can be seen
in the right lower ${\rm sig}$ plot in Figs. \ref{7TeVbbzpole} and
\ref{14TeVbbzpole}. From Fig. \ref{7TeV14TeVdMbb} we can see that
$\sigma^A$ is dominated by the contribution from EW processes, while
the total cross section $\sigma$  arises mostly from QCD processes.

In order to do the cross-check with the measurements at the LEP, the
clear understanding of the QCD contributions is necessary. This can
be carried out by the measurements in a wide $b\bar{b}$ energy
interval discussed in Sec. \ref{three} in which $A_{\rm OFB}$ is
dominated by the QCD contribution, or in a small $M_{b\bar{b}}$
region far away from the Z-pole.

It is quite interesting to estimate how much integrated luminosity
is needed in order to achieve the similar precision of $A_{\rm
FB}^b$ at the LEP. We only give a rough estimation here and a
precise study needs complicated real detector simulation, which is
beyond the current discussion. As a $e^+e^-$ collider, $A_{\rm
FB}^b$ arises mainly from EW interaction. However, $A_{\rm OFB}^b$
at the LHC have both EW and QCD contributions. Noticing that the EW
and QCD contributions have different distribution shapes as shown in
Fig.~\ref{7TeV14TeVdMbb}, the QCD induced asymmetric and symmetric
events can be removed as a continuous background and pure EW induced
asymmetry can be defined as $A_{\rm EWOFB}^b=N^A_{\rm EW}/N_{\rm
EW}$. According to the error propagation formula, the statistical
fluctuation of $A_{\rm EWOFB}^b$ can be expressed as $\sigma(A_{\rm
EWOFB}^b)=\sqrt{4N^F_{\rm EW} N^B_{\rm EW}/(N^F_{\rm EW}+N^B_{\rm
EW})^3}$, where $N^F_{\rm EW}$/$N^B_{\rm EW}$ are EW induced
forward/backward events. For a comparable precision, it can be
required that the statistical fluctuation of $A_{\rm EWOFB}^b$ at
the LHC should be of the same order $O(0.001)$ as that of $A_{\rm
FB}^b$ at the LEP. By taking the relative data in
Fig.~\ref{7TeV14TeVdMbb} into $\sigma(A_{\rm EWOFB}^b)$, the
effective luminosity is calculated to be $\epsilon{\cal{L}}=9.8~
\mbox{fb}^{-1}$ for 7 TeV and $\epsilon
{\cal{L}}=5.0~\mbox{fb}^{-1}$ for 14 TeV. Here $\epsilon$ is the $b$
quark selecting efficiency.

\section{Conclusions and discussions \label{five}}

Forward backward asymmetry $A_{\rm FB}$ is a tool to study the
nature of couplings, even the quantum structure in the SM and/or
BSM. In the past measurements at the LEP and Tevatron, the
deviations from the SM predictions have inspired extensive studies
in the SM/BSM. However as the proton-proton collider, the LHC does
not have the preferred direction contrary to her counterpart, namely
LEP and Tevatron.  By utilizing the property that the momentum of
valence quark is usually larger than that of sea quark, the
preferred direction at parton level can be kept. For the top pair
production at LHC, we have proposed to apply cut on $z$ direction
momentum of the top quark pair in order to keep the forward backward
asymmetry at partonic level, dubbed as one-side forward backward
asymmetry $A_{\rm OFB}$. In this paper we extend our studies to the
charged leptons and bottom quarks as the final states. Our numerical
results show that at the LHC $A_{\rm OFB}$ can be utilized to study
the nature of the couplings once enough events are collected.

There are some points we should emphasize: (1) Once the preferred
direction at the LHC can be defined, $A_{\rm OFB}$ for any precisely
measured final state particle can be utilized as a tool to study the
structure in the SM and/or BSM; (2) Our studies, especially on the
QCD induced  $A_{\rm OFB}$, indicate that the validity of the
observable $A_{\rm OFB}$ does not depend on whether the higher order
effects is included or not; (3) The backgrounds to the specific
final states are not included in our study, however in the realistic
analysis this issue should be investigated in detail.

{\em Acknowledgements:} SHZ would thank Y.~J. Mao and B. Zhu of CMS
collaboration for the stimulating discussions. YKW would like to
thank Xia Wan for the valuable discussions. This work was supported
in part by the Natural Sciences Foundation of China (No. 10775001,
No. 10635030, and No. 11075003).

\end{document}